\documentclass[12pt]{article}
\usepackage{amsmath,amssymb,epsfig,amsfonts}
\usepackage{graphicx,subfigure}
\usepackage[usenames, dvipsnames]{color}
\usepackage{cite}

\makeatletter

\newcommand{\be}{\begin{equation}}
\newcommand{\ee}{\end{equation}}
\newcommand{\ba}{\begin{aligned}}
\newcommand{\ea}{\end{aligned}}

\def\doty{\cdot_{\tilde{Y}_4}}

\def\unit{{1\kern-.65ex {\rm l}}}
\def\1{{1\kern-.65ex {\rm l}}}

\newcount\hour \newcount\minute
\hour=\time \divide \hour by 60
\minute=\time
\def\now{%
\ifnum \hour<13
  \ifnum \hour=0 \advance \hour by 12 \number\hour:\else \number\hour:\fi%
     \ifnum \minute<10 0\fi%
     \number\minute%
\ A.M.%
\else \advance \hour by -12 \number\hour:%
  \ifnum \minute<10 0\fi%
  \number\minute%
  \ P.M.%
\fi%
}

\makeatother

\begin{document}

\baselineskip=18pt  
\numberwithin{equation}{section}  
\allowdisplaybreaks  



%
%


\thispagestyle{empty}

\vspace*{-2cm}
\begin{center}
{\tt EFI-11-24} \qquad\qquad  
{\tt KCL-MTH-11-15}\qquad \qquad 
{\tt NSF-KITP-11-166}
\end{center}

\vspace*{1.2cm}
\begin{center}
 {\LARGE Yukawas, $G$-flux, and Spectral Covers from \\ Resolved Calabi-Yau's\\}
 \vspace*{1.5cm}
 Joseph Marsano$^1$ and Sakura Sch\"afer-Nameki$^2$\\
 \vspace*{1.0cm}
 
 { \it$^1$ Enrico Fermi Institute, University of Chicago\\}
{ \it 5640 S Ellis Avenue, Chicago, IL 60637 USA\\}
{\tt marsano uchicago.edu \\}
\smallskip

{ \it $^2$ Department of Mathematics, King's College, University of London \\ }
{ \it The Strand, WC2R 2LS, London, UK\\ }
{ \it and \\ }
{ \it Kavli Institute for Theoretical Physics \\ }
{\it University of California, Santa Barbara, CA 93106, USA \\ }

{\tt ss299 theory.caltech.edu\\ }

\vspace*{0.8cm}
\end{center}
\vspace*{.5cm}

\noindent
We use the resolution procedure of Esole and Yau \cite{Esole:2011sm} to study Yukawa couplings, $G$-flux, and the emergence of spectral covers from elliptically fibered Calabi-Yau's with a surface of $A_4$ singularities.  We provide a global description of the Esole-Yau resolution and use it to explicitly compute Chern classes of the resolved 4-fold, proving the conjecture of \cite{Blumenhagen:2009yv} for the Euler character in the process.  We comment on the physical implications of the surprising singular fibers in codimension 2 and 3 in \cite{Esole:2011sm} and emphasize a group theoretic interpretation based on the $A_4$ weight lattice.  We then construct explicit $G$-fluxes by brute force in one of the 6 birationally equivalent Esole-Yau resolutions, quantize them explicitly using our result for the second Chern class, and compute the spectrum and flux-induced 3-brane charges, finding agreement with results and conjectures of local models in all cases.  Finally, we provide a precise description of the spectral divisor formalism in this setting and sharpen the procedure described in \cite{Marsano:2011nn} in order to explicitly demonstrate how the Higgs bundle spectral cover of the local model emerges from the resolved Calabi-Yau geometry.  Along the way, we demonstrate explicitly how the quantization rules for fluxes in the local and global models are related.

\newpage
\setcounter{page}{1} 



\tableofcontents
\newpage

\section{Introduction and Summary}

Some of the most important issues in F-theory model building center on the singularity structure of the compactification geometry.  This includes not only the superpotential, which depends on homology classes of vanishing cycles at the singularities, but also the $G$-fluxes that generate chiral matter and geometrically induced corrections to the flux quantization rule and 3-brane tadpole.
The predominant approach to dealing with these things in the literature is to adopt a somewhat indirect approach.  Subtleties connected to singularities of the Calabi-Yau are encoded in the physics near those singularities, which admits a nice alternative description in terms of the worldvolume theory on a stack of 7-branes \cite{Donagi:2009ra}.  
Global models \cite{Andreas:2009uf,Marsano:2009ym,Collinucci:2009uh,Blumenhagen:2009up,Marsano:2009gv,Blumenhagen:2009yv,Marsano:2009wr,Grimm:2009yu,Cvetic:2010rq,Chen:2010ts,Chen:2010tp,Chung:2010bn,Chen:2010tg,Knapp:2011wk,Knapp:2011ip} are then built via a sort of hybrid procedure wherein one constructs a singular Calabi-Yau, supplements it with some worldvolume data that is supposed to describe the $G$-flux, and appeals to the worldvolume theory to answer any questions related to singularity structure.  Heterotic/F-theory duality \cite{Vafa:1996xn,Morrison:1996na,Morrison:1996pp,Friedman:1997yq,Bershadsky:1996nh,Bershadsky:1997zs,Curio:1998bva,Donagi:2008ca,Hayashi:2008ba,Donagi:2008kj,Hayashi:2009ge,Tatar:2009jk} supports much of the intuition that we gain from the worldvolume perspectve in this way \cite{Donagi:2009ra,Tatar:2009jk,Marsano:2009gv,Hayashi:2009bt} and even suggests that spectral cover methods in the worldvolume theory can capture contributions from the singularities to the geometrically induced 3-brane charge \cite{Blumenhagen:2009yv}.   

This approach to building global models, though, is only as useful as our ability to match worldvolume data with bulk data of the compactification.  For local geometric data the dictionary is well-known \cite{Donagi:2009ra} but things are more intricate when it comes to $G$-flux.  Worldvolume physics depends only on a suitable `local flux' that must be obtained, in a proper global model, from a well-defined global flux.  Understanding this relationship requires one to first develop a framework for describing global flux in a useful way.  The type of flux that affects worldvolume physics is a $(2,2)$-form that integrates nontrivially over holomorphic surfaces containing vanishing cycles.  In principle we should be able to construct such a $(2,2)$-form by specifying its Poincare dual holomorphic surface in a Calabi-Yau resolution $\tilde{Y}_4$ of the singular Calabi-Yau $Y_4$.   

Explicitly describing $G$-fluxes, however, can be problematic for many reasons.  Firstly, $G$-fluxes in F-theory compactifications must be orthogonal to all surfaces that are pulled back from horizontal or vertical surfaces in $Y_4${\footnote{By horizontal and vertical we mean surfaces that sit inside the section or that consist of an elliptic fibration over a curve in $B_3$.}}.  This often means that the projection of $G$ to $Y_4$ is homologically trivial, though recent work has emphasized the possibility of using cycles that are homologically nontrivial even in $Y_4$ \cite{Braun:2011zm}.  An even more serious problem, however, is describing how to build a surface $G$ that has the right intersections with vanishing cycles.  For this we must know something about the resolution.  A framework for doing this without explicitly resolving $Y_4$ was proposed in \cite{Marsano:2010ix} and refined in \cite{Marsano:2011nn} based on a type of global object referred to as a `spectral divisor', ${\cal{D}}_{\text{spectral}}$.  The logic behind this formalism is to take advantage of the fact that the local geometry of $Y_4$ near $S_2$ is equivalent to that of a $dP_9$ fibration over $S_2$ whose generic fiber exhibits an $A_4$ singularity.  Intersection theory on $dP_9$ allows us to say something about how certain divisors and surfaces in the $dP_9$ fibration interact with vanishing cycles and the spectral divisor ${\cal{D}}_{\text{spectral}}$ is introduced to encapsulate that knowledge in a useful way.  This allows us to say quite a lot about $G$-fluxes and their impact on the physics without ever having to introduce a specific resolution of $Y_4$.

This situation may not completely satisfactory, however, as many things are not as sharp as one would like.  We have a good idea for how $G$-fluxes constructed from spectral divisors are related to data of the worldvolume theory by looking at a sort of ``limiting behavior" of the spectral divisor but, in the singular geometry, this is not particularly rigorously defined \cite{Marsano:2011nn}.  We also have a  good idea for how quantization conditions of the two approaches should be related \cite{Andreas:1999ng,Marsano:2011nn} but a completely explicit demonstration of this is lacking.  Heterotic/F-theory duality gives credence to all these things but to really sharpen them and gain complete confidence in the spectral divisor formalism, we must stop dancing around the core issue: we need to properly resolve the singular Calabi-Yau 4-fold $Y_4$.

\subsection{Esole-Yau Resolution and Topological Computations}

Fortunately, Esole and Yau recently wrote a beautiful paper \cite{Esole:2011sm} which outlines a specific procedure for explicitly resolving generic Calabi-Yau 4-folds $Y_4$ that exhibit a surface of $A_4$ singularities above a divisor $S_2$ inside the base $B_3$.  The procedure relies on a standard realization of $Y_4$ as a hypersurface inside a $\mathbb{P}^2$-bundle $X_5$ and proceeds via a series of 4 relatively simple blow-ups in $X_5$ that produce a new 5-fold $\tilde{X}_5$.  The proper transform $\tilde{Y}_4$ of $Y_4$ under these blow-ups is a smooth Calabi-Yau hypersurface of $\tilde{X}_5$ that is precisely the resolution we seek.  Esole and Yau actually find six different combinations of blow-ups that one can do in $\tilde{X}_5$ which give rise to six different resolutions that are connected by flop transitions.  For much of this paper, we will focus our attention on just one of these resolutions for simplicity.

One remarkable application of the Esole-Yau resolution is to the computation of topological data of the resolved Calabi-Yau 4-fold $\tilde{Y}_4$.  Previously, the only complete resolutions in the literature relied on the realization of the singular 4-fold as a complete intersection in some ambient toric space \cite{Blumenhagen:2009yv,Grimm:2009yu,Knapp:2011wk,Knapp:2011ip}.  The resolution could be performed by blowing up the ambient space and topological data of the resolved 4-fold could be studied effectively because of the vast reservoir of knowledge about the geometry of toric varieties.  The Esole-Yau resolution, by contrast, does not require the presence of an ambient toric space so it can be applied quite generally, including to 4-folds built on base 3-folds $B_3$ like the one of \cite{Marsano:2009ym} that does not sit nicely inside some toric variety.  Even though the ambient space is not toric, we still know a lot about it because it is very simple: it is just a $\mathbb{P}^2$-bundle over $B_3$ that has been blown up four times.  Further, the resolved 4-fold is always a hypersurface in this space, so determining topological invariants is in principle very easy.  We are only limited by things that we might not know about the structure of $\tilde{X}_5$.

Our first order of business in this paper is to promote the local description of the resolution of \cite{Esole:2011sm} to a global one.  We emphasize that a factorization structure crucial to resolving codimension 2 and 3 singularities in $B_3$ extends globally beyond the coordinate patch studied in \cite{Esole:2011sm} to the level of holomorphic sections.  This allows a very simple global description of the blow-ups in $X_5$ that are needed to fully resolve $Y_4$ in terms of global holomorphic sections rather than local coordinates.  With this description, we explicitly compute both the Euler character and the second Chern class of the resolved 4-fold $\tilde{Y}_4$.  Our result for the Euler character indicates a shift compared to that of smooth Calabi-Yau hypersurfaces of $X_5$.  This shift precisely agrees with a conjecture of \cite{Blumenhagen:2009yv}, thereby proving it in full generality by direct computation{\footnote{Of course the same assumptions as in \cite{Blumenhagen:2009yv} apply.  The computation assumes a generic $Y_4$ with a surface of $A_4$ singularities and the result is expected to be modified when $Y_4$ is nongeneric.  Examples of nongeneric $Y_4$'s are those that engineer extra $U(1)$ symmetries \cite{Marsano:2009gv,Blumenhagen:2009yv,Marsano:2009wr,Grimm:2009yu,Dolan:2011iu} or additional non-Abelian gauge groups.  Subtleties associated with the engineering of extra $U(1)$'s have been discussed in \cite{Hayashi:2010zp,Grimm:2010ez,Marsano:2010ix,Marsano:2011nn}.}}.  Our result for the second Chern class enters crucially in the quantization of $G$-flux, which we discuss momentarily.

\subsection{Yukawa couplings and codimension 3 singularities}

Before getting to $G$-fluxes, however, we would like to clarify some issues related to the physics of singularities that sit above points of codimension 3 in $B_3$.  Esole and Yau found that the singular fibers behaved a bit differently than people have naively assumed in the physics literature.  This has led to some speculation that there might be problems with the generation of top Yukawa couplings in F-theory GUT models.

To address this, we can recall a few simple properties of the Esole-Yau resolution.  Firstly, the fact that it took precisely 4 blow-ups in $X_5$ to resolve $Y_4$ is completely intuitive.  As a well-behaved smooth hypersurface, the divisors of $\tilde{Y}_4$ are inherited from those of the blown-up space $\tilde{X}_5$.  Each blow-up of $X_5$ introduces a new exceptional divisor so in total the resolution adds 4 new divisors to $\tilde{Y}_4$.  Of course we should have expected this from the outset; resolving the $A_4$ singularity above $S_2$ yields 4 divisors 
obtained by fibering a $\mathbb{P}^1$ from the $A_4$ fiber over $S_2$.
We denote these by ${\cal{D}}_{-\alpha_i}$ following the standard notation $\alpha_i$ for the roots of $A_4$ and refer to them as the \emph{Cartan divisors} since their dual $(1,1)$-forms give rise to the $U(1)$ Cartan generators of $SU(5)$.

Since we get exactly 4 new divisors, we should expect on general grounds to obtain exactly 4 new holomorphic curve classes.  These curves are orthogonal to all divisors pulled back from $Y_4$ so they are uniquely determined by their intersections with the ${\cal{D}}_{-\alpha_i}$.  This is physically sensible since we assign to each degenerate 2-cycle a physical state from a wrapped M2 brane and that state is determined by the 2-cycle's homology class.  The statement that the homology class is specified by intersections with the ${\cal{D}}_{-\alpha_i}$ is just a geometric way of saying that the state is determined by its $U(1)$ Cartan charges.

This type of reasoning leads to a small puzzle.  Above generic points on $S_2$, the singularity type is $A_4$ with its four new holomorphic curves.  Above loci of higher codimension in $B_3$, however, the singularity type is known to enhance and the singular fiber is expected to have more than 4 components.  In codimension 2 we get the ``$D_5$" and ``$A_5$" enhancements above ``matter curves" where $\mathbf{10}$ and $\mathbf{\overline{5}}$ matter fields are expected to localize, respectively.  In codimension 3, we get the ``$E_6$" and ``$D_6$" enhancements above points where matter curves intersect.  One naively pictures each enhancement as the addition of a new component to the singular fiber and interprets that component as a root of the indicated higher rank gauge group.
A completely new holomorphic curve class cannot appear out of thin air, though; we know that the net number of new curve classes is exactly four.  What is really happening is that one or more of the $A_4$ roots splits into multiple components.  The homology classes of those components, or equivalently their $U(1)$ charges, correspond to weights of states in the $\mathbf{10}$ or $\mathbf{\overline{5}}$ representation.  On a group theoretic level, this means that we should not think about some larger root lattice but instead in terms of the $SU(5)$ weight lattice.  It is this lattice that corresponds to the new elements of $H_2(\tilde{Y}_4,\mathbb{Z})$ that emerge after the resolution.  Above generic points in $S_2$, the simple roots generate the cone of effective curves in the fiber.  When the singularity type worsens, some of the weights become effective and the lattice of effective curves in the fiber changes.

With this perspective, let us now ask what we should expect of the singular fibers in codimension 2 and 3 in order to agree with our physical expectations.  
Above the $\mathbf{10}$ ($\mathbf{\overline{5}}$) matter curve where the singularity type is supposed to enhance to ``$D_5$'' (``$A_5$''), we expect that some curves with $\mathbf{10}$ ($\mathbf{\overline{5}}$) weights become effective.  Given our global description of $\tilde{Y}_4$, it is a simple matter to compute the Cartan charges of the effective curves in the singular fibers wherever we like.  In the case of codimension 2 singularities, our approach is similar to that of \cite{Intriligator:1997pq} and the behavior that we ultimately see bears a strong resemblance to the examples studied recently in \cite{Morrison:2011mb}.

Upon specializing to one of the Esole-Yau resolutions, we find that two of the $SU(5)$ roots, $-\alpha_2$ and $-\alpha_4$, split above the $\mathbf{10}$ matter curve as
\begin{equation}\begin{split}(1,-2,1,0)&\rightarrow (1,-1,1,-1)+(0,-1,0,1) \\
-\alpha_2&\rightarrow \left[-\mu_{10}+\alpha_1+\alpha_2+\alpha_3\right] + \left[\mu_{10}-\alpha_1-2\alpha_2-\alpha_3\right]\\
\end{split}\end{equation}
and
\begin{equation}\begin{split}
(0,0,1,-2)&\rightarrow (1,-1,1,-1)+(1,0,0,-1)+(-2,1,0,0) \\
-\alpha_4&\rightarrow  \left[-\mu_{10}+\alpha_1+\alpha_2+\alpha_3\right] + \left[\mu_{10}-\alpha_2-\alpha_3-\alpha_4\right]+\left[-\alpha_1\right]
\end{split}\end{equation}
where we identify states both by their Cartan charges as well as the conventional notation $\alpha_m$ for $SU(5)$ roots and $\mu_{10}$ for the highest weight of the $\mathbf{10}$.

Note that the net number of distinct curves, counting the roots $-\alpha_1,-\alpha_3$ and the extended node $-\alpha_0$, is 6 as opposed to 5 for the $A_4$ fiber.  In this sense, the singular fiber appears to enhance in rank.  Note also that some curves appear more than once so that the fiber is not expected to be of a standard Kodaira type.  Above the $\mathbf{\overline{5}}$ matter curve we find something similar in that the root $-\alpha_2$ splits as
\begin{equation}\begin{split}(1,-2,1,0)&\rightarrow (0,-1,1,0)+(1,-1,0,0) \\
-\alpha_2&\rightarrow \left[\mu_5-\alpha_1-\alpha_2\right] + \left[-\mu_5 + \alpha_1\right]\end{split}\end{equation}
where $\mu_5$ is the highest weight of the $\mathbf{5}$ representation.  Again the number of distinct curves is 6 so the fiber appears to have enhanced in rank.  All components appear exactly once so the fiber appears standard here.

We now turn to Yukawa couplings from codimension 3 singularities. Here let us emphasize the results  related to `$E_6$' points, whose singular fibers appeared most mysterious in \cite{Esole:2011sm}.
We expect on physical grounds, both from Heterotic/F-theory duality and the worldvolume gauge theory description, to obtain $\mathbf{10}\times\mathbf{10}\times\mathbf{5}+\text{cc}$ Yukawa couplings from these points.  To be consistent with this, we expect that a pair of $\mathbf{10}$ weights can be connected to a $\mathbf{5}$ weight by a 3-chain that degenerates above the `$E_6$' point.  To study the structure of the singular fiber above an `$E_6$' point we can look at what happens as we approach it along the $\mathbf{10}$ matter curve.  When we finally reach the `$E_6$' point, one of the weights splits further
\begin{equation}\begin{split}(1,-1,1,-1)&\rightarrow (1,0,0,-1)+(0-1,1,0) \\
\left[-\mu_{10}+\alpha_1+\alpha_2+\alpha_3\right] &\rightarrow \left[\mu_{10}-\alpha_2-\alpha_3-\alpha_4\right] + \left[\mu_5-\alpha_1-\alpha_2-\alpha_3\right]
\end{split}\end{equation}
Note that the new $\mathbf{10}$ weight is just a second copy of one that we already had above the $\mathbf{10}$ matter curve.  The splitting therefore changes the classes of the components of the singular fiber but does not change the number of distinct classes that we find.  In that sense, the `rank' of the fiber does not change.

Despite this lack of any `rank enhancement', it is easy to see explicitly that a Yukawa coupling can be generated.  For this, let us try to approach the `$E_6$' point instead from along the $\mathbf{\overline{5}}$ matter curve.  The splitting of weights here is more complicated and we focus only on what the $\mathbf{\overline{5}}$ weight does
\begin{equation}\begin{split}(1,-1,0,0) & \rightarrow (1,0,0,-1)+(0,-1,0,1) \\
-\mu_5+\alpha_1 &\rightarrow \left[\mu_{10}-\alpha_2-\alpha_3-\alpha_4\right] + \left[\mu_{10}-\alpha_1-2\alpha_2-\alpha_3\right]
\end{split}\end{equation}
Because the $\mathbf{\overline{5}}$ weight splits into a pair of $\mathbf{10}$ weights above the `$E_6$' point, we can construct a 3-chain connecting an M2 in the $\mathbf{5}$ wrapping $-(1,-1,0,0)\ ${\footnote{In other words, the M2 wraps $(1,-1,0,0)$ with opposite orientation.}} above the $\mathbf{\overline{5}}$ matter curve with a pair of $\mathbf{10}$'s wrapping $(1,0,0,-1)$ and $(0,-1,0,1)$ above the $\mathbf{10}$ matter curve.  This 3-chain passes through one of the `$E_6$' points and it completely degenerates when the M2's all move to one single `$E_6$' point.  This is precisely the behavior we need to be consistent with generation of a Yukawa coupling.

Note that we have not  said anything in this discussion about intersections between pairs of 2-cycles that sit in singular fibers above codimension 2 and codimension 3 singularities.  The number of homologically distinct 2-cycles in these fibers, which is a number that one might naively call the `rank' of the fiber, has also not played any role even though we mentioned it in passing for comparison with \cite{Esole:2011sm}.  Neither of these things matter for physics.  The reason we talk about the intersection matrix and rank of the fiber in codimension 1 is because of a rather nice coincidence special to that situation.  There each curve $\Sigma$ also specifies a divisor of $\tilde{Y}_4$ (in fact a Cartan divisor) obtained by fibering $\Sigma$ over the codimension 1 singular locus.  The number of components of the singular fiber therefore tells us both the number of new divisors, each giving rise to a $U(1)$, and the number of new holomorphic curves, each giving rise to a wrapped M2-brane state.  Intersections of curves in the singular fiber carry a global meaning: they compute intersections of the curves in the fiber with the Cartan divisors.  They also carry a physical meaning: the intersection matrix of the singular fiber specifies the $U(1)$ charges of the wrapped M2-brane states.  

For fibers above codimension 2 and codimension 3 loci where the singularity type enhances, the intersections that are globally meaningful and physically relevant are intersections between curves in those fibers and the Cartan divisors constructed from the $SU(5)$ roots.  Intersections between distinct curves in the enhanced singular fiber do not carry a global or physical meaning and neither does the number of distinct curve classes that happen to sit in that fiber.  Above $E_6$ points, for instance, we saw a group theoretic understanding for the lack of ``rank enhancement" and, furthermore, that this did not have any effect on the generation of a top type Yukawa coupling.

We believe that a focus on root lattices of larger groups as opposed to the weight lattice of $SU(5)$ is responsible for many of the naive expectations that have been expressed in the literature concerning the nature of the singular fibers in codimension 3.  Thinking in terms of weight lattices makes the very interesting and intricate behavior seen by Esole and Yau completely intuitive and in line with what we expect from group theory.  The different Esole-Yau resolutions should correspond to the different inequivalent ways that $SU(5)$ roots can split into weight vectors above various codimension 2 and 3 loci in a way that is consistent between them.  We further expect that the $D_6$ symmetry group that permutes the Esole-Yau resolutions should admit a pure group theoretic explanation along these lines.

\subsection{$G$-fluxes and spectral covers}

Perhaps the most interesting application of the resolution procedure of \cite{Esole:2011sm} is that it allows a very direct construction of $G$-fluxes.  For simplicity, we focus on one particular resolution of \cite{Esole:2011sm} though the generalization to the rest should be completely straightforward.  Once we have a resolved geometry $\tilde{Y}_4$, we can understand the set of allowed $G$-fluxes by simply exploring the set of holomorphic surfaces.  Restricting our attention to a nice class of surfaces that descend from 3-folds in $\tilde{X}_5\ ${\footnote{It would be interesting to study surfaces that do not arise in this way as advocated in \cite{Braun:2011zm}.}}, we find a 1-parameter family of fluxes that satisfy all of the conditions we require of $G$-flux in F-theory and do not break $SU(5)$.  We use our explicit computation of $c_2(\tilde{Y}_4)$ to quantize this flux and compute the chiral spectrum and flux-induced 3-brane charge.  For the chiral spectrum, we find exactly the result of \cite{Donagi:2009ra} for local models corresponding to generic 4-folds with $A_4$ singuarities.  For the flux-induced 3-brane charge, we reproduce exactly the expression from local models that has been conjectured to capture this quantity \cite{Blumenhagen:2009yv}.

Given these successes, we finally come back to the idea of connecting global $G$-flux directly with the spectral cover construction of the 7-brane worldvolume theory.  As we mentioned earlier, the spectral divisor formalism \cite{Marsano:2010ix,Marsano:2011nn} provides a natural way to do this.  A rough idea for how the connection can work was described in \cite{Marsano:2011nn} and, with an explicit resolution $\tilde{Y}_4$ in hand, we are in a position to sharpen that proposal and test it. 

The hallmark feature of the spectral divisor ${\cal{D}}_{\text{spectral}}$ is that it behaves, near the surface of $A_4$ singularities, like a sum of 5 exceptional lines in $dP_9$.  These exceptional lines are distinguished in that they intersect the $A_4$ roots in a way specified by the Cartan charges of a $\mathbf{10}$ representation.  By studying the proper transform of ${\cal{D}}_{\text{spectral}}$ in $\tilde{Y}_4$ we can in fact verify that the intersection of its 5 sheets with the Cartan roots is consistent with the highest weight of the $\mathbf{10}$ in that each intersects only one $A_4$ root, $-\alpha_2$ with intersection number 1{\footnote{The structure of ${\cal{D}}_{\text{spectral}}$ is somewhat more subtle above points where the singularity type is enhanced.}}.  The `limiting behavior' of ${\cal{D}}_{\text{spectral}}$ near the $A_4$ singularities that was advocated in \cite{Marsano:2011nn} as a way to see the Higgs bundle spectral cover now lifts to something very well-defined in $\tilde{Y}_4$.  We simply intersect the proper transform of ${\cal{D}}_{\text{spectral}}$ with the Cartan divisor ${\cal{D}}_{-\alpha_2}$ corresponding to the root $-\alpha_2$.  The result is a 5-sheeted covering of $S_2$ that is exactly the Higgs bundle spectral cover ${\cal{C}}_{\text{Higgs,loc}}$ of the 7-brane worldvolume theory.

As described in detail in \cite{Marsano:2011nn}, the spectral divisor formalism allows us to describe a certain set of holomorphic surfaces in $\tilde{Y}_4$ in terms of surfaces that sit inside ${\cal{D}}_{\text{spectral}}$ in the singular $Y_4$.  The basic procedure for connecting the two 
is to start with a surface $S_1$ inside ${\cal{D}}_{\text{spectral}}$ and form a linear combination $S_1-S_0$ for some surface $S_0$ that is homologous to $S_1$ inside $Y_4$ but does not sit inside ${\cal{D}}_{\text{spectral}}$.  One then lifts this to $\tilde{Y}_4$ by taking the proper transform of $S_1$ but the total transform of $S_0$.  The result is a nontrivial surface in $\tilde{Y}_4$ whose nontriviality is essentially localized in the resolved singular fibers.  Intersections of these surfaces with one another and with the matter surfaces only depends on what $S_1$ is doing near singular fibers.  Since $S_1$ sits inside ${\cal{D}}_{\text{spectral}}$, though, all of this information is carried in the restriction of $S_1$ to the Cartan divisor ${\cal{D}}_{-\alpha_2}$.  This specifies a certain curve inside the Higgs bundle spectral cover ${\cal{C}}_{\text{Higgs,loc}}$ so we obtain an explicit mapping from holomorphic surfaces in $\tilde{Y}_4$ to curves in ${\cal{C}}_{\text{Higgs},loc}$.

To make everything explicit we adopt this procedure to assign a particular holomorphic surface ${\cal{S}}_{\Sigma}$ in $\tilde{Y}_4$ to each curve $\Sigma$ in ${\cal{C}}_{\text{Higgs},loc}$ {\footnote{Simply requiring that ${\cal{S}}_{\Sigma}$ intersect ${\cal{C}}_{\text{Higgs,loc}}$ in the curve $\Sigma$ does not uniquely specify ${\cal{S}}$.  We made some very specific assumptions about the intersections of ${\cal{S}}$ with Cartan divisors ${\cal{D}}_{-\alpha_i}$ in \cite{Marsano:2011nn}, however, and imposing these additional conditions determines ${\cal{S}}$.}}.  In \cite{Marsano:2011nn} we claimed that intersections of the ${\cal{S}}_{\Sigma}$'s with one another could be computed in the Higgs bundle spectral cover and we verify this explicitly, reproducing all the relevant formulae from \cite{Marsano:2011nn}.  We then construct explicitly the surface corresponding to the `inherited flux' of generic $SU(5)$ local models \cite{Donagi:2009ra} which, not surprisingly, is the same $G$-flux that we found earlier by the brute force approach.  Finally, we determine the surface that corresponds to the ramification divisor $r$ of the Higgs bundle spectral cover and explicitly verify that the odd part of this surface exactly corresponds to the odd part of $c_2(\tilde{Y}_4)$.  This is the first completely explicit connection between the quantization rule for fluxes in local models and the $G$-flux quantization rule of Witten \cite{Witten:1996md}.  

In the end, we are able to explicitly verify that the spectral divisor formalism of \cite{Marsano:2010ix,Marsano:2011nn} works as advertised.  We demonstrate in a very concrete way that the spectral divisor is able to capture all aspects of the resolved geometry for which it was engineered.  The emergence of the spectral cover construction of the worldvolume gauge theory is also completely manifest and straightforward.  We believe this lends significant credence to the application of spectral divisor techniques to less generic geometries, such as those that engineer $U(1)$ symmetries \cite{Hayashi:2010zp,Grimm:2010ez,Marsano:2010ix,Marsano:2011nn}, though it would certainly be interesting to study explicit resolutions in those cases as well.

\subsection{Outline}

The rest of this paper is organized as follows.  In section \ref{sec:generalities}, we review the Esole-Yau resolution procedure \cite{Esole:2011sm} in our global language and use it to compute Chern classes of the resolved 4-fold $\tilde{Y}_4$.  We then specialize to one of the six Esole-Yau resolutions and use it in section \ref{sec:yukawas} to provide global descriptions of the $\mathbf{10}$ and $\mathbf{\overline{5}}$ matter surfaces, investigate the nature of the codimension 3 singular fibers, and comment on the physics of these singularities.  We turn to $G$-fluxes in section \ref{sec:Gflux}, again in one of the Esole-Yau resolutions.  We build a $G$-flux by brute force that does not break $SU(5)$, use our result for $c_2(\tilde{Y}_4)$ to quantize it, and compute the spectrum and induced 3-brane charge.  We then elaborate on the spectral divisor formalism and the emergence of the Higgs bundle spectral cover.  Several computational details can be found in Appendix \ref{app:MatterYuk} while useful properties of the resolved spaces $\tilde{X}_5$ and $\tilde{Y}_4$ are listed in Appendix \ref{app:properties}.  We comment briefly on the connection between our global holomorphic sections and local coordinates of \cite{Esole:2011sm} in the most important coordinate patch in Appendix \ref{app:EsoleYau} and review some basic properties of local models in Appendix \ref{app:local}.


\section{Resolution: Generalities}
\label{sec:generalities}

In this section, we look at general topological properties of elliptically fibered Calabi-Yau 4-folds with section that can be used to engineer $SU(5)$ GUT models in F-theory.  We begin by reviewing features of smooth 4-folds before presenting a global description of the Esole-Yau resolutions of generic 4-folds that exhibit a surface of $A_4$ singularities.  With this description, we compute basic topological data of the resolved 4-folds that are important for model building.  We prove the conjecture of \cite{Blumenhagen:2009yv} for the shift in Euler character by direction computation and derive a result for the second Chern class of the resolved 4-folds that will be used to explicitly quantize $G$-flux in section \ref{sec:Gflux}.

\subsection{Smooth Elliptic Calabi-Yau 4-folds}

Before turning to the singular 4-folds of interest, we begin by reviewing some basic properties of smooth elliptically fibered Calabi-Yau's with section.  To construct such a 4-fold, we choose a base $B_3$ for the fibration and consider the 5-fold
\begin{equation}X_5=\mathbb{P}\left({\cal{O}}\oplus K_{B_3}^{-2}\oplus K_{B_3}^{-3}\right) \,.
\end{equation}
As the space $X_5$ is a $\mathbb{P}^2$ bundle, divisors on $X_5$ consist of pullbacks of divisors on $B_3$ under the projection
\begin{equation}\pi_X:X_5\rightarrow B_3
\end{equation}
and a new divisor, $\sigma$, inherited from the hyperplane of the $\mathbb{P}^2$ fiber. 
The projective coordinates $w$, $x$, and $y$ on the $\mathbb{P}^2$ fiber of $X_5$ are sections of the following bundles on $X_5$
\begin{equation}\begin{array}{c|c}
\text{Section} & \text{Bundle} \\ \hline
w & {\cal{O}}(\sigma) \\
x & {\cal{O}}(\sigma+2c_1) \\
y & {\cal{O}}(\sigma+3c_1)
\end{array}\label{wxysections}\end{equation}
Here, and in what follows, we use the shorthand $c_n$ for pullbacks of Chern classes $c_n(B_3)$ of $B_3$
\begin{equation}c_n\equiv \pi_X^*c_n(B_3)\end{equation}
and abuse notation by using $c_n$ to refer both to an $(n,n)$ and its Poincare dual.  Note that this forces a slight change in notation from the standard literature on local models.  In the latter, $c_1$ is typically used to represent the first Chern class of the surface $S_2$.  Here, our emphasis on global 4-folds means that it is much more convenient to take $c_1$ to denote $c_1(B_3)$.  When needed, we will write the first Chern class of $S_2$ explicitly as $c_1(S_2)$.  Common notation for local models can be found in Appendix \ref{app:local}.

Throughout this paper we will utilize a few other notational shortcuts to avoid clutter.  If $D_i$ is a divisor on $B_3$ we will use the same notation $D_i$ to indicate its pullback to $X_5$ under $\pi_X$.
\begin{equation}D_i\leftrightarrow \pi_X^*D_i \,.
\end{equation}
We will also not distinguish between divisors on $X_5$ and their restrictions to a hypersurface.

Given our nice space $X_5$, any smooth anti-canonical divisor represents a nice Calabi-Yau 4-fold.  Such a divisor will generically take the form of a cubic in $w$, $x$, and $y$ and, without loss of generality, can be written in the `Tate form' \cite{Bershadsky:1996nh}
\begin{equation}wy^2 = x^3 + a_0 w^3 + a_2 xw^2 + a_3yw^2 + a_4 wx^2 + a_5wxy \,.
\label{generaltate}\end{equation}
The resulting Calabi-Yau 4-fold, $Y_4$, has the structure of an elliptic fibration with section.  As a divisor in $X_5$, it is in the class
\begin{equation}Y_4 = 3\sigma + 6c_1 \,.
\end{equation}
As described in \cite{Sethi:1996es}, this realization of elliptically fibered 4-folds is particularly convenient for topological computations since we know a lot about the ambient space $X_5$.  For starters, divisors on $Y_4$ are inherited from divisors on $X_5 \ ${\footnote{Again, we emphasize that we will not notationally distinguish between a divisor on $X_5$ or its restriction to $Y_4$.}}.  As $X_5$ is a rather simple space, its total Chern class is easy to determine
\begin{equation}c(X_5) = c(B_3)\left(1+\sigma\right)\left(1+\sigma+2c_1\right)\left(1+\sigma + 3c_1\right) \,.
\end{equation}
This allows us to compute Chern classes of $Y_4$ by adjunction
\begin{equation}c(Y_4) = \left.\frac{c(X_5)}{1+3\sigma + 6c_1}\right|_{Y_4} \,.
\label{cY4}\end{equation}
Since $c_1(Y_4)=0$ (which can be easily verified from \eqref{cY4}), all topological invariants determined by Chern classes depend only on $c_2(Y_4)$ and $c_4(Y_4)$.  Physically, each of these has a very important meaning.  The second Chern class, $c_2(Y_4)$, tells us the quantization rule for $G$-flux \cite{Witten:1996md}
\begin{equation}G + \frac{1}{2}c_2(Y_4)\in H^4(Y_4,\mathbb{Z}) \,,
\label{Gfluxquant}\end{equation}
while $c_4(Y_4)$ just tells us the Euler character $\chi$ which, in turn, determines the geometrically induced 3-brane charge
\begin{equation}n_{D3,\text{geometric}} = -\frac{\chi}{24} \,.
\label{geomD3}\end{equation}
For our smooth elliptically fibered Calabi-Yau 4-fold with section, these quantities are readily determined.  We have
\begin{equation}\begin{split}c_2(Y_4) &= \left.\left(c_2 + 11c_1^2 + 4\sigma\cdot c_1\right)\right|_{Y_4} \\
\chi(Y_4) &= 12\int_{B_3}c_1 (c_2+30c_1^2)\,.
\end{split}
\label{chernsmooth}\end{equation}
For this computation we used
\begin{equation}\sigma(\sigma+2c_1)(\sigma+3c_1)=0\,,\end{equation}
which follows from the fact that $w$, $x$, and $y$ are projective coordinates on the fiber.  From \eqref{wxysections} we see that this means
\begin{equation}\sigma\cdot_{Y_4} (\sigma+3c_1) = 0 \,.
\end{equation}
One thing we can check with these results is that the Todd genus of $Y_4$ is as we expect.  Recall that
\begin{equation}\int_{Y_4}\text{Td}(Y_4) = \sum_{q=0}^4 (-1)^q h^{0,q}\,,\end{equation}
where the Todd class is
\begin{equation}\begin{split}\text{Td}(W) &= 1+\frac{1}{2}c_1(W) + \frac{1}{12}\left(c_1(W)^2+c_2(W)\right) + \frac{1}{24}c_1(W)c_2(W) \\
&\qquad + \frac{1}{720}\left(-c_1(W)^4+4c_1(W)^2c_2(W)+c_1(W)c_3(W)+3c_2(W)^2-c_4(W)\right) + \ldots \,.
\end{split}\end{equation}
The Todd genus must be 2 for a Calabi-Yau 4-fold with exactly $SU(4)$ holonomy since $h^{0,0}=h^{0,4}=1$ and $h^{0,1}=h^{0,2}=h^{0,3}=0$.  To verify this for our 4-fold $Y_4$, we compute
\begin{equation}\begin{split}\int_{Y_4}\text{Td}(Y_4) &= \frac{1}{720}\int_{Y_4}\left(3c_2(Y_4)^2-c_4(Y_4)\right) \\
&= \frac{1}{720}\int_{B_3}\left[3\times\left(120c_1^3+24c_1c_2\right)-3c_1\left(4c_2+120c_1^2\right)\right] \\
&= \frac{1}{12}\int_{B_3}c_1c_2\\
&= 2\int_{B_3}\text{Td}(B_3) \,.
\end{split}\end{equation}
The Todd genus of $Y_4$ is 2 provided the Todd genus of our base, $B_3$, is 1.  Any suitable base manifold $B_3$ must satisfy this property because any holomorphic $(p,q)$-form on $B_3$ will pull back to a $(p,q)$ form on $Y_4$ under the elliptic fibration.  An appropriate $B_3$ must have $h^{0,1}=h^{0,2}=h^{0,3}=0$ and hence a Todd genus of 1.  Equivalently,
\begin{equation}\int_{B_3}c_1c_2 = 24 \,.\end{equation}

In addition to the Todd genus, we are often interested in the divisibility properties of $\chi$ and $c_2(Y_4)$.  A failure of $\chi(Y_4)$ to be divisible by 24 leads via \eqref{geomD3} to a non-integer geometric 3-brane charge.  A net fractional 3-brane charge is of course nonsensical and the $G$-flux quantization condition \eqref{Gfluxquant} precisely ensures that the geometric and flux-induced contributions sum to an integer
\begin{equation}n_{D3,\text{induced}} = -\frac{\chi}{24} + \frac{1}{2}G^2 \,.
\end{equation}
This is easily verified using the fact that $Y_4$ has Todd genus 2
\begin{equation}1440 = \int_{Y_4}3c_2(Y_4)^2- \chi(Y_4)\end{equation}
to write
\begin{equation}n_{D3,\text{induced}} = -60 + \frac{1}{2}\left[\alpha^2 - \alpha c_2(Y_4)\right]\label{D3induced}\end{equation}
where
\begin{equation}\alpha = G+\frac{1}{2}c_2(Y_4)\in H^4(Y_4,\mathbb{Z})\end{equation}
Integrality of the second term in \eqref{D3induced} is a general property of intersections involving $c_2(Y_4)$ when $\alpha$ is an integral class \cite{Witten:1996md}.

Though we are always ensured an integral induced 3-brane charge, it is important to know if $c_2(Y_4)$ is an even class because that determines whether a nonzero, half-integral $G$-flux must be introduced.  It is a general result that $c_2(Y_4)$ is even when $Y_4$ is a smooth elliptically fibered Calabi-Yau.  This is almost manifest from \eqref{chernsmooth} except for the piece $c_2+c_1^2$.  In \cite{Collinucci:2010gz}, it was shown that $c_1^2+c_2$ is an even class in $B_3$ and hence its pull-back is even on $Y_4$.  Correspondingly, $\chi(Y_4)$ is given by
\begin{equation}\chi(Y_4) = 4\int_{B_3}c_1c_2+ 120\int_{B_3}c_1^3 = 24\left(3+4\int_{B_3}c_1^3\right)\end{equation}
which is nicely divisible by 24.

\subsection{Esole-Yau resolution of singular 4-folds with surface of $A_4$ singularities}

To engineer an $SU(5)$ GUT model, we need an elliptically fibered 4-fold that exhibits a surface of $A_4$ singularities.  We can do this by choosing a distinguished divisor $S_2$ inside $B_3$ to support the singularities (or equivalently a stack of 5 7-branes) that is defined by the vanishing of a holomorphic section $z$ of the bundle ${\cal{O}}(S_2)$.  We then specialize the form \eqref{generaltate} to{\footnote{Because we are dealing with $SU(5)$ the subtleties of \cite{Katz:2011qp} do not apply and we can use a Tate model without loss of generality.}}
\begin{equation}wy^2 = x^3 + b_0 w^3 z^5 + b_2 xw^2 z^3 + b_3 yw^2 z^2 + b_4 x^2 w z + b_5 xyw \,.\label{SU5tate}\end{equation}
The various objects in \eqref{SU5tate} are sections of the indicated bundles on $X_5$
\begin{equation}\begin{array}{c|c}
\text{Section} & \text{Bundle} \\ \hline
w & {\cal{O}}(\sigma) \\
x & {\cal{O}}(\sigma + 2c_1) \\
y & {\cal{O}}(\sigma + 3c_1) \\
z & {\cal{O}}(S_2) \\
b_m & {\cal{O}}([6-m]c_1-[5-m]S_2)
\end{array}\end{equation}
It is a nontrivial condition on $B_3$ that the bundles ${\cal{O}}([6-m]c_1-[5-m]S_2)$ all admit nontrivial holomorphic sections so that suitable $b_m$'s actually exist.  We will always assume that this condition is satisfied and, moreover, that the $b_m$'s are suitably generic{\footnote{Nongeneric choices of $b_m$'s can engineer extra $U(1)$ symmetries and are often more interesting for phenomenology (see for instance \cite{Marsano:2008jq,Heckman:2008qt,Marsano:2009wr,Dolan:2011iu}).  In such cases, we expect that the Esole-Yau procedure  will result in a 4-fold that is still singular and requires additional resolution.}}.

It is well-known that geometries of the type \eqref{SU5tate} exhibit a surface of $A_4$ singularities along $x=y=z=0$ that enhance along the loci
\begin{equation}\begin{split}``D_5":&\quad z=b_5=0\\
A_5 &\quad z=P=0 \,,
\end{split}\end{equation}
where
\begin{equation}P\equiv b_0b_5^2-b_2b_3b_5+b_3^4b_4 \,.\end{equation}
The nature of the singular fibers was recently explored by Esole and Yau \cite{Esole:2011sm}, who introduced a beautiful procedure for explicitly resolving all singularities of \eqref{SU5tate}.  Significant attention was paid in \cite{Esole:2011sm} to the structure of singular fibers above isolated points in $B_3$ where the curves of ``$D_5$" and ``$A_5$" singularities intersect.  We will turn to a discussion of these singular fibers and their implications for physics in the next section.  For now, we focus on using the resolution of \cite{Esole:2011sm} to compute the physically relevant quantities $c_2(\tilde{Y}_4)$ and $\chi(\tilde{Y}_4)$ for the fully resolved 4-fold $Y_4$.

It is remarkable that all singularities of \eqref{SU5tate} can be resolved by performing a series of 4 blow-ups in the ambient space $X_5$ and passing from the singular $Y_4$ to its smooth proper transform $\tilde{Y}_4$.  We are interested in a global description of this procedure so we describe the steps in a bit of detail below.  We describe how our global sections are related to the local coordinates in the most important patch of \cite{Esole:2011sm} in Appendix \ref{app:EsoleYau}.

\subsubsection{Step 1: Blow up $x=y=z=0$}

The first step is to blow up $X_5$ along the codimension 3 locus $x=y=z=0$ to get the once blown-up space $X_5^{(1)}$.  Homologically, this means we are blowing up a surface in the class
\begin{equation}(\sigma+2c_1)(\sigma+3c_1)S_2\,.\end{equation}
This introduces an exceptional divisor $E_1$ that is defined in the blown-up space by the vanishing of a holomorphic section $\zeta$ of ${\cal{O}}(E_1)$.  In the blown-up space we also get new holomorphic sections $x_1$, $y_1$, and $\tilde{z}$ satisfying
\begin{equation}x = \zeta x_1\,,\qquad y = \zeta y_1\,,\qquad z = \zeta\tilde{z}\,.
\end{equation}
For convenience, we list the new sections and their corresponding bundles
\begin{equation}\begin{array}{c|c}\text{Section} & \text{Bundle} \\ \hline
x_1 & {\cal{O}}(\sigma+2c_1 - E_1) \\
y_1 & {\cal{O}}(\sigma+3c_1 - E_1) \\
\tilde{z} & {\cal{O}}(S_2-E_1) \\
\zeta & {\cal{O}}(E_1)
\end{array}\end{equation}
The sections $x_1$, $y_1$, and $\tilde{z}$ correspond to homogeneous coordinates on the $\mathbb{P}^2$ that the blow-up introduces along the surface $x=y=z=0$.  As such,
\begin{equation}x_1=y_1=\tilde{z}=0\text{ has no solutions}\end{equation}
or, homologically
\begin{equation}(\sigma+2c_1-E_1)(\sigma+3c_1-E_1)(S_2-E_1)=0 \,.\end{equation}
The defining equation \eqref{SU5tate} of $Y_4$ becomes, after the blow-up
\begin{equation}\zeta^2\left(-wy_1^2 + \zeta x_1^2 + \left[b_0\tilde{z}^5 w^3\zeta^3 + b_2\tilde{z}^3 x_1w^2\zeta^2 + b_3\tilde{z}^2 x_1 w^2 \zeta^2 + b_4\tilde{z}x_1^2w\zeta + b_5x_1y_1w\right]\right) = 0\,.\end{equation}
The canonical class of $X_5^{(1)}$ shifts by $+2E_1$ relative to that of $X_5$ so the proper transform $Y_4^{(1)}$ of $Y_4$, given by the equation in $(\,)$'s, is an anti-canonical divisor of $X_5^{(1)}$.  $Y_4^{(1)}$ represents a partially resolved Calabi-Yau 4-fold but it remains singular so we blow up again.

\subsubsection{Step 2: Blow up $x_1=y_1=\zeta=0$}

The second step of the Esole-Yau procedure is to blow-up $X_5^{(1)}$ along the codimension 3 surface $x_1=y_1=\zeta=0$ to get the twice blown-up space $X_5^{(2)}$.  Homologically this means we are blowing up a surface in the class
\begin{equation}(\sigma+2c_1-E_1)(\sigma+3c_1-E_1)E_1\,.
\end{equation}
We get a new exceptional divisor $E_2$ that is defined by the vanishing of a section $\alpha$ of the bundle ${\cal{O}}(E_2)$.  We also get new sections $\tilde{x}$, $\tilde{y}$, and $\tilde{\zeta}$ satisfying
\begin{equation}x_1 = \tilde{x}\alpha \,,\qquad y_1 = \tilde{y}\alpha \,,\qquad \zeta = \tilde{\zeta}\alpha \,.
\end{equation}
We again list all new sections and their corresponding bundles
\begin{equation}\begin{array}{c|c}\text{Section} & \text{Bundle} \\ \hline
\tilde{x} & {\cal{O}}(\sigma+2c_1-E_1-E_2) \\
\tilde{y} & {\cal{O}}(\sigma+3c_1-E_1-E_2) \\
\tilde{\zeta} & {\cal{O}}(E_1-E_2) \\
\alpha & {\cal{O}}(E_2)
\end{array}\end{equation}
The sections $x_1$, $y_1$, and $\tilde{z}$ correspond to homogeneous coordinates on the $\mathbb{P}^2$ that is introduced by the blow-up along $x_1=y_1=\zeta=0$.  As such,
\begin{equation}\tilde{x}=\tilde{y}=\tilde{\zeta}=0\text{ has no solutions}\end{equation}
or, homologically
\begin{equation}(\sigma+2c_1-E_1-E_2)(\sigma+3c_1-E_1-E_2)(E_1-E_2)=0\,.\end{equation}
The defining equation \eqref{SU5tate} of $Y_4$ becomes, after the first two blow-ups
\begin{equation}(\tilde{\zeta}\alpha)^2\alpha^2\left(-w\tilde{y}^2 + \tilde{x}^3\tilde{\zeta}\alpha^2+\left[b_0\tilde{z}^5w^3\tilde{\zeta}^3\alpha + b_2\tilde{z}^3\tilde{x}w^2\tilde{\zeta}^2\alpha + b_3\tilde{z}^2 w^2\tilde{y}\tilde{\zeta} + b_4\tilde{z}w\tilde{x}^2\tilde{\zeta}\alpha+b_5w\tilde{x}\tilde{y}\right]\right)\label{Y42} \,.
\end{equation}
The canonical class of $X_5^{(2)}$ shifts by $+2E_2$ relative to that of $X_5^{(1)}$ so the proper transform $Y_4^{(2)}$ of $Y_4$ given by the equation in $(\,)$'s is an anti-canonical divisor of $X_5^{(2)}$.  

\subsubsection{The final two blow-ups}
\label{subsubsec:finalblowups}

The space $Y_4^{(2)}$ remains singular so we have more work to do.  It may not be obvious from \eqref{Y42} that $Y_4^{(2)}$ remains singular but we can make this fact more manifest by writing the equation in $(\,)$'s as
\begin{equation}w\tilde{y}\left(\tilde{y}-b_3w\tilde{z}^2\tilde{\zeta}-b_5\tilde{x}\right) = \alpha\tilde{\zeta}\left(b_0w^3\tilde{z}^5\tilde{\zeta}^2 + b_2\tilde{z}^3\tilde{\zeta}w^2\tilde{x} + b_4\tilde{z}w\tilde{x}^2 + \alpha\tilde{x}^3\right) \,.
\end{equation}
This has the form
\begin{equation}wv_1v_2 = u_1u_2u_3\label{factstruct}\end{equation}
with
\begin{equation}\begin{split}\label{vsus}
v_1 &= \tilde{y} \\
v_2 &= \tilde{y}-b_3w\tilde{z}^2\tilde{\zeta}-b_5\tilde{x} \\
u_1 &= \alpha \\
u_2 &= \tilde{\zeta} \\
u_3 &= b_0w^3\tilde{z}^5\tilde{\zeta}^2 + b_2\tilde{z}^3\tilde{\zeta}w^2\tilde{x} + b_4\tilde{z}w\tilde{x}^2 + \alpha\tilde{x}^3 \,.
\end{split}\end{equation}
A local factorization structure of the form \eqref{factstruct} played a crucial role in the resolution described by Esole and Yau.  We emphasize here that the structure \eqref{factstruct} extends globally at the level of holomorphic sections on $X_5^{(2)}$.  This allows a simple global description of the last two steps of the resolution and, correspondingly, the computation of topological invariants of the fully resolved 4-fold.

As evidenced by \eqref{factstruct}, the singularities of $Y_4^{(2)}$ lie along the intersections of the codimension 2 loci described by
\begin{equation}\label{FinalBlowupLoci}
v_i=u_a\qquad i=1,2,\,\,\,\,\,a=1,2,3 \,.
\end{equation}
These can be resolved by blowing up $X_5^{(2)}$ along two of those loci
\begin{equation}v_1=u_a=0\,,\qquad v_2=u_b=0 \,,\end{equation}
for some choice of $a$ and $b$.  There are six choices in all corresponding to three choices of the pair $(a,b)$ and two different ways to assign elements of that pair to $v_1$ and $v_2$.  After the blow-ups, we get two new exceptional divisors, $E_3$ and $E_4$, defined by the vanishing of holomorphic sections $\delta_3$ and $\delta_4$ of the bundles ${\cal{O}}(E_3)$ and ${\cal{O}}(E_4)$, respectively.  We also get new holomorphic sections $[V_1,U_a]$ and $[V_2,U_b]$ descending from homogeneous coordinates on the $\mathbb{P}^1$'s that we grow along $v_1=u_a=0$ and $v_2=u_b=0$.  These satisfy
\begin{equation}\label{FinalBlowups1}
v_1 = \delta_3 V_1 \,,\qquad u_a = \delta_3 U_a\end{equation}
and
\begin{equation}\label{FinalBlowups2}
v_2 = \delta_4 V_2\,,\qquad u_b = \delta_4 U_b\,.\end{equation}
For convenience, we summarize the new sections introduced by all of the blow-ups and their corresponding bundles below{\footnote{We omit sections like $x_1$ and $y_1$ that are products of sections in the table \eqref{lotsofsections}.}}
\begin{equation}\begin{array}{c|c}\text{Section} & \text{Bundle} \\ \hline
\tilde{z} & {\cal{O}}(S_2-E_1) \\
\tilde{\zeta} & {\cal{O}}(E_1-E_2) \\
\tilde{x} & {\cal{O}}(\sigma+2c_1-E_1-E_2) \\
\tilde{y} & {\cal{O}}(\sigma+3c_1-E_1-E_2) \\
V_1 & {\cal{O}}(\sigma+3c_1-E_1-E_2-E_3) \\
U_a & {\cal{O}}([u_a] - E_3) \\
\delta_3 & {\cal{O}}(E_3) \\
V_2 & {\cal{O}}(\sigma+3c_1-E_1-E_2-E_4) \\
U_b & {\cal{O}}([u_b]-E_4) \\
\delta_4 & {\cal{O}}(E_4)
\end{array}\label{lotsofsections}\end{equation}
As usual
\begin{equation}V_1=U_a=0\text{ and }V_2=U_b=0\text{ do not admit any solutions}\end{equation}
so that
\begin{equation}(\sigma+3c_1-E_1-E_2-E_3)([u_a]-E_3)=0\end{equation}
and
\begin{equation}(\sigma+3c_1-E_1-E_2-E_4)([u_b]-E_4)=0 \,.\end{equation}
We also note for later use the divisor classes corresponding to the $u_a$
\begin{equation}\begin{split}[u_1] &= E_2 \\
[u_2] &= E_1-E_2 \\
[u_3] &= 3(\sigma+2c_1-E_1-E_2)+E_2\,.
\end{split}\end{equation}
We will refer to the 5-fold that we get after performing all 4 blow-ups as ${\tilde{X}}_5$.  The defining equation of $Y_4$ becomes
\begin{equation}\label{Y4Smooth}
\tilde{\zeta}^2\alpha^4\delta_3\delta_4\left(-wV_1V_2+U_aU_bu_c\right)  =0\,.
\end{equation}
where $a,b,c=1,2,3$ are all distinct.  The proper transform $\tilde{Y}_4$ of $Y_4$ is defined by the equation in parentheses and represents a divisor of $\tilde{X}_5$ in the class
\begin{equation}\tilde{Y}_4 = Y_4 - 2E_1-2E_2-E_3-E_4 \,.\end{equation}
Because $Y_4$ was an anti-canonical divisor of $X_5$ and
\begin{equation}K_{\tilde{X}_5} = K_{X_5} + 2E_1+2E_2+E_3+E_4\end{equation}
we see that $\tilde{Y}_4$ is a smooth anti-canonical divisor of $\tilde{X}_5$.  It therefore represents a smooth Calabi-Yau resolution of $Y_4$.

\subsection{Chern Classes of $\tilde{Y}_4$}

Our ability to realize $\tilde{Y}_4$ as a smooth hypersurface of the relatively simple space $\tilde{X}_5$ means that it is almost a trivial matter to determine the physically important quantities $c_2(\tilde{Y}_4)$ and $\chi(\tilde{Y}_4)$.  The only obstacle to this is determining the Chern classes of the 4-times blown-up space $\tilde{X}_5$.

To proceed, we can make use of a general mathematical result that relates Chern classes of a space $X$ to those of a space $\tilde{X}$ obtained by blowing up along a subvariety $A$ of codimension $d$.  We denote the exceptional divisor of this blow-up by $E_A\cong\mathbb{P}(N_{A/X})$ and define the commutative diagram
\be
\ba
E \quad & \stackrel{j}{\longrightarrow} \quad \tilde{X} \cr
\stackrel{g\, }{\, }\downarrow \quad & \qquad    \,  \quad \downarrow \stackrel{\, f}{\,} \cr
A \quad & \stackrel{i}{\longrightarrow}  \quad X  
\ea
\ee
With these definitions, the Chern classes of $\tilde{X}$ are shifted from those of $X$ according to \cite{aluffi-blowups,Fulton}
\begin{equation}c(\tilde{X})-f^*c(X) = j_*\left[g^*c(A)\cdot \rho\right]\label{chernshift}\,,\end{equation}
where
\begin{equation}\rho = \frac{1}{\nu}\sum_{i=0}^d\left[\left(1-(1-\nu)[1+\nu]^i\right)\right]g^*c_{d-i}(N_{A/X})\,.\end{equation}
Here $\nu$ is the hyperplane of the $\mathbb{P}^{d-1}$ that grows during the blow-up.  This, in turn, is the restriction of $(-E_A)$ to $E_A$ inside $\tilde{X}$
\begin{equation}\nu = -E_A|_{E_A}\,.\end{equation}
Using \eqref{chernshift}, then, it is a straightforward matter to compute Chern classes of $\tilde{X}_5$ from which we can get those of $\tilde{Y}_4$ by adjunction.  Our results satisfy a few nice sanity checks.  First, we verified explicitly that
\begin{equation}\int_{\tilde{Y}_4}c_2(\tilde{Y}_4)^2 - \frac{\chi(\tilde{Y}_4)}{3} = 480\label{toddcheck}\end{equation}
holds for all 6 resolutions so that the Todd genus remains 2 in all cases as expected.  Given \eqref{toddcheck}, it is enough to work only with $c_2(\tilde{Y}_4)$ from this point onward.  We write this as
\begin{equation}\begin{split}c_2(\tilde{Y}_4) &= \left(3\sigma^2+13\sigma\cdot c_1 + c_2 + 11c_1^2\right) \\
& +\left([u_a]+[u_b]-2E_3-2E_4\right)\left(\sigma+3c_1-E_1-E_2\right)-7c_1(E_1+E_2)+3E_1E_2+E_3E_4+E_1S_2 \\
&= c_2(Y_4) +\left([u_a]+[u_b]-2E_3-2E_4\right)\left(\sigma+3c_1-E_1-E_2\right)\\
&\qquad -7c_1(E_1+E_2)+3E_1E_2+E_3E_4+E_1S_2 \end{split}\label{c2Y4gen}\end{equation}
Though $c_2(\tilde{Y}_4)$ explicitly depends on the choice of resolution, it is straightforward to verify that $\int_{\tilde{Y}_4}c_2(\tilde{Y}_4)^2$ and hence $\chi(\tilde{Y}_4)$ does not. This is another nice check as the resolutions are related to one another by flop transitions that should not change the Euler character.  The result that we obtain for $\chi(\tilde{Y}_4)$ in all cases takes a relatively simple form that can be written as an integral over $B_3$
\begin{equation}\begin{split}\chi(\tilde{Y}_4) &= 3\int_{B_3}\left(120c_1^3-250c_1^2S_2-40S_2^3+4c_1c_2+175c_1S_2^2\right) \\
&= \chi(Y_4) - 15 \int_{B_3}S_2\left(50 c_1^2 - 35c_1S_2+8S_2^2\right) \,,
\end{split}\label{eulercharacter}\end{equation}
where we explicitly show the shift in the Euler character from that of the smooth elliptically fibered 4-fold $Y_4$.  Because this shift takes the form $S_2\cdot_{B_3}(\ldots)$ we can write it in terms of intersections on $S_2$.  Using the notation for local models reviewed in Appendix \ref{app:local}, \eqref{eulercharacter} becomes 
\begin{equation}\chi(\tilde{Y}_4)-\chi(Y_4) = -15\int_{S_2}\left(488 c_1(S_2)^2 - 211c_1(S_2)\eta + 23\eta^2\right) \,,
\end{equation}
which precisely reproduces the shift that was conjectured by the authors of \cite{Blumenhagen:2009yv} using Heterotic/F-theory duality.  With the Esole-Yau resolution, we have been able to explicitly prove that conjecture for all generic 4-folds that engineer $SU(5)$ GUTs by a simple direct computation.  It is also clear that, as suggested by \cite{Blumenhagen:2009yv}, the result will change in the presence of additional singularities{\footnote{This potentially includes singularities of all codimension including the new curves of singularities that are introduced when the $b_m$ are engineered in such a way that an additional $U(1)$ is present.}} since those will in general necessitate further blow-ups.


\section{Matter Surfaces and Yukawa Couplings}
\label{sec:yukawas}

Having now described the general structure of the resolved Calabi-Yau fourfold, we will present details of one of the six birationally equivalent
resolutions.  We use this to explicitly construct the Cartan divisors and corresponding Cartan 2-cycles as well as study the structure of higher codimension singularities.
In \cite{Esole:2011sm} it was pointed out that the fibers in codimension 2 and 3 are not necessarily of Kodaira type in the sense that the intersection matrix of the blown-up $\mathbb{P}^1$s are not necessarily (extended) ADE Dynkin diagrams. 
The physical relevance of this observation has been somewhat obscure thus far. With the explicit equations for the resolved fourfold it is possible to study in detail how the various fibers for matter representations (from codimension 2) can `interact' to form Yukawa couplings (from codimension 3) that are perfectly consistent with the interactions that had been claimed to arise in these geometries.


\subsection{Sample Blow-up}
\label{subsec:sampleblowup}

Let us begin by specializing to one of the 6 birationally equivalent Esole-Yau resolutions.  
Recall that there are six distinct choices for the last two blow-ups of $X_5$ that are characterized by blowing up the subloci \eqref{FinalBlowupLoci}.  This yields six distinct blow-ups characterized by (\ref{FinalBlowups1},\ref{FinalBlowups2})
\begin{equation}
\ba
v_1 &= \delta_3 V_1\,, \qquad u_a = \delta_3 U_a \cr
v_2 &= \delta_4 V_2 \,,\qquad u_b= \delta_4 U_b \,.
\ea
\label{blowupssec3}\end{equation}
Once we do the blow-ups, the total transform of our original equation \eqref{SU5tate} for $Y_4$ takes the form (\ref{Y4Smooth})
\be
\tilde{\zeta}^2\alpha^4\delta_3\delta_4\left(-wV_1V_2+U_aU_bu_c\right) =0 \,.
\ee
Of the six different choices for $(a,b)$ with $a, b=1, 2, 3$, we will consider the following example: with the identifications of sections in (\ref{vsus}) we resolve the locus corresponding to $(a,b) =(1,2)$
\be\ba
\tilde{y}=\alpha &=0 \cr
-\tilde{y}+ b_3  {w}\tilde{z}^2\tilde{\zeta}+b_5\tilde{x}=\tilde{\zeta}&=0 \,.
\ea\ee
The blow-up equations \eqref{blowupssec3} take the form 
\begin{equation}\label{ExampleBlowUp}
\ba
\tilde{y} = \bar{y}\delta_3\qquad &\alpha = \bar{\alpha}\delta_3\cr
-\tilde{y} + b_3  {w}\tilde{z}^2\tilde{\zeta}+b_5\tilde{x} = \hat{y}\delta_4\qquad &\tilde{\zeta} = \bar{\zeta}\delta_4 \,,
\ea
\end{equation}
where $\delta_3,\delta_4$ are sections of the new bundles ${\cal{O}}(E_3)$, ${\cal{O}}(E_4)$.  For convenience we summarize all of the basic holomorphic sections in $\tilde{X}_5$ and their corresponding bundles
\begin{equation}\begin{array}{c|c}
\text{Section} & \text{Bundle} \\ \hline
\bar{y} & {\cal{O}}(\sigma +3c_1 - E_1 - E_2 - E_3) \\
\bar{\alpha} & {\cal{O}}(E_2-E_3) \\
\hat{y} & {\cal{O}}(\sigma+3c_1-E_1-E_2-E_4) \\
\bar{\zeta} & {\cal{O}}(E_1-E_2-E_4) \\
\delta_3 & {\cal{O}}(E_3) \\
\delta_4 & {\cal{O}}(E_4) \\
\tilde{x} & {\cal{O}}(\sigma+2c_1-E_1-E_2) \\
\tilde{z} & {\cal{O}}(S_2-E_1)
\end{array}
\end{equation}

Note that using the blowup equations we obtain a nontrivial relation among these sections 
\begin{equation}\label{b5blowup}
b_5\tilde{x} = \hat{y}\delta_4 + \bar{y}\delta_3 - b_3  {w}\tilde{z}^2\bar{\zeta} \delta_4  \,.
\end{equation}
The final smooth 4-fold is then described by the equation
\be
 w\bar{y}\hat{y} + \bar{\alpha } \bar{\zeta } \left({b_0}
   \delta_4^2 \bar{\zeta}^2 w^3 \tilde{z}^5+{b_2} {\delta_4} \bar{\zeta
   } w^2 \tilde{x} \tilde{z}^3+{b_4} w \tilde{x}^2 \tilde{z}+\bar{\alpha} {\delta_3} \tilde{x}^3\right)=0 \,.
\ee


\subsection{Codimension One: Cartan Divisors}

To understand the resolved geometry and all the physics implications from it, it is crucial to study the structure of the fibers after the blowup.  As mentioned in the introduction, we are particularly interested in the homology classes of the 2-cycles of various singular fibers.  To address this, we first need to describe the new divisors in our 4-fold.  These will arise from the fact that the divisor $z=0$, which described the $A_4$ singularity fibered over $S_2$, splits into multiple components in the resolution.  We see this explicitly by noting that the holomorphic section $z$ splits in $\tilde{X}_5$ as
\be\label{zexpanded}
z=  \bar{\zeta}\tilde{z}\bar{\alpha}\delta_3\delta_4 =0\,.
\ee
The divisors that are components of this locus in $\tilde{X}_5$ are obtained simply by restricting the factors of $z=0$ in \eqref{zexpanded} to the equations of the resolved 4-fold $\tilde{Y}_4$.  We list the components and their classes along with some notation for each component that will prove convenient
\begin{equation}\begin{array}{c|c|c}\label{ComponentsZeroz}
\text{Component} & \text{Class in }\tilde{Y}_4 & \text{Notation} \\ \hline
\left(\bar{\zeta}=0\right)|_{\tilde{Y}_4} & \left(E_1-E_2-E_4\right) & {\cal{D}}_{-\alpha_1} \\
\left(\delta_3=0\right)|_{\tilde{Y}_4} & E_3 & {\cal{D}}_{-\alpha_2} \\
\left(\bar{\alpha}=0\right)|_{\tilde{Y}_4} & \left(E_2-E_3\right) & {\cal{D}}_{-\alpha_3} \\
\left(\delta_4=0\right)|_{\tilde{Y}_4} & E_4 & {\cal{D}}_{-\alpha_4} \\
\left(\tilde{z}=0\right)|_{\tilde{Y}_4} & \left(S_2-E_1\right) & {\cal{D}}_{-\alpha_0} \\
\end{array}\end{equation}
Note that, to avoid unnecessary clutter, we do not notationally distinguish between divisors in $\tilde{X}_5$ and their restriction to $\tilde{Y}_4$.  For each component in \eqref{ComponentsZeroz}, it is possible to check at the level of equations that it is indeed irreducible as a divisor in $\tilde{Y}_4$.

Each of these divisors takes the form of an $A_4$ root fibered over $S_2$: $\alpha_i\rightarrow S_2$.  For this reason, we will refer to them as {\it Cartan divisors}.  It is amusing to see the emergence of the $A_4$ Dynkin diagram homologically as follows.  From each of our divisors above we can construct a curve by intersecting it with $D_1\cdot D_2$ for any two divisors $D_1,D_2$ in $B_3$ so that
\begin{equation}D_1\cdot_{B_3}D_2\cdot_{B_3}S_2=1 \,.
\end{equation}
By intersecting with $D_1\cdot D_2$ we are restricting ourselves to $\pi^*(\text{point on }S_2)$.  Restricting further to $\tilde{Y}_4$ we expect to find the 2-cycles corresponding to $A_4$ roots.  Let us explicitly write these curves as
\begin{equation}\ba\label{Cartans}
\Sigma_{\alpha_1} &= {\cal{D}}_{-\alpha_1}\doty D_1\doty D_2 \cr
\Sigma_{\alpha_2} &= {\cal{D}}_{-\alpha_2}\doty D_1\doty D_2 \cr
\Sigma_{\alpha_3} &= {\cal{D}}_{-\alpha_3}\doty D_1\doty D_2 \cr
\Sigma_{\alpha_4} &= {\cal{D}}_{-\alpha_4}\doty D_1\doty D_2 \cr
\Sigma_{\alpha_0} &= {\cal{D}}_{-\alpha_0}\doty D_1\doty D_2 \,.
\ea\end{equation}
It is straightforward to compute the intersection matrix ${\cal{D}}_{-\alpha_i}\doty \Sigma_{\alpha_j}$ explicitly since it is just a complete intersection in the space $\tilde{X}_5$.  We describe some helpful simplifications in Appendix \ref{app:properties}.  The result of this is as expected.  We find
\begin{equation}{\cal{D}}_{-\alpha_i}\doty \Sigma_{\alpha_j} = \begin{pmatrix}-2 & 1 & 0 & 0 & 1 \\
1 & -2 & 1 & 0 & 0 \\
0 & 1 & -2 & 1 & 0 \\
0 & 0 & 1 & -2 & 1 \\
1 & 0 & 0 & 1 & -2
\end{pmatrix}\,,
\end{equation}
which is exactly ($-1$ times) the intersection matrix of the extended $A_4$ Dynkin diagram.  Since we are dealing with a codimension 1 singularity in $B_3$, this computation also tells us the intersection matrix of $\mathbb{P}^1$'s in the singular fiber above generic points on $S_2$.  

It is also amusing to use our explicit description of the $\Sigma_{\alpha_i}$'s to compute their genus.  We do this by using adjunction
\begin{equation}c(\Sigma) = \frac{c(X)}{\prod (1+{\cal{D}}_a)} \,,
\end{equation}
for the suitable divisors ${\cal{D}}_a$ that appear in the definition of each $\Sigma_{\alpha_i}$.  This yields
\begin{equation}c_1(\Sigma_{\alpha_1}) = \left[c_1(\tilde{X}) - (E_1-E_2-E_4)-D_1-D_2-\tilde{Y}_4\right]_{\Sigma_{\alpha_1}}\,,\end{equation}
and similar for the other $\Sigma_{\alpha_i}$.  We find explicitly that 
\begin{equation}c_1(\Sigma_{\alpha_i})=2\qquad i=1,\ldots,5 \,,
\end{equation}
consistent with the fact that we know the curves $\Sigma_{\alpha_i}$ are $\mathbb{P}^1$s.


\subsection{Codimension Two Singularities: Matter}

We expect charged matter to localize where the singularity type jumps in codimension 2 \cite{Katz:1996xe,Bershadsky:1996nh}{\footnote{We focus on the engineering of rather ordinary representations in this paper for which an understanding at the level of \cite{Katz:1996xe,Bershadsky:1996nh} is sufficient.  Recent work that delves more deeply into the engineering of charged matter, including more exotic representations, can be found in \cite{Morrison:2011mb}.}}.  To understand this better, we recap what happens to the singular fiber as we move from generic points on $S_2$ to the matter curves where $\mathbf{10}$'s and $\mathbf{\overline{5}}$'s are expected to localize.
The main difference in our approach compared the discussion of codimension 2 singularities in \cite{Esole:2011sm} is that we do not study intersections of different components of the singular fiber above a matter curve with one another.  Rather, we focus on the homology classes of the components of these fibers which are determined by their intersections with the Cartan divisors ${\cal{D}}_{-\alpha_i}$.  The discussion is naturally formulated in the language of $SU(5)$ weights and this will allow us to see directly how the geometry is consistent with the generation of Yukawa couplings from singularities in codimension 3.


\subsubsection{{\bf 10} Matter}

We start with the $\mathbf{10}$ matter curve $\Sigma_{\mathbf{10}}$, which is characterized by the sublocus 
\be
b_5=0, \quad z=0\,.
\label{10matt}\ee
in $B_3$.  In $\tilde{X}_5$, this implies by the relation (\ref{b5blowup})  that the sections $\bar{y}$ and $\hat{y}$ are related 
\begin{equation}
\left(\hat{y}-b_3w\tilde{z}^2\bar{\zeta}\right)\delta_4 + \bar{y}\delta_3=0 \,.
\end{equation}
We study what happens to the singular fiber by looking at the reducible surface specified by the equations \eqref{10matt} in $\tilde{Y}_4$.  Above a generic curve $\Sigma$ in $S_2$ we should have that $z=0$ splits into 4 components of the form $(\alpha_i\rightarrow \Sigma)$.  Above $\Sigma_{\mathbf{10}}$, however, we find more components than this since some of the 2-cycles $\alpha_i$ split into smaller irreducible components.  By intersecting each surface with a divisor from $B_3$ that picks out a single point on $\Sigma_{\mathbf{10}}$, we can study the individual curve components that result from the splitting of various $\alpha_i$'s and compute their $U(1)$ Cartan charges (or equivalently their homology classes).  We provide details of this analysis in Appendix \ref{app:MatterYuk}.  We list the surfaces into which $z=b_5=0$ decomposes along with the $U(1)$ Cartan charges of the 2-cycle associated to each surface in the table below
\begin{equation}\begin{array}{c|c|c|c}
\text{Components of }\eqref{10matt} & \text{Multiplicity} & \text{Cartan charges} & \text{Weight} \\ \hline
(E_1-E_2-E_4)\cdot c_1 & 2 & (-2,1,0,0) & -\alpha_1 \\
(E_2-E_3)\cdot c_1 & 1 & ( 0,1,-2,1) & -\alpha_3 \\
E_3\cdot E_4 & 2 & (1,-1,1,-1) & -(\mu_{10}-\alpha_1-\alpha_2-\alpha_3) \\
E_3\cdot (c_1-E_4) & 1 & (0,-1,0,1) & \mu_{10}-\alpha_1-2\alpha_2-\alpha_3 \\
E_4\cdot (c_1-E_3)-(E_1-E_2-E_4)\cdot c_1 & 1 & (1,0,0,-1)& \mu_{10}-\alpha_2-\alpha_3-\alpha_4
\end{array}\label{10table}\end{equation}
In \eqref{10table} we have also listed the multiplicity with which each component appears in the fiber and also indicate the $SU(5)$ weight vector in standard notation as utilized in Appendix \ref{app:MatterYuk}.
If we add all the weights in \eqref{10table} together with multiplicities we find
\begin{equation}-2\alpha_1-\alpha_3-2(\mu_{10}-\alpha_1-\alpha_2-\alpha_3)+(\mu_{10}-\alpha_1-2\alpha_2-\alpha_3)+(\mu_{10}-\alpha_2-\alpha_3-\alpha_4) = -\alpha_1 -\alpha_2 -\alpha_3 -\alpha_4\end{equation}
which is consistent with the fact that the net class of our singular fiber in general is minus the sum of the $A_4$ roots{\footnote{Actually this is the net class of the singular fiber less the extended node that meets the section, which we always omit since it is a harmless spectator.}}\footnote{Since each degenerate 2-cycle is the negative of a root in this normalization--this is related to the fact that the intersection matrix is minus 1 times the Cartan--we can always redefine our Cartan charges in order to get the signs that we want.}.  The surface containing the extended node of the $A_4$ Dynkin diagram, which corresponds to the class $(S_2-E_1)$, does not split.

What has happened here is that as we move over $b_5=0$ the singular fiber reorganizes itself from four distinct roots into two roots and three states from the $\mathbf{10}$ or $\mathbf{\overline{10}}$ with multiplicities.  If our 2-cycles were the roots $-\alpha_i$ combined with $\mu_{10}$ then we would have exactly the $D_5$ Dynkin diagram and a typical $D_5$ fiber.  We do not get an explicit 2-cycle corresponding to $\mu_{10}$, though, as some different states appear instead.  The states that we do get, however, are special in the following sense.  The cone of effective curves in the fiber consisting of positive linear combinations of the components (along with -1 times these linear combinations to include M2's with opposite orientation) contains all states in the $\mathbf{10}$ and $\mathbf{\overline{10}}$ representation.  This is enough for us to be convinced that quantizing the wrapped M2 branes will yield $\mathbf{10}$'s and $\mathbf{\overline{10}}$'s by the same reasoning typically used in geometric engineering \cite{Katz:1996xe}.

The surfaces that correspond to $\mathbf{10}/\mathbf{\overline{10}}$ weights are quite special and will play a crucial role in determining the chiral spectrum.  We will adopt the general terminology of  {\it matter surface} for surfaces like these that provide us with charged matter from M2 branes wrapping their $\mathbb{P}^1$ fibers.  


\subsubsection{${\bf \bar{5}}$ Matter}

The ${\bf\bar{5}}$ matter curve $\Sigma_{\mathbf{\overline{5}}}$, which represents the remaining codimension 2 locus in $B_3$ where the singularity type enhances, is characterized by
\be
P\equiv b_0 b_5^2 - b_2 b_3 b_5 + b_3^2 b_4=0,\quad z=0 \,,
\label{5barcurve}\ee
which is in the class
\be
[P] = (8 c_1 - 5 S_2)S_2 \,.
\ee
Solutions to $P=0$ will generically not be subloci of $b_5=0$ (this will become relevant for the codimension 3 case, i.e. Yukawa couplings).  As in the case of $\mathbf{10}$'s, the surface in $\tilde{Y}_4$ described by the equations \eqref{5barcurve} splits into more than 4 components because some of the $\alpha_i$ split over $\Sigma_{\mathbf{\overline{5}}}$.  Again we can explicitly determine the homology classes of the various components of \eqref{5barcurve} and compute the Cartan charges of the 2-cycle associated to each component.  The details are provided in Appendix \ref{app:MatterYuk} and the results are summarized in the table below
\begin{equation}\begin{array}{c|c|c}\label{5barMatter}
\text{Components of }\eqref{5barcurve} & \text{Cartan charges} & \text{Weight} \\\hline
(E_1-E_2-E_4)\cdot (8c_1-5S_2) & (-2,1,0,0) & -\alpha_1 \\
(E_2-E_3)\cdot (8c_1-5S_2)& (0,1,-2,1) & -\alpha_3 \\
E_3\cdot (\sigma+3c_1-E_1-2E_2+E_3-E_4) & (0,-1,1,0) & \mu_5-\alpha_1-\alpha_2 \\
E_3\cdot (5c_1-5S_2-\sigma+E_1+2E_2-E_3+E_4)& (1,-1,0,0) & -(\mu_5-\alpha_1) \\
E_4\cdot (8c_1-5S_2) & (0,0,1,-2) & -\alpha_4 \\
\end{array}\end{equation}
Again we see that an $SU(5)$ root has split as we move over $P=0$.  In this case, the splitting is simply
\begin{equation}-\alpha_2\rightarrow (\mu_5-\alpha_1-\alpha_2) + (-\mu_5+\alpha_1)\,.\end{equation}
Note that in this case every component appears with multiplicity 1.  This is consistent with the fact that Esole and Yau find an ordinary $A_5$ fiber above $P=0$ \cite{Esole:2011sm}.  Further, the cone of effective curves in the fiber contains all states in the $\mathbf{5}$ and $\mathbf{\overline{5}}$ representations.


\subsection{Codimension Three Singularities: Yukawas}

The structure of the singular fibers above codimension three singularities governs the Yukawa couplings of the $SU(5)$ GUT model.  The worldvolume perspective tells us that we should expect top and bottom type Yukawas from points where the fiber enhances to ``$E_6$" and ``$D_6$" respectively.  What we really mean by this is that we expect top Yukawas from intersections of the $\mathbf{10}$ and $\mathbf{\overline{5}}$ matter curves where $b_5=b_4=0$ and bottom Yukawas from intersections where $b_5=b_3=0$.  One puzzle that emerged in \cite{Esole:2011sm} is the absence of actual honest Kodaira $E_6$ fibers which are supposed to yield the top Yukawas.  This is important to understand because the ability to generate top Yukawa couplings is one of the main motivations for studying F-theory models versus perturbative type IIB orientifolds.
What we need to get a top Yukawa coupling, however, is not necessarily an $E_6$ fiber but rather the ability of wrapped M2-branes to suitably interact at the codimension 3 locus.  Geometrically, this means that the 2-cycles wrapped by the M2's corresponding to $\mathbf{10}$'s and $\mathbf{5}$'s that participate in a given coupling should be connected by a 3-chain that degenerates in the fiber over the corresponding codimension 3 singularity.  We will see that this property is indeed satisfied.


\subsubsection{$E_6$ Points}

Points that are supposed to yield $\mathbf{10}\times\mathbf{10}\times\mathbf{5}+\text{cc}$ type top Yukawa couplings are characterized by
\be
b_4=b_5=z=0 \,.
\label{E6points}\ee
In $\tilde{Y}_4$, the equations \eqref{E6points} define a collection of curves that comprise the singular fiber over the ``$E_6$" point.  It is straightforward to determine the homology classes of these curves and compute their $U(1)$ Cartan charges.  The details are again in Appendix \ref{app:MatterYuk}.  We present only the results here
\begin{equation}\begin{array}{c|c|c|c}
\text{Curve}   & \text{Multiplicity}  & \text{Cartan charges} & \text{Weight} \\ \hline
\Sigma_{\bar{\zeta}}^{E6} & 2& (-2,1,0,0) & -\alpha_1  \\
\Sigma_{\bar{\alpha}}^{E6} & 1& (0,1,-2,1) & -\alpha_3  \\
\Sigma_{34\bar{y}}^{E6}  & 2& (1,0,0,-1) & \mu_{10}-\alpha_2-\alpha_3-\alpha_4  \\
\Sigma_{34\hat{y}}^{E6} & 2& (0,-1,1,0) & \mu_5 - \alpha_1-\alpha_2-\alpha_3  \\
\Sigma_3^{E6} & 1& (0,-1,0,1) & \mu_{10}-\alpha_1-2\alpha_2-\alpha_3  
\end{array}\end{equation}
We have again dropped the extended node, referred to as $\Sigma_{\tilde{z}}$ in Appendix \ref{app:MatterYuk}, since it is a harmless spectator that does not split.

We can now describe what seems to be happening to 2-cycles along the $\mathbf{10}$ and $\mathbf{\overline{5}}$ matter curves as we move to an ``$E_6$" point.  First we recall the Cartan charges of 2-cycles associated to the $\mathbf{10}$ matter curve $\Sigma_{\mathbf{10}}$
\begin{equation}\label{10Multi}
\begin{array}{c|c}
\text{Charges} & \text{Multiplicity} \\ \hline
(-2,1,0,0) & 2 \\
(0,1,-2,1) & 1 \\
(1,-1,1,-1) & 2 \\
(0,-1,0,1) & 1 \\
(1,0,0,-1) & 1
\end{array}\end{equation}
Now recall the situation for the ``$E_6$" point
\begin{equation}\label{E6Multi}
\begin{array}{c|c}
\text{Charges} & \text{Multiplicity} \\ \hline
(-2,1,0,0) & 2 \\
(0,1,-2,1) & 1 \\
(1,0,0,-1) & 3 \\
(0,-1,1,0) & 2 \\
(0,-1,0,1) & 1
\end{array}\end{equation}
As studied explicitly in Appendix \ref{app:MatterYuk}, what happens as we move toward an ``$E_6$" point along $\Sigma_{\mathbf{10}}$ is that the curve with weights $(1,-1,1,-1)$ splits according to
\begin{equation}(1,-1,1,-1)\rightarrow (1,0,0,-1)+(0,-1,1,0)\,.\end{equation}
Notice that one of the components into which $(1,-1,1,-1)$ splits is something that we already had, which now just appears with a higher multiplicity.  As a result, the number of distinct components of the singular fiber does not change and it is in this sense that the rank does not enhance.

We can also try to think about this from the perspective of the $\mathbf{\overline{5}}$ matter curve $\Sigma_{\mathbf{\overline{5}}}$, though it is a bit more complicated.  Recall that the components there had Cartan charges
\begin{equation}\label{5barMulti}
\begin{array}{c|c}
\text{Charges} & \text{Multiplicity} \\ \hline
(-2,1,0,0) & 1 \\
(0,1,-2,1) & 1 \\
(0,-1,1,0) & 1 \\
(1,-1,0,0) & 1 \\
(0,0,1,-2) & 1
\end{array}\end{equation}
As we move toward the ``$E_6$" point along $\Sigma_{\mathbf{\overline{5}}}$, both an $SU(5)$ root and an $SO(10)$ weight split according to
\begin{equation}\begin{split}(1,-1,0,0) &\rightarrow (1,0,0,-1)+(0,-1,0,1) \\
(0,0,1,-2) &\rightarrow (-2,1,0,0)+(0,-1,1,0)+2\times (1,0,0,-1) \,.
\end{split}\end{equation}
Now, suppose we have two $\mathbf{10}$ states wrapping $(1,0,0,-1)$ and $(0,-1,0,1)$.  If we have a $\mathbf{5}$ wrapping $-(1,-1,0,0)$ then, at the $E_6$ point, our $\mathbf{5}$ will degenerate into $(-1,0,0,1)+(0,1,0,-1)$ which is exactly -1 times the curves wrapped by our $\mathbf{10}$ states.  This is exactly what we need from the geometry to generate a top $\mathbf{10}\times\mathbf{10}\times\mathbf{5}$ Yukawa coupling even though there is no `rank enhancement' to $E_6$ in the fiber.


\subsubsection{$D_6$ points}

The ``$D_6$" points are characterized by $b_3 =b_5=0$. Again we look at the irreducible components inside $\tilde{Y}_4$ of
$b_3=b_5 =z=0$. The class of $b_3$ is $3 c_1 - 2 S_2$. We find (see Appendix \ref{app:MatterYuk} for details)
\begin{equation}\begin{array}{c|c|c|c}
\text{Curve}   				& \text{Multiplicity}  & \text{Cartan charges} & \text{Weight} \\ \hline
\Sigma_{\bar{\zeta}}^{D6} 	& 2& (-2,1,0,0) & -\alpha_1  \\
\Sigma_{\bar{\alpha}}^{D6} 	& 2& (0,1,-2,1) & -\alpha_3  \\
\Sigma_{34}^{D6}  			& 2& (1,-1,1,-1) & -\mu_{10}+\alpha_1+\alpha_2+\alpha_3  \\
\Sigma_{3}^{D6}			& 2^*& (0,-1,1,0) & \mu_5 - \alpha_1-\alpha_2  \\
\Sigma_4^{D6} 			& 1& (1,0,0,-1) & \mu_{10}-\alpha_2-\alpha_3-\alpha_4  
\end{array}\end{equation}
A few remarks about this: the multiplicities of the curves are all inherited from the $D_5$ matter surfaces, except $\Sigma_3^{D_6}$, which is really a reducible curve with Cartan charge $(0, -2, 2, 0)$. 

From the charges of the ${\bf 10}$ matter we can infer, see the table (\ref{10Multi}), that the splitting is
\be\label{10Split}
(0, -1, 0,1)  \rightarrow  2 \times (0, -1, 1, 0) + (0, 1, -2, 1)\,.
\ee
Similarly from the ${\bf \bar{5}}$ matter surfaces and multiplicities (\ref{5barMulti}) the splitting is
\be
\ba
(1, -1, 0, 0) & \rightarrow (1, -1, 1, -1) + (0, 1, -2, 1) + (0, -1, 1, 0) \cr
(0, 0, 1, -2) & \rightarrow (1, 0, 0, -1) + (1, -1, 1, -1) + (-2, 1, 0, 0) \,.
\ea
\ee
Note that the relation \eqref{10Split} indicates that the M2 corresponding to a $\mathbf{10}$ state decomposes into the sum of M2's $(0,-1,1,0)$ and $[(0,-1,1,0)+(0,1,-2,1)]=(0,0,-1,1)$ that correspond to $\mathbf{5}$ states.  This is exactly what we need from the geometry to generate a bottom type Yukawa coupling $\mathbf{10}\times\mathbf{\overline{5}}\times\mathbf{\overline{5}}$ despite the fact that the fiber is not a standard Kodaira fiber yielding the Dynkin diagram for extended $D_6$.


\section{$G$-Flux: Generalities and Spectral Covers}
\label{sec:Gflux}

Perhaps the most interesting application of explicit resolutions of singular Calabi-Yau's for model-building is that they allow a very direct study of the $G$-fluxes that are needed to induce a chiral spectrum.  We adopt two different approaches to the problem in this section.  First, we proceed directly in $\tilde{Y}_4$ to construct holomorphic surfaces with the right properties to yield a $G$-flux that does not break $SU(5)$ and induces a net chirality of $\mathbf{10}$'s and $\mathbf{\overline{5}}$'s.  We use our explicit expression for $c_2(\tilde{Y}_4)$ to quantize the $G$-flux and compute the spectrum, finding agreement with the spectrum of $SU(5)$ local models with generic spectral covers \cite{Donagi:2009ra}.  We also directly compute the flux-induced 3-brane charge and find agreement with local results \cite{Blumenhagen:2009yv}.

All of this agreement with local models suggests a more direct connection.  This is provided by the spectral divisor formalism introduced in \cite{Marsano:2010ix} and refined in \cite{Marsano:2011nn}.  In \cite{Marsano:2011nn} the connection between $G$-flux and curves in the spectral cover ${\cal{C}}_{\text{Higgs,loc}}$ of the local model was described in terms of the limiting behavior of a global object called the spectral divisor near the surface of $A_4$ singularities.  In general there can be many spectral divisors since we really only care about the behavior near $S_2$.  The work \cite{Marsano:2011nn} used a special choice called the `Tate divisor' ${\cal{C}}_{\text{Tate}}$ and we will use this choice here just to be concrete.

With an explicit resolution $\tilde{Y}_4$ in hand, we can now make the prescription for relating ${\cal{C}}_{\text{Tate}}$ with the Higgs bundle spectral cover ${\cal{C}}_{\text{Higgs,loc}}$ completely precise.  The notion of limiting behavior lifts to the restriction of the proper transform of ${\cal{C}}_{\text{Tate}}$ with one of the Cartan divisors.  This intersection naturally yields the Higgs bundle spectral cover ${\cal{C}}_{\text{Higgs,loc}}$ and, with this, we can make explicit the connection between curves in ${\cal{C}}_{\text{Higgs,loc}}$ and certain holomorphic surfaces in $\tilde{Y}_4$.  We verify the intersection formulae of \cite{Marsano:2011nn}, which described how the intersections of various surfaces could be computed in the local model, and explicitly construct the $G$-flux corresponding to the traceless flux of the local model.  Not surprisingly this reproduces exactly the $G$-flux obtained by the direct approach in section \ref{subsec:Gfluxgen}.  The ability of the local model to capture intersections on $\tilde{Y}_4$ as outlined in \cite{Marsano:2011nn} also explains why the local model can successfully compute the flux-induced 3-brane charge and we verify this agreement with an explicit computation.  Perhaps more interestingly, we can also make a direct and explicit connection between the quantization of the gauge bundle in the local model and the quantization of $G$-flux in the global one.  In particular, the ramification divisor $r$ of ${\cal{C}}_{\text{Higgs,loc}}$ corresponds to a particular surface in $\tilde{Y}_4$ according to \cite{Marsano:2011nn} and it is easy to see, using the dictionary we present here, that the odd part of that surface coincides exactly with the odd part of $c_2(\tilde{Y}_4)$.  This example makes completely explicit the workings of the spectral divisor formalism and how it captures the relation of spectral covers to geometry and $G$-flux in generic singular Calabi-Yau 4-folds that exhibit a surface of $A_4$ singularities.

\subsection{$G$-fluxes: General approach}
\label{subsec:Gfluxgen}

Before turning to spectral covers, let us address in full generality the question of $G$-fluxes in $\tilde{Y}_4$ that induce net chiralities of $\mathbf{10}$'s and $\mathbf{\overline{5}}$'s.  Our $\mathbf{10}$'s and $\mathbf{\overline{5}}$'s are associated to M2 branes that wrap vanishing cycles that are effective only over curves inside $B_3$.  Given the matter surface associated with such a cycle, the multiplet that lives there is determined by the Cartan charges of the vanishing 2-cycle or equivalently its intersections with the Cartan divisors.  In the previous section we identified matter surfaces associated to components of the $\mathbf{10}$ and $\mathbf{\overline{5}}$ representations.  Let us identify two in particular
\begin{equation}\ba
{\cal{S}}_{\mathbf{10}} &= E_3\cdot E_4\cr
{\cal{S}}_{\mathbf{\overline{5}}} &=E_3\cdot (3c_1-E_1-2E_2+E_3-E_4)\,.
\label{mattersurfs}
\ea
\end{equation}
The matter fields couple to a flux obtained by integrating $G$ over the vanishing cycle.  It follows that the net chirality is simply the integral of $G$ over the matter surface.  Now, the matter surfaces above only tell us about the net chiralities of the particular states to which they correspond.  Provided $G$ does not break $SU(5)$, the net chirality of one state in the $\mathbf{10}$ associated to ${\cal{S}}_{\mathbf{10}}$ determines the net chirality of $\mathbf{10}$'s in general by $SU(5)$ invariance and similar for $\mathbf{\overline{5}}$'s{\footnote{If we identify $\mu_{10}$ as the highest weight of a $\mathbf{10}$ rather than that of a $\mathbf{\overline{10}}$ and similar for $\mu_5$ then ${\cal{S}}_{\mathbf{10}}$ actually describes a state in the $\mathbf{\overline{10}}$ and ${\cal{S}}_{\mathbf{\overline{5}}}$ describes a state in the $\mathbf{5}$.  In what follows, we will adopt the convention that $\mu_{10}$ instead describes a state in the $\mathbf{\overline{10}}$ and $\mu_5$ a state in the $\mathbf{\overline{5}}$.}}
\begin{equation}
n_{\mathbf{10}}-n_{\mathbf{\overline{10}}}=\int_{{\cal{S}}_{\mathbf{10}}}G_4\,, \qquad n_{\mathbf{\overline{5}}}-n_{\mathbf{5}} = \int_{ {\cal{S}}_{\mathbf{\overline{5}}}}G_4 \,.
\end{equation}

\subsubsection{Conditions on $G$}

Since we want a $G$-flux that integrates nontrivially over holomorphic surfaces in $\tilde{Y}_4$ the type of $G$-flux we are after is a $(2,2)$-form.  We will construct it by specifying a holomorphic surface in $\tilde{Y}_4$.  Any $G$-flux in F-theory must satisfy 2 crucial properties, though: it must be orthogonal to any surface that sits in the section of the elliptic fibration and it must be orthogonal to any surface that takes the form of an elliptic fibration over a curve in $B_3$.  In other words, it must be orthogonal to the pullbacks of all vertical and horizontal surfaces in $Y_4$ to $\tilde{Y}_4$
\begin{equation}G\cdot_{\tilde{Y}_4} \sigma\cdot_{\tilde{Y}_4} D_1 = G\cdot_{\tilde{Y}_4}D_1\cdot_{\tilde{Y}_4}D_2=0\label{verthoriz}\end{equation}
for $D_1$ and $D_2$ the pullbacks of two divisors in $B_3$.

In addition, we would like to focus for now on $G$-fluxes that do not break $SU(5)$.  For this, we must ensure that
\begin{equation}G\cdot_{\tilde{Y}_4} {\cal{D}}_{-\alpha_i}\cdot_{\tilde{Y}_4} D=0\label{cartancond}\end{equation}
for any $D$ that is the pullback of a divisor in $B_3$.

\subsubsection{Building Blocks for $G$}
\label{subsubsec:buildingblocks}

We can now proceed to construct holomorphic surfaces that satisfy all of these properties.  We focus here on inherited surfaces that arise as linear combinations of complete intersections of 2 divisors in $\tilde{X}_5$ with $\tilde{Y}_4$.  It would be interesting to study surfaces that do not arise in this way.  So-called non-universal fluxes such as those used to build models in \cite{Marsano:2009gv,Marsano:2009wr} should arise like this.  A study of this type of $G$-flux was recently undertaken in geometries with low rank singularities in \cite{Braun:2011zm}.

The general conditions \eqref{verthoriz} already restrict us to two types of surfaces.  The first are Cartan fluxes
\begin{equation}E_i\cdot_{\tilde{Y}_4} D\label{cartanflux}\end{equation}
for $D$ a divisor pulled back from $B_3$.  The second are pairwise intersections of exceptional divisors
\begin{equation}E_j\cdot_{\tilde{Y}_4}E_j\,.
\label{pairwise}\end{equation}
The Cartan fluxes \eqref{cartanflux} will intersect Cartan divisors and break $SU(5)$.  In general the pairwise intersections \eqref{pairwise} will also do this, however, so we will need to take suitable combinations of both to ensure that our net flux does not break $SU(5)$.  It is easy to check that the pairwise intersections \eqref{pairwise} always have intersections with Cartan surfaces in \eqref{cartancond} that are proportional to linear combinations of
\begin{equation}S_2\cdot_{B_3}D\cdot_{B_3}A\,,
\end{equation}
where $A$ is either $c_1$ or $S_2$.  Canceling these contributions generically restricts our attention to the Cartan fluxes
\begin{equation}E_i\cdot_{\tilde{Y}_4} c_1\quad\text{and}\quad E_i\cdot_{\tilde{Y}_4}S_2 \,.
\label{c1S2fluxes}\end{equation}
Together with the 10 pairwise fluxes from \eqref{pairwise} we find an 18-parameter family of fluxes on which we will impose the conditions \eqref{cartancond}.  Doing this in a completely naive fashion, one finds a 7-parameter family of solutions of which a 6-parameter subspace does not affect the net chirality of $\mathbf{10}$'s or $\mathbf{\overline{5}}$'s.  In fact it is easy to see that the 6-parameter subspace does not appear to participate in any intersections and for good reason.  Because of the blow-ups and the nature of $\tilde{Y}_4$, there are a number of nontrivial relations between the pairwise fluxes \eqref{pairwise} and the Cartan fluxes in \eqref{c1S2fluxes}.  In fact there are 8 such relations which we can use to eliminate 8 of the 10 pairwise fluxes in \eqref{pairwise} in favor of the remaining two and the 8 Cartan fluxes in \eqref{c1S2fluxes}.  This is described in detail in Appendix \ref{app:properties}.  Here, we simply note that a convenient choice of two pairwise fluxes \eqref{pairwise} to keep is
\begin{equation}E_3\cdot_{\tilde{Y}_4}E_4\quad \text{ and }\quad E_2\cdot_{\tilde{Y}_4}E_4 \,.
\label{twopairwise}\end{equation}
We will express the rest in terms of \eqref{twopairwise} and \eqref{c1S2fluxes}.  Our generic $G$-flux will therefore be a linear combination of \eqref{c1S2fluxes} and \eqref{twopairwise}.

\subsubsection{Quantizing $G$}

Before proceeding to construct the general linear combination and apply the constraints \eqref{cartancond}, let us build $G$-flux quantization into the computation.  Using the relations \eqref{Y4Rels} of Appendix \ref{app:properties} we can write a simple expression for $c_2(\tilde{Y}_4)$ \eqref{c2Y4gen} for the specific resolution under consideration
\begin{equation}\begin{split}c_2(\tilde{Y}_4) &= \left(3\sigma^2+13\sigma c_1 + c_2 + 11c_1^2\right) \\
& \quad -E_3\cdot E_4 - c_1\cdot \left(8E_1+3E_2+6E_3-2E_4\right) + S_2\cdot \left(2E_1+4E_3-2E_4\right) \,.
\end{split}\end{equation}
The first line is just $c_2(Y_4)$ which we recall is an even class.  This means that
\begin{equation}c_2(\tilde{Y}_4) = E_3\cdot E_4 + E_2\cdot c_1 + \text{even} \,.
\label{oddc2}\end{equation}
To build a properly quantized $G$-flux, now, we start with
\begin{equation}G = \frac{1}{2}\left(E_3\doty E_4 + E_2\doty c_1\right) + \sum_i E_i\doty \left(a_i c_1 +b_i S_2\right) + p E_3\doty E_4 + q E_2\doty E_4\,.\label{Gansatz}\end{equation}
In order for $G+\frac{1}{2}c_2(\tilde{Y}_4)$ to be an integral class we require that the $a_i$, $b_i$, $p$, and $q$ are integers.

\subsubsection{The Properly Quantized $G$}
\label{subsubsec:quantizedG}

We are now ready to apply the constraints \eqref{cartancond} to the ansatz \eqref{Gansatz}.  We find a one-parameter family of solutions with
\begin{equation}\begin{split}a_1 &= -1-2a_2 \\
a_3 &= 1+2a_2 \\
a_4 &= 2+4a_2 \\
b_i &= 0 \\
p &= -3-5a_2 \\
q &= 0 \,.
\end{split}\end{equation}
We can write our net $G$-flux then as
\begin{equation} G = -\frac{1}{2}\left(1+2a_2\right)\left(5E_3\doty E_4 + c_1\doty\left[2E_1-E_2-2E_3-4E_4\right]\right)\,.
\label{bruteGflux}\end{equation}
It is a straightforward matter to intersect this with the matter surfaces ${\cal{S}}_{\mathbf{10}}$ and ${\cal{S}}_{\mathbf{\overline{5}}}$.  The result is
\begin{equation}G\doty {\cal{S}}_{\mathbf{10}} = G\doty {\cal{S}}_{\mathbf{\overline{5}}} = -\frac{1}{2}\left(2a_2+1\right)S_2\cdot_{B_3}c_1\cdot_{B_3}\left(6c_1-5S_2\right)\,.
\end{equation}
Quite nicely, this result reduces to an intersection in $S_2$.  Writing that result in the standard notation of local models reviewed in Appendix \ref{app:local} we find
\begin{equation}G\doty {\cal{S}}_{\mathbf{10}} = G\doty {\cal{S}}_{\mathbf{\overline{5}}} = -\frac{1}{2}\left(2a_2+1\right)\eta\cdot_{S_2}(\eta-5c_1(S_2))\,,\qquad a_2\in\mathbb{Z}\,,\label{brutespectrum}\end{equation}
where we remind ourselves that $a_2$ is forced to be an integer by the quantization condition.  We recognize this as \emph{exactly} the spectrum of generic local $SU(5)$ models \cite{Donagi:2009ra} with even the quantization reproduced correctly.  We briefly review the local model computation in Appendix \ref{app:local}.

We can also check the 3-brane tadpole induced by this $G$-flux.  Using
\begin{equation}n_{D3,induced} = \frac{1}{2}G\cdot_{\tilde{Y}_4}G\end{equation}
we find
\begin{equation}n_{D3,induced} = \frac{5}{8}\left(1+2a_2\right)^2c_1\cdot_{B_3}(6c_1-5S_2)\cdot_{B_3}S_2\,.\end{equation}
We can again write this as an intersection in $S_2$.  Expressed in the conventional local model language we have
\begin{equation}n_{D3,induced} = \frac{5}{8}\left(1+2a_2\right)^2 \eta\cdot_{S_2}(\eta-5c_1(S_2))\,.\end{equation}
This also agrees with the expression typically conjectured for this quantity in local models \cite{Blumenhagen:2009yv}.  We briefly review the local model computation of this quantity also in Appendix \ref{app:local}.



\subsection{Spectral Covers} 

Finally we would like to understand how the globally resolved fourfold relates to the local spectral cover description \cite{Donagi:2009ra}.
Let us recall how we expect spectral covers to emerge \cite{Marsano:2011nn}.  We start with the Tate divisor
\begin{equation}w\left(b_0w^2z^5 + b_2wxz^3 + b_3wyz^2 + b_4 wx^2 z + b_5 wxy\right)=0
\label{Tatedivisor0}\end{equation}
in the singular 4-fold $Y_4$ \eqref{SU5tate}
\begin{equation}wy^2 = x^3 + w\left(b_0w^2z^5 + b_2wxz^3 + b_3wyz^2 + b_4 wx^2 z + b_5 wxy\right)\,.\end{equation} 
Then we note that the meromorphic section $y/x$ is holomorphic when restricted to this divisor.  We call this $t$ and use the fact that $y^2=x^3$ along the Tate divisor (in the $w=1$ patch) to write
\begin{equation}b_0z^5 + b_2z^3t^2 + b_3z^2t^3 + b_4zt^4 + b_5t^5=0 \,.
\end{equation}
In the limit $t\rightarrow 0$, $z\rightarrow 0$, with $s=z/t$ fixed this turns into
\begin{equation}t^5\left[b_0s^5 + b_2s^3 + b_3s^2 + b_4 s + b_5\right] \,.\end{equation}
We remove the $t^5$, which is expected to capture the effect of taking a proper transform, and what remains is precisely the Higgs bundle spectral cover ${\cal{C}}_{\text{Higgs,loc}}$.

This limiting behavior can be turned into something much more precise in the resolved 4-fold $\tilde{Y}_4$.  Recall that, after the resolution, we have
\begin{equation}\begin{split}
x &= \tilde{x}\bar{\alpha}^2\delta_3^2\delta_4\bar{\zeta} \\
y &= \tilde{y}\bar{\alpha}^2\delta_3^3\delta_4\bar{\zeta} \\
z &= \tilde{z}\bar{\alpha}\delta_3\delta_4\bar{\zeta}
\end{split}\end{equation}
so that
\begin{equation}\frac{y}{x}\rightarrow \frac{\bar{y}\delta_3}{\tilde{x}}\qquad \hbox{and } \qquad 
\frac{zx}{y}\rightarrow \frac{\tilde{x}\tilde{z}\bar{\alpha}\bar{\zeta}\delta_4}{\bar{y}} \,.
\end{equation}
It is now clear how to send $y/x\rightarrow 0$ while holding $zx/y$ fixed.  This simply amounts in the resolved geometry to sending
\be
\delta_3\rightarrow 0 \,.
\ee


\subsubsection{The Tate divisor in $\tilde{Y}_4$}

In the limit $\delta_3 \rightarrow 0$ we will now show that the Tate divisor in the resolved fourfold limits to the spectral cover ${\cal{C}}_{\text{Higgs,loc}}$ of the Higgs bundle in the worldvolume theory.  First, we have to properly describe the proper transform of ${\cal{C}}_{\text{Tate}}$ under the resolution.
To do this, recall that the resolved fourfold $\tilde{Y}_4$ is defined by
\begin{equation}w\bar{y}\hat{y} + \bar{\alpha}\bar{\zeta}\left[b_0w^3\tilde{z}^5(\delta_4\bar{\zeta})^2+b_2w^2\tilde{x}\tilde{z}^3(\delta_4\bar{\zeta})+b_4w\tilde{x}^2\tilde{z}+\bar{\alpha}\delta_3\tilde{x}^3\right)\,,\end{equation}
where
\begin{equation}\hat{y}\delta_4+\bar{y}\delta_3 = b_3w\tilde{z}^2(\delta_4\bar{\zeta})+b_5\tilde{x}\end{equation}
and the total transform of the equation \eqref{Tatedivisor0} for the Tate divisor is
\begin{equation}w\bar{\alpha}^4\delta_3^5\delta_4^2\bar{\zeta}^2\left(b_5\tilde{x}\bar{y}+b_4\tilde{z}\tilde{x}^2\bar{\alpha}(\bar{\zeta}\delta_4)+b_3w\tilde{z}^2\bar{y}(\delta_4\bar{\zeta})+b_2\tilde{z}^3w\tilde{x}\bar{\alpha}(\bar{\zeta}\delta_4)^2 + b_0\tilde{z}_5w^2\bar{\alpha}(\delta_4\bar{\zeta})^3\right) \,.\label{Tatedivisor1}\end{equation}
We will keep the term in parentheses as the proper transform.  Actually we have to be a bit careful because there is another factor of $\delta_3$ hiding\footnote{To see this, let us plug the relations
\begin{equation}b_5\tilde{x} + b_3w\tilde{z}^2\delta_4\bar{\zeta} = \hat{y}\delta_4+\bar{y}\delta_3\end{equation}
and
\begin{equation}\bar{\alpha}\bar{\zeta}\left(b_0w^3\tilde{z}^5(\delta_4\bar{\zeta})^2+b_2w^2\tilde{x}\tilde{z}^2(\delta_4\bar{\zeta})+b_4w\tilde{x}^2\tilde{z}\right) = -w\bar{y}\hat{y} - \bar{\alpha}^2\bar{\zeta}\delta_3\tilde{x}^3\end{equation}
into the equation for the spectral divisor.}  and we find a dramatic simplification that leads to
\begin{equation}\bar{\alpha}^4\delta_3^6\delta_4^2\bar{\zeta}^2\left(w\bar{y}\left[\hat{y}\delta_4+\bar{y}\delta_3\right]- \delta_4\left[w\bar{y}\hat{y}+\bar{\alpha}^2\bar{\zeta}\delta_3\tilde{x}^3\right]\right) =0
\end{equation}
or
\begin{equation}\bar{\alpha}^4\delta_3^6\delta_4^2\bar{\zeta}^2\left(w\bar{y}^2-\bar{\alpha}^2\bar{\zeta}\delta_4\tilde{x}^3\right) =0\,.
\end{equation}
So we should really define the Tate divisor as the restriction of
\begin{equation}w\bar{y}^2 - \bar{\alpha}^2\bar{\zeta}\delta_4\tilde{x}^3
\label{specdivprop}\end{equation}
and note that the term in $(\,)$'s in \eqref{Tatedivisor1}, $b_5\tilde{x}\bar{y}+\ldots$, is always satisfied on this locus.  But we are still not done.  The divisor obtained by restricting $w\bar{y}^2-\bar{\alpha}^2\bar{\zeta}\delta_4\tilde{x}^3$ to our 4-fold is reducible!  There is one component $\bar{y}=\bar{\zeta}=0$ and then there is the remainder.  What we are after is the remainder so we must remove by hand this $\bar{y}=\bar{\zeta}=0$.  Nicely, $\bar{y}=\bar{\zeta}=0$ is just the restriction of $\bar{\zeta}=0$ to $\tilde{Y}_4$ so we can describe the Tate divisor by
\begin{equation}{\cal{C}}_{\rm Tate}:\qquad \left[w\bar{y}^2-\bar{\alpha}^2\bar{\zeta}\delta_4\tilde{x}^3=0\right]\cdot \tilde{Y}_4 \quad - \quad \left[\bar{\zeta}=0\right]\cdot\tilde{Y}_4\,,\end{equation}
which is in the class
\begin{equation}{\cal{C}}_{\rm Tate}= \left[3\sigma + 6c_1 - 3E_1 -E_2 - 2E_3 + E_4\right]_{\tilde{Y}_4}\,.\end{equation}
One neat thing to compute with this is the intersection of ${\cal{C}}_{\rm Tate}$ with the Cartan roots \eqref{Cartans}.  We find
\begin{equation}{\cal{C}}_{\rm Tate}\doty \Sigma_{\alpha_i} =(0,1,0,0)\times 5\end{equation}
consistent with the fact that each of the five sheets of ${\cal{C}}_{\rm Tate}$ restricted to $z=0$ intersects the roots as $(0,1,0,0)$.  Note that this is the highest weight of the $\mathbf{10}$, consistent with claims that the 5 sheets of ${\cal{C}}_{\rm Tate}$ locally behave like a sum of five exceptional lines of $dP_9$ that transform as $\mathbf{10}$'s of $SU(5)_{\rm GUT}$ \cite{Marsano:2011nn}.

\subsubsection{Emergence of ${\cal{C}}_{\text{Higgs,loc}}$}

We now return to the task at hand, which is to see the emergence of the Higgs bundle spectral cover.  For this, we restrict ${\cal{C}}_{\text{Tate}}$ to $\delta_3=0$.  
\begin{equation}{\cal{C}}_{\text{Higgs,loc}}={\cal{C}}_{\rm Tate}\cdot_{\tilde{Y}_4} E_3 \,.
\label{higgsspeccov}\end{equation}
Things look unnecessarily messy if we use directly the equations for the fully resolved $\tilde{Y}_4$ inside $\tilde{X}_5$. This mess is largely due to the final blow-up that gives $E_4$.  Since we are looking at a restriction to the divisor $E_3$ and are always taking proper transforms with respect to $E_4$, we can describe \eqref{higgsspeccov} just as well if we blow $E_4$ back down.  Let us do that, then, and return to the 3 times blown up space $X_5^{(3)}$ and the proper transform of $Y_4^{(3)}$ under the first 3 blow-ups.  The equations describing the divisor $E_3$ inside $Y_4^{(3)}$ are
\begin{equation}\begin{split}0 &= \delta_3 \\
&=w\bar{y}\left(b_3w\tilde{z}^2\tilde{\zeta} + b_5\tilde{x}\right) + \bar{\alpha}\tilde{\zeta}\left(b_0w^3\tilde{z}^5\tilde{\zeta}^2 + b_2\tilde{z}^3\tilde{\zeta}w^2\tilde{x}+b_4\tilde{z}w\tilde{x}^2\right)\,,
\end{split}\label{E3inY4simple}\end{equation}
while our equation for the Tate divisor is
\begin{equation}w\bar{y}^2-\bar{\alpha}^2\tilde{\zeta}\tilde{x}^3=0\qquad\text{less}\qquad \bar{y}=\tilde{\zeta}=0\,.\label{Tatesimple}\end{equation}
Now, we can make a few simplifications.  Firstly, when $\delta_3=0$ we are forced to sit at the point
\begin{equation}[w,x,y]=[1,0,0]\end{equation}
on the original $\mathbb{P}^2$ fiber of $X_5$.  Further, we are forced to sit at the point  
\begin{equation}[x_1,y_1,\tilde{z}]=[0,0,1]\end{equation}
on the $\mathbb{P}^2$ corresponding to the first blow-up.  We are confined to a $\mathbb{P}^1$ inside the $\mathbb{P}^2$ corresponding to the second blow-up parametrized by
\begin{equation}[\tilde{x},\tilde{\zeta}]\label{xzetaP1}\end{equation}
and of course we have sections describing coordinates on the $\mathbb{P}^1$ of the third blow-up
\begin{equation}[\bar{y},\bar{\alpha}]\,.\end{equation}
Let us further note that $\bar{\alpha}$ can never be zero inside the Tate divisor because that would imply $\bar{y}=0$ as well.  This means we are really forced to sit on the coordinate patch
\begin{equation}[\bar{y},1]\end{equation}
of this last $\mathbb{P}^1$.  Using all of this information we can rewrite the equations \eqref{E3inY4simple} and \eqref{Tatesimple} as{\footnote{Note that our equations are not homogeneous in these sets of coordinates because the blow-ups are intertwined.  Our three times blown up space can be viewed as a submanifold of a $\mathbb{P}^1$ bundle over a $\mathbb{P}^2$ bundle over a $\mathbb{P}^2$ bundle over $B_3$, not a $\mathbb{P}^1\times\mathbb{P}^2\times\mathbb{P}^2$ bundle over $B_3$.}}
\begin{equation}\begin{split}0 &= \delta_3 \\
&= \bar{y}\left(b_3\tilde{\zeta}+b_5\tilde{x}\right) + \tilde{\zeta}\left(b_0\tilde{\zeta}^2 + b_2\tilde{\zeta}\tilde{x}+b_4\tilde{x}^2\right) \\
&= \bar{y}^2-\tilde{\zeta}\tilde{x}^3\,.
\end{split}\label{higgseqns}\end{equation}
In the $\tilde{x}=1$ patch of \eqref{xzetaP1} we can substitute $\tilde{\zeta}=\bar{y}^2$ to convert the second equation into
\begin{equation}b_0\bar{y}^5 + b_2\bar{y}^3 + b_3\bar{y}^2 + b_4\bar{y}+b_5\,,\label{higgsfinally}\end{equation}
which we immediately recognize as the noncompact Higgs bundle spectral cover of the local model!  Note that $\tilde{x}=0$ implies $\delta_3=\bar{y}=b_0=0$ which represents a single point that compactifies \eqref{higgsfinally}.

Let us point out a couple of other nice features.  Given our explicit description of the matter surfaces \eqref{mattersurfs} in Appendix \ref{app:MatterYuk} it is easy to see that the $\mathbf{10}$ matter surface ${\cal{S}}_{\mathbf{10}}$ of \eqref{mattersurfs} meets ${\cal{C}}_{\text{Higgs,loc}}={\cal{C}}_{\text{Tate}}\doty E_3$ precisely along the curve
\begin{equation}\bar{y}=b_5=0\end{equation}
which we recognize as the $\mathbf{10}$ matter curve of local models.  Further, the $\mathbf{\overline{5}}$ matter surface ${\cal{S}}_{\mathbf{\overline{5}}}$ of \eqref{mattersurfs} meets ${\cal{C}}_{\text{Higgs,loc}}={\cal{C}}_{\text{Tate}}\doty E_3$ along the restriction of $b_3w\tilde{z}^2\tilde{\zeta}+b_5\tilde{x}$ to (\ref{E3inY4simple},\ref{Tatesimple}){\footnote{We must go back to the full resolution in order to see this effectively.  In that setting, what we have more concretely is that the intersection ${\cal{C}}_{\text{Tate}}\doty E_3\doty {\cal{S}}_{\mathbf{\overline{5}}}$ is the restriction of $\hat{y}$ to ${\cal{C}}_{\text{Tate}}\doty E_3$.  When we blow down $E_4$ $\hat{y}$ passes to $b_3w\tilde{z}^2\tilde{\zeta}+b_5\tilde{x}$.}}.  This is exactly the $\mathbf{\overline{5}}$ matter curve of local models.

\subsubsection{Spectral Cover Fluxes}

Now it is clear how to construct surfaces of the type described in \cite{Marsano:2011nn} that are supposed to correspond to curves inside ${\cal{C}}_{\text{Higgs,loc}}$.  First we start with $p^*\Sigma$ where $p$ is the projection map
\begin{equation}p:{\cal{C}}_{\text{Higgs,loc}}\rightarrow S_2\end{equation}
and $\Sigma$ is the restriction of a divisor $D$ in $B_3$ to $S_2$.  We are instructed to first write
\begin{equation}{\cal{S}}_{p^*D} = {\cal{C}}_{\rm Tate}\cdot D\end{equation}
and then subtract classes inherited from $Y_4$ so that ${\cal{S}}_{p^*\Sigma}$ is orthogonal to all horizontal and vertical surfaces.  The result is
\begin{equation}{\cal{S}}_{p^*D} = {\cal{C}}_{\rm Tate}\cdot D - (3\sigma+6c_1)\cdot D\,.\label{SpD}\end{equation}

One can explicitly verify that this misses all of the Cartan roots except for the root $\alpha_2$ from $E_3$.  Now, we turn to ${\cal{S}}_{\sigma\cdot {\cal{C}}}$ where $\sigma$ here refers to the section of the $\mathbb{P}^1$ bundle in the local model as described in Appendix \ref{app:local}.  It is clear that basically what we want is a surface that contains the curve $\bar{y}=b_5=0$ inside $\delta_3=0$.  The actual surface class that we identify with ${\cal{S}}_{\sigma\cdot{\cal{C}}}$, however, is 
\begin{equation}
{\cal{S}}_{\sigma\cdot {\cal{C}} } = [\bar{y}]\doty [\delta_4] - [b_5]\doty E_1 
=  c_1\doty (E_4-E_1)-E_3\doty E_4\,,
\label{SsigC}\end{equation}
where we used the intersection relations (\ref{Y4Rels}) in order to write this in terms of the surfaces in (\ref{surfbas1}, \ref{surfbas2}).
The thing that restricts on $E_3={\cal{D}}_{-\alpha_2}$ to the $\mathbf{10}$ matter curve itself is $\bar{y}\cdot \delta_4$.  We have to make a subtraction from this, however, because $[\bar{y}]\cdot [\delta_4]$ meets Cartan surfaces ${\cal{D}}_{-\alpha_i}\cdot D$ other than the one with $i=2$.  The reason this is not inconsistent with our claims that the spectral divisor sheets only meet $\Sigma_{\alpha_2}$ is that the restriction of $[\bar{y}]\cdot [\delta_4]$ to $E_3={\cal{D}}_{-\alpha_2}$ is a particular curve that sits above $b_5=0$.  This is a special locus where the singular fiber is not just $A_4$.  The implication of this is that $[\bar{y}]\cdot [\delta_4]$ does not have quite the intersection properties that are assumed of ${\cal{S}}_{\sigma\cdot {\cal{C}}}$ from the local model{\footnote{For one thing, we expect that taking a suitable `traceless' combination of ${\cal{S}}_{\sigma\cdot {\cal{C}}}$ and ${\cal{S}}_{p^*D}$ will yield a flux that does not break $SU(5)_{\rm GUT}$.  This implicitly assumes that each of these intersects only the Cartan surfaces ${\cal{D}}_{-\alpha_2}\cdot D$ in a way that is inherited from the intersection of ${\cal{C}}_{\text{Tate}}$ with $\Sigma_{\alpha_2}$.  Any departure from this must be corrected.}}.  We can fix this by adding a Cartan flux constructed from a linear combination of surfaces ${\cal{D}}_{-\alpha_i}\doty D$.  The $[b_5]\doty E_1$ is just such a correction.

With these definitions of ${\cal{S}}_{p^*D}$ and ${\cal{S}}_{\sigma\cdot{\cal{C}}}$ in place, we can explicitly verify the intersection formulae derived in \cite{Marsano:2011nn}.  We find that
\begin{equation}\begin{split}{\cal{S}}_{\sigma\cdot{\cal{C}}}^2 &= -c_1S_2^2 \\
S_{\sigma\cdot{\cal{C}}}\cdot S_{p^*D} &= -6c_1S_2 D\\
S_{p^*D}^2 &= -30S_2D^2\,.\end{split}\end{equation}
Written in the standard notation of local models reviewed in Appendix \ref{app:local}, these exactly reproduce the results from \cite{Marsano:2011nn}.

Armed with the identifications \eqref{SpD} and \eqref{SsigC} we can now construct a $G$-flux that corresponds to the ``traceless flux" of generic $SU(5)$ local models \cite{Donagi:2009ra} which we review briefly in Appendix \ref{app:local}.  We have
\begin{equation}\begin{split}G_{\text{local model}} &= 5{\cal{S}}_{\sigma} - {\cal{S}}_{p^*c_1}\\
&= -\left(5E_3\doty E_4 + c_1\doty\left[2E_1-E_2-2E_3-4E_4\right]\right)\,.
\end{split}\end{equation}
This is exactly proportional to the flux \eqref{bruteGflux} that we constructed by brute force in section \ref{subsubsec:quantizedG}.  Further, we know from local models that this flux should be half-integer quantized so we get exactly \eqref{bruteGflux} which we already know induces a spectrum consistent with local models.

Actually, we can connect flux quantization in the local and global models more directly as follows.  Recall that in local models the flux is quantized in such a way that the curve $\gamma$ in ${\cal{C}}_{\text{Higgs,loc}}$ satisfies
\begin{equation}\gamma+\frac{1}{2}r \in H_2({\cal{C}}_{\text{Higgs,loc}},\mathbb{Z})\,,\end{equation}
where $r$ is the ramification divisor of the 5-sheeted covering ${\cal{C}}_{\text{Higgs,loc}}\rightarrow S_2$.  Recall further that
\begin{equation}r = {\cal{C}}_{\text{Higgs,loc}}\cdot \left({\cal{C}}_{\text{Higgs},loc}-\sigma-\sigma_{\infty}\right)\,.\end{equation}
Translating to the notation of this paper we find that the surface in $\tilde{Y}_4$ corresponding to $r$ according to the identifications \eqref{SpD} and \eqref{SsigC} is
\begin{equation}{\cal{S}}_r = 3S_{\sigma}+S_{p^*(5c_1-4S_2)}\end{equation}
This allows us to verify explicitly that the odd parts of ${\cal{S}}_r$ and $c_2(\tilde{Y}_4)$ agree
\begin{equation}\ba
{\cal{S}}_r^{(odd)} &= c_1E_2+E_3E_4 \cr
c_2(\tilde{Y}_4)^{(odd)}&= c_1E_2+E_3E_4  \,, 
\ea
\end{equation}
To our knowledge, this is the first explicit demonstration of how the quantization rules in the local and global models are connected.

\section*{Acknowledgements}

We thank M.~Esole, H.~Jockers, N.~Saulina, and S.~Sethi for helpful discussions and D.~Morrison for very helpful comments on a draft of this paper.  We are especially grateful to H.~Clemens for a number of helpful and inspiring conversations about spectral covers.  SSN thanks the Caltech theory group and the Kavli Institute for Theoretical Physics for hospitality during this work.  We are both grateful to the Aspen Center for Physics for hospitality as we completed this work.  The work of JM is supported by DOE grant DE-FG02-90ER-40560 and NSF grant PHY-0855039.  The work of SSN was supported in part by DARPA under Grant No.
HR0011-09-1-0015 and by the National Science Foundation under Grant No. PHY05-51164.

\newpage

\appendix


\section{Details on Matter and Yukawas}
\label{app:MatterYuk}

In this Appendix we give details on the description of various singular fibers in section \ref{sec:yukawas}. 

\subsection{Basic Properties of $\tilde{X}_5$ and $\tilde{Y}_4$}

We begin by collecting a few useful things about the space $\tilde{X}_5$ and the smooth Calabi-Yau 4-fold $\tilde{Y}_4$ that we obtain for the particular Calabi-Yau resolution described in section \ref{subsec:sampleblowup}.  The fundamental holomorphic sections are
\begin{equation}\begin{array}{c|c}
\text{Section} & \text{Bundle} \\ \hline
\bar{y} & {\cal{O}}(\sigma +3c_1 - E_1 - E_2 - E_3) \\
\bar{\alpha} & {\cal{O}}(E_2-E_3) \\
\hat{y} & {\cal{O}}(\sigma+3c_1-E_1-E_2-E_4) \\
\bar{\zeta} & {\cal{O}}(E_1-E_2-E_4) \\
\delta_3 & {\cal{O}}(E_3) \\
\delta_4 & {\cal{O}}(E_4) \\
\tilde{x} & {\cal{O}}(\sigma+2c_1-E_1-E_2) \\
\tilde{z} & {\cal{O}}(S_2-E_1)
\end{array}
\label{app:sections}\end{equation}

On $\tilde{X}_5$ there is one nontrivial relation among these sections that descends from the fourth blow-up 
\begin{equation}
b_5\tilde{x} = \hat{y}\delta_4 + \bar{y}\delta_3 - b_3  {w}\tilde{z}^2\bar{\zeta} \delta_4  \,.
\label{app:b5rel}\end{equation}
The 4-fold $\tilde{Y}_4$ is now given by
\be
 w\bar{y}\hat{y} + \bar{\alpha } \bar{\zeta } \left({b_0}
   \delta_4^2 \bar{\zeta}^2 w^3 \tilde{z}^5+{b_2} {\delta_4} \bar{\zeta
   } w^2 \tilde{x} \tilde{z}^3+{b_4} w \tilde{x}^2 \tilde{z}+\bar{\alpha} {\delta_3} \tilde{x}^3\right)=0 \,.
\label{app:Y4}
\ee
We also list the sets of sections that come from projective coordinates of various blow-ups
\begin{equation}\begin{array}{ll}\text{Blow-up 1} & [x_1,y_1,\tilde{z}] = [\bar{\alpha}\tilde{x}\delta_3,\bar{\alpha}\bar{y}\delta_3^2,\tilde{z}] \\
\text{Blow-up 2} & [\tilde{x},\tilde{y},\tilde{\zeta}] = [\tilde{x},\bar{y}\delta_3,\bar{\zeta}\delta_4] \\
\text{Blow-up 3} & [\bar{y},\bar{\alpha}] \\
\text{Blow-up 4} & [\hat{y},\bar{\zeta}]
\end{array}\label{bcoords}\end{equation}

\subsection{``$D_5$" singularity}

We now describe the structure of singular fibers above $\mathbf{10}$ matter curves.  To do this, we study the surface obtained by restricting
\begin{equation}z=b_5=0\label{app:10surf}\end{equation}
to \eqref{app:Y4}.  We naively expect 5 components since $z$ splits into 5 factors
\begin{equation}z = \tilde{z}\bar{\zeta}\bar{\alpha}\delta_3\delta_4\,.\end{equation}
When $b_5=0$, though, something interesting happens.  The relation \eqref{app:b5rel} on $\tilde{X}_5$ becomes
\begin{equation}\left(\hat{y}-b_3w\tilde{z}^2\bar{\zeta}\right)\delta_4 + \bar{y}\delta_3=0\,.\label{b5zero}\end{equation}
This really does not affect the restriction of $\bar{\alpha}=0$ or $\bar{\zeta}=0$ to $b_5=0$ but it does have implications for the restrictions of $\delta_3=0$ and $\delta_4=0$.  Inserting $\delta_3=0$ into \eqref{b5zero} we find
\begin{equation}\left(\hat{y}-b_3w\tilde{z}^2\bar{\zeta}\right)\delta_4=0\,.\end{equation}
The 3-fold $\delta_3=b_5=0$ is apparently reducible in $\tilde{X}_5${\footnote{Note that from \eqref{bcoords} we see that $\delta_3=\delta_4=0$ implies $b_5=0$ via \eqref{app:b5rel} because $\tilde{x}$ cannot be zero there.  The second component is not just $\delta_3=\hat{y}-b_3w\tilde{z}^2\bar{\zeta}\delta_4=0$ because this will include a component where \eqref{app:b5rel} is solved by setting $\tilde{x}=0$ instead of $b_5=0$.}}  
\begin{equation}[\delta_3]\cdot[ b_5] = [\delta_3]\cdot [\delta_4] + [\delta_3]\cdot ([b_5]-[\delta_4]])\,.\end{equation}
In terms of homology classes, we can write this as
\begin{equation}E_3\cdot c_1 = E_3\cdot E_4 + E_3\cdot (c_1-E_4)\,.\end{equation}

We turn now to the restriction of $\delta_4=0$ to $b_5=0$.  In that case, \eqref{app:b5rel} becomes
\begin{equation}\bar{y}\delta_3=0\end{equation}
We find a second copy of our old friend $\delta_3=\delta_4=0$.  The remaining component
is just $\delta_4=\bar{y}=0$.  This automatically implies that $b_5=0$ since $\tilde{x}$ cannot be zero there from \eqref{bcoords}.  To summarize, we have
\begin{equation}[\delta_4]\cdot [b_5] = [\delta_4]\cdot [\delta_3] + [\delta_4]\cdot [\bar{y}]\end{equation}
or
\begin{equation}\begin{split}E_4\cdot c_1 &= E_4\cdot E_3 + E_4\cdot (\sigma+3c_1-E_1-E_2-E_3) \\
& =E_3\cdot E_4 + E_4\cdot (c_1-E_3)
\end{split}
\end{equation}
where the last expression follows from
\begin{equation}(\sigma+2c_1-E_1-E_2)\cdot E_4=0 \,.\end{equation}
This is just the statement that $\delta_4$ and $\tilde{x}$ cannot simultaneously vanish.  To see why not, suppose that they were both zero.  In this case we would have also $\bar{y}\delta_3=0$ but this would contradict the fact that from \eqref{bcoords} not all of $\tilde{x}$, $\bar{y}\delta_3$, and $\bar{\zeta}\delta_4$ can vanish simultaneously.

In the end, we see that $z=b_5=0$ splits into 6 components rather than 5{\footnote{Note that this split is already at the level of 3-folds in $\tilde{X}_5$.  It is easy to see that there is no further split when we restrict to $\tilde{Y}_4$.}}.  Each of these components has a $\mathbb{P}^1$ fiber and it is a straightforward matter to compute the Cartan charges of the $\mathbb{P}^1$.  We summarize everything in the following table

\begin{equation}\label{MatterTable}
\begin{array}{c|c|c|c}\text{Component of }(z=b_5=0)|_{\tilde{Y}_4}  & \text{Equations in } \tilde{Y}_4  & \text{Multiplicity} & \text{Cartan charges}  \\ \hline
(S_2-E_1)  \doty c_1 &  \tilde{z}=0  & 1  & (1, 0, 0, 1) \\
&  b_5=0  & &  \\ \hline
(E_1-E_2-E_4) \doty c_1  &  \bar{\zeta}=0 & 2 & (-2, 1, 0, 0)  \\
&  b_5=0  & & \\
& (\Rightarrow  \bar{y} = 0 ) & & \\ \hline
(E_2-E_3) \doty c_1 & \bar{\alpha}=0 &1  &(0,1, -2,1)  \\
&  b_5=0  && \\
& (\Rightarrow  \hat{y} = 0 )& &  \\ \hline
E_3   \doty  E_4 &  \delta_3=0  & 2 & (1, -1, 1, -1 )\\
&  \delta_4=0  & & \\ \hline
E_3   \doty  ( c_1- E_4) &  \delta_3=0  & 1  & (0, -1, 0, -1 )\\
& \hat{y}-b_3w\tilde{z}^2\bar{\zeta} = 0   & & \\ \hline
E_4 \doty (c_1 - E_3)  - (E_1 - E_2 - E_4) \cdot c_1 &  \delta_4=0  & 1  & (1, 0, 0, -1)\\
&  \bar{y}=0 & &  \\
 \end{array}\end{equation}

We can now use the tables of section \ref{app:subsec:weights} to associate each set of Cartan charges to an $SU(5)$ weight vector in the standard notation
\begin{equation}\begin{array}{c|c|c|c}
\text{Component of }(z=b_5=0)|_{\tilde{Y}_4} & \text{Multiplicity} & \text{Cartan charges} & \text{Weight} \\ \hline
(E_1-E_2-E_4)\doty c_1 & 2 & (-2,1,0,0) & -\alpha_1 \\
(E_2-E_3)\doty c_1 & 1 & ( 0,1,-2,1) & -\alpha_3 \\
E_3\doty E_4 & 2 & (1,-1,1,-1) & -(\mu_{10}-\alpha_1-\alpha_2-\alpha_3) \\
E_3\doty (c_1-E_4) & 1 & (0,-1,0,1) & \mu_{10}-\alpha_1-2\alpha_2-\alpha_3 \\
E_4\doty (c_1-E_3)-(E_1-E_2-E_4)\cdot c_1 & 1 & (1,0,0,-1)& \mu_{10}-\alpha_2-\alpha_3-\alpha_4
\end{array}\end{equation}


\subsection{``$A_5$" singularity}

Let us now turn to the structure of singular fibers above the $\mathbf{\overline{5}}$ matter curves.  For this, we study the surface obtained by restricting
\be
P\equiv b_0 b_5^2 - b_2 b_3 b_5 + b_3^2 b_4=0\qquad z=0  
\label{app:5barcurve}
\ee
to \eqref{app:Y4}.  For later use note that the class of $P$ is
\be
[P] = 8 c_1 - 5 S_2 \,.
\ee
The condition $P=0$ tells us when the equations
\begin{equation}b_0s^4 + b_2s^2 + b_4=0\quad\text{and}\quad b_3s^2 + b_5=0\end{equation}
admit simultaneous solutions.  It seems reasonable to expect, then, that any funny behavior associated with $P=0$ will have to do with an intersection between
\begin{equation}b_3w\tilde{z}^2(\bar{\zeta}\delta_4)+b_5\tilde{x}\quad\text{and}\quad b_0w^2\tilde{z}^4(\bar{\zeta}\delta_4)^2+b_2w\tilde{z}^2(\delta_4\bar{\zeta})\tilde{x} + b_4\tilde{x}^2\,.\end{equation}
It is further clear that, among the Cartan roots $\Sigma_{\alpha_i}$, the one that will be sensitive to this is the one $\Sigma_{\alpha_2}$ that describes the $\mathbb{P}^1$ fiber of $\delta_3=0$.  So, what we expect is that the restriction of three of the Cartan divisors to \eqref{app:5barcurve} will remain irreducible but the restriction of $E_3\cdot\tilde{Y}_4$ to $P=0$ will split.  Let's look at this splitting more closely.  When we restrict to $\delta_3\cdot \tilde{Y}_4$ we get a 3-fold defined by the equations
\begin{equation}\begin{split}0 &= \delta_3 \\
0 &= w\bar{y}\hat{y} + \bar{\alpha}\bar{\zeta} w\tilde{z} \left(b_0w^2(\delta_4\bar{\zeta})^2\tilde{z}^4 + b_2 w (\delta_4\bar{\zeta})\tilde{x}\tilde{z}^2 + b_4\tilde{x}^2\right)\,,
\end{split}\end{equation}
where $\hat{y}$ is determined by the equation
\begin{equation}\hat{y}\delta_4 = b_3w\tilde{z}^2(\bar{\zeta}\delta_4) + b_5\tilde{x} \,.\end{equation}
Let us now intersect this with $\hat{y}=0$.  The result is a surface defined by
\begin{equation}\begin{split}0 &= \delta_3 \\
0 &= \bar{\alpha}\bar{\zeta}w\tilde{z}  \left(b_0w^2(\delta_4\bar{\zeta})^2\tilde{z}^4 + b_2 w (\delta_4\bar{\zeta})\tilde{x}\tilde{z}^2 + b_4\tilde{x}^2\right) \\
\hat{y} &= 0\,,\end{split}\end{equation}
where also
\begin{equation}b_3w\tilde{z}^2(\bar{\zeta}\delta_4) + b_5\tilde{x}=0\,.\label{b3blahzero}\end{equation}
Since
\begin{equation}\bar{\zeta}|_{\hat{y}=0}=1\qquad w|_{\delta_3=0}=1\qquad \tilde{z}|_{\delta_3=0}=1\end{equation}
this is a sum of two surfaces, one with $\delta_3=\bar{\alpha}=\hat{y}=0$ and the other with $\delta_3=\hat{y}=(b_0w^2(\delta_4\bar{\zeta})^2\tilde{z}^4+\ldots)=0$.  The condition \eqref{b3blahzero} tells us that the second of these always sits inside $\tilde{Y}_4\cdot E_3\cdot P=0$ so that is the irreducible component of $\tilde{Y}_4\cdot E_3\cdot P=0$ that we want.  As a result, we see that
\begin{equation}[E_3]\cdot [\tilde{Y}_4]\cdot P = [E_3]\cdot [\tilde{Y}_4]\cdot \left\{\left( [\hat{y}]-[\bar{\alpha}]\right) - \left([P]-[\hat{y}]+[\bar{\alpha}]\right)\right\}\,.\end{equation}

The degenerate curves are now in classes given by intersecting the surfaces with some divisor $D_1$ such that $D_1\cdot_{B_3}P\cdot_{B_3}S_2=1$.  We now list the components of the restriction of \eqref{app:5barcurve} to $\tilde{Y}_4$ and the Cartan charges of the $\mathbb{P}^1$'s associated to those components
\begin{equation}\begin{array}{c|c|c|c}
\text{Component of }(z=P=0)|_{\tilde{Y}_4} & \text{Equations} & \text{Mult} & \text{Cartan charges} \\ \hline 
(E_1-E_2-E_4)\doty (8c_1-5S_2) & \bar{\zeta}=0 & 1 & (-2,1,0,0) \\
& P=0 & & \\
& \tilde{Y}_4=0 & & \\ \hline
(E_2-E_3)\doty (8c_1-5S_2) & \bar{\alpha}=0 & 1 & (0,1,-2,1) \\ 
& P=0 & & \\
& \tilde{Y}_4=0 & & \\ \hline 
E_3\doty (\sigma+3c_1-E_1-2E_2+E_3-E_4) & \delta_3=0 & 1 & (0,-1,1,0) \\
& \bar{\alpha}=0 & & \\
& \hat{y}=0 & & \\ \hline
E_3\doty (5c_1-5S_2-\sigma+E_1+2E_2-E_3+E_4) & \delta_3=0 & 1 & (1,-1,0,0) \\
& \hat{y}=0 & & \\
& b_0w^2(\delta_4\bar{\zeta})^2\tilde{z}^4+b_2w(\delta_4\bar{\zeta})\tilde{x}\tilde{z}^2 & & \\
& +b_4\tilde{x}^2=0 & & \\ \hline
E_4\doty (8c_1-5S_2) & \delta_4=0 & 1 & (0,0,1,-2) \\ 
& P=0 & & \\
& \tilde{Y}_4=0 & &
\end{array}\end{equation}
As indicated, the multiplicities are all 1.  We write $\tilde{Y}_4=0$ when one of the three equations we can use to define the given component as a complete intersection in $\tilde{X}_5$ is the defining equation of $\tilde{Y}_4$.  As before, we can use the tables of section \ref{app:subsec:weights} to identify each set of Cartan charges with a particular $SU(5)$ weight vector in the standard notation
\begin{equation}\begin{array}{c|c|c|c}
\text{Component of }(z=P=0=0)|_{\tilde{Y}_4} & \text{Multiplicity} & \text{Cartan charges} & \text{Weight} \\ \hline
(E_1-E_2-E_4)\doty (8c_1-5S_2) & 1 & (-2,1,0,0) & -\alpha_1 \\
(E_2-E_3)\doty (8c_1-5S_2) & 1 & ( 0,1,-2,1) & -\alpha_3 \\
E_3\doty (\sigma+3c_1-E_1-2E_2+E_3-E_4) & 1 & (1,-1,1,-1) & \mu_5-\alpha_1-\alpha_2 \\
E_3\doty (5c_1-5S_2-\sigma+E_1+2E_2-E_3+E_4) & 1 & (0,-1,0,1) & -(\mu_5-\alpha_1) \\
E_4\doty (8c_1-5S_2) & 1 & (1,0,0,-1)& -\alpha_4
\end{array}\end{equation}


\subsection{``$E_6$" points}

Let us now consider what happens above the ``$E_6$" points where $b_4=b_5=0$.  Really, when we restrict $\tilde{Y}_4$ to $b_4=b_5=z=0$ we expect to find a curve in general.  What we will find is in fact a reducible collection of curves.  It is easy to see that there are six components in general.  We can write these as
\begin{equation}
\begin{array}{c|c|c}\text{Name for curve }\Sigma & \text{Equations for curve }\Sigma & \text{Sections that take the value ``1" on }\Sigma \\ \hline
\Sigma_{\bar{\zeta}} & \bar{\zeta}=0 & w=1 \\
& \bar{y}=0 & \hat{y}=1 \\
& \delta_4=0 & \\ \hline
\Sigma_{\bar{\alpha}} & \bar{\alpha}=0 & w=1 \\
& \hat{y}=0 & \bar{y}=1 \\
& \delta_3-b_3\delta_4\bar{\zeta}=0 & \tilde{z}=1 \\ \hline
\Sigma_{34\bar{y}} & \delta_3=0 & w=1 \\
& \delta_4=0 & \tilde{z}=1 \\
& \bar{y}=0 & \tilde{x}=1 \\
& & \bar{\alpha}=1 \\ \hline
\Sigma_{34\hat{y}} & \delta_3=0 &  w=1 \\
& \delta_4=0 & \tilde{z}=1 \\
& \hat{y}=0 & \tilde{x}=1 \\
& & \bar{\zeta}=1 \\ \hline
\Sigma_3 & \delta_3=0 & w=1 \\
& \hat{y}-b_3\bar{\zeta}=0 & \tilde{z}=1 \\
& b_3\bar{y}+\bar{\alpha}\delta_4(b_2\tilde{x}+b_0\delta_4)=0 & \\ \hline
\Sigma_{\tilde{z}}& \tilde{z}=0 & \\
& \bar{y}\delta_3+\hat{y}\delta_4=0 & \\
& w\bar{y}\hat{y}+\tilde{x}^3\bar{\alpha}^2\delta_3\bar{\zeta}=0 & 
\end{array}\end{equation}
We have written each of these as 3 equations in $\tilde{X}_5$ that are supplemented by $b_4=b_5=0$.  One would naively expect for each set of 5 equations like this to specify a point but in fact each one specifies an irreducible curve that moreover sits nicely inside $\tilde{Y}_4$.  We have also given names to each of these curves in order to simplify the presentation that follows.

Our next objective is to determine how to write each of these as the restriction of some surface in $\tilde{X}_5$ to a curve in $\tilde{Y}_4$.  The first few are easy
\begin{equation}\begin{split}
\Sigma_{\bar{\zeta}} &= \left([b_4]\cdot [b_5] \cdot [\bar{\zeta}]\right)_{\tilde{Y}_4} \\
\Sigma_{\bar{\alpha}} &= \left([b_4]\cdot [b_5]\cdot [\bar{\alpha}]\right)_{\tilde{Y}_4} \\
\Sigma_{\tilde{z}} &= \left([b_4]\cdot [b_5] \cdot [\tilde{z}]\right)|_{\tilde{Y}_4} \,.
\end{split}\end{equation}
The rest require a bit more care.  To address them, let us note that
\begin{equation}\left([b_4]\cdot [\delta_4]\cdot [\bar{y}]\right)_{\tilde{Y}_4} = \Sigma_{\bar{\zeta}}+\Sigma_{34\bar{y}}\,,\end{equation}
while
\begin{equation}\left([\delta_3]\cdot [\delta_4]\cdot [b_4]\right)_{\tilde{Y}_4} = \Sigma_{34\bar{y}}+\Sigma_{34\hat{y}}\,.\end{equation}
Finally,
\begin{equation}\left([\delta_3]\cdot [\hat{y}-b_3w\bar{\zeta}]\cdot [b_4]\right)|_{\tilde{Y}_4} = \Sigma_3+\Sigma_{\tilde{x}}\,,\end{equation}
where $\Sigma_{\tilde{x}}$ is given by the equations
\begin{equation}\Sigma_{\tilde{x}}: \delta_3=\hat{y}-b_3w\bar{\zeta} = b_4 = b_3\bar{y}+\bar{\alpha}\delta_4(b_2\tilde{x}+b_0\delta_4)=\tilde{x}=0\,.\end{equation}
The curve $\Sigma_{\tilde{x}}$ is actually writeable as a complete intersection itself
\begin{equation}\Sigma_{\tilde{x}} = \left([\delta_3]\cdot [b_4]\cdot [\tilde{x}]\right)_{\tilde{Y}_4}\,.\end{equation}
We can therefore write
\begin{equation}\begin{split}\Sigma_{34\bar{y}} &= \left([b_4]\cdot [\delta_4]\cdot [\bar{y}] - [b_4]\cdot [b_5]\cdot [\bar{\zeta}]\right)_{\tilde{Y}_4} \\
\Sigma_{34\hat{y}} &= \left([b_4]\cdot [\delta_4]\cdot ([\delta_3]-[\bar{y}]) + [b_4]\cdot [b_5]\cdot [\bar{\zeta}]\right)_{\tilde{Y}_4} \\
\Sigma_3 &= \left([\delta_3]\cdot [b_4]\cdot ([\hat{y}-b_3w\bar{\zeta}]-[\tilde{x}])\right)_{\tilde{Y}_4} \,.
\end{split}\end{equation}
We summarize this in the table
\begin{equation}\begin{array}{c|c|c|c|c}\text{Curve} & \text{Equations in }\tilde{Y}_4 & \text{Homological class} & \text{Cart Chges} & \text{Mult} \\ \hline
\Sigma_{\bar{\zeta}} & \bar{\zeta}=0 & (E_1-E_2-E_4)\cdot c_1\cdot (2c_1-S_2) & (-2,1,0,0) & 2 \\
& \bar{y}=0 & &  \\
& \delta_4=0 & &  \\ \hline
\Sigma_{\bar{\alpha}} & \bar{\alpha}=0 & (E_2-E_3)\cdot c_1\cdot (2c_1-S_2) & (0,1,-2,1) & 1 \\
& \hat{y}=0 & & \\
& \bar{y}\delta_3-b_3 w\tilde{z}^2\delta_4\bar{\zeta}=0 & & \\ \hline
\Sigma_{34\bar{y}} & \delta_3=0 & (2c_1-S_2) & (1,0,0,-1) & 3 \\
& & \cdot \left\{E_4\cdot \left(\sigma+3c_1-E_1-E_2-E_3\right)\right. & \\
& & \left.-(E_1-E_2-E_4)\cdot c_1\right\} & &\\
& \delta_4=0 & & \\
& \bar{y}=0 & & \\ \hline
\Sigma_{34\hat{y}} & \delta_3=0 & (2c_1-S_2) & (0,-1,1,0) & 2 \\
& & \cdot \left\{E_4\cdot\left(E_3-[\sigma+3c_1-E_1-E_2-E_3]\right)\right. & \\
& & \left. +c_1(E_1-E_2-E_4)\right\} & \\
& \delta_4=0 & & \\
& \hat{y}=0 & & \\ \hline
\Sigma_3 & \delta_3=0 & E_3\cdot (2c_1-S_2)\cdot (c_1-E_4) & (0,-1,0,1) & 1 \\
& \hat{y}-b_3w\tilde{z}^2\bar{\zeta}=0 & & \\
& \bar{y}\hat{y}+ w\tilde{z}^3\bar{\alpha}\delta_4\bar{\zeta}^2(b_2\tilde{x}+b_0w\tilde{z}^2\delta_4\bar{\zeta}) & & \\ \hline
\Sigma_{\tilde{z}} & \tilde{z}=0 & (S_2-E_1)\cdot c_1\cdot (2c_1-S_2) & (1,0,0,1) & 1 \\
& \bar{y}\delta_3+\hat{y}\delta_4=0 & & \\
& w\bar{y}\hat{y}+\tilde{x}^3\bar{\alpha}^2\delta_3\bar{\zeta}=0 & & \\ 
\end{array}\end{equation}

$\Sigma_{\tilde{z}}$ is just the extended node which we have been leaving out so far because it is a harmless spectator.  For the rest, we use the tables of section \ref{app:subsec:weights} to identify each set of Cartan charges with a particular $SU(5)$ weight vector in the standard notation
\begin{equation}\begin{array}{c|c|c|c}\text{Curve} & \text{Cartan charges} & \text{Weight} & \text{Multiplicity} \\ \hline
\Sigma_{\bar{\zeta}} & (-2,1,0,0) & -\alpha_1 & 2 \\
\Sigma_{\bar{\alpha}} & (0,1,-2,1) & -\alpha_3 & 1 \\
\Sigma_{34\bar{y}} & (1,0,0,-1) & \mu_{10}-\alpha_2-\alpha_3-\alpha_4 & 2 \\
\Sigma_{34\hat{y}} & (0,-1,1,0) & \mu_5 - \alpha_1-\alpha_2-\alpha_3 & 2 \\
\Sigma_3 & (0,-1,0,1) & \mu_{10}-\alpha_1-2\alpha_2-\alpha_3 & 1 
\end{array}\end{equation}


\subsection{``$D_6$" points}

The ``$D_6$" points are characterized by $b_3 =b_5=0$. Again we expect curves as restrictions inside $\tilde{Y}_4$ to
$b_3=b_5 =z=0$. The class of $b_3$ is $3 c_1 - 2 S_2$.  We find

\begin{equation}
\begin{array}{c|c|c|c}
\text{Curve }\Sigma & \text{Equations for curve $\Sigma$ on $[b_3]$}& \text{Homology class} & \text{Mult/Cartan}  \cr\hline
\Sigma^{D_{6}}_{\bar{\zeta}} & \bar{\zeta}=0 & [b_3] \cdot [b_5] \cdot[\bar{\zeta}]\cdot [\tilde{Y}_4]&     2\cr
						& \bar{y}=0    && (-2, 1, 0 ,0) \cr
						& \delta_4 =0  &&\cr\hline
\Sigma^{D_6}_{\bar{\alpha}} & \bar{\alpha} =0 & [b_3 ] \cdot [b_5] \cdot [\bar{\alpha}] \cdot  [\tilde{Y}_4] & 2 \cr 		
						& \hat{y} =0&& (0, 1, -2, 1)\cr
						& \delta_3 =0 &&	 \cr\hline
\Sigma^{D_6}_{34} & \delta_3 = 0 & [\delta_3]\cdot[\delta_4]\cdot [b_3] \cdot  [\tilde{Y}_4]& 2  \cr
				& \delta_4=0 &&(1, -1, 1, -1)\cr
				& w \bar{y}\hat{y} + \bar{\alpha} \bar{\zeta} b_4 w \tilde{x}^2 \tilde{z} =0 	&& \cr\hline		
\Sigma^{D_6}_{3} & 	 	\delta_3 = 0 & [\delta_3]\cdot ([b_5] - [\delta_4]) \cdot [b_3]\cdot    [\tilde{Y}_4]& 2^*  \cr
				& \hat{y}=0 &   - [b_3] \cdot [b_5] \cdot \bar{\alpha} \cdot [{\tilde{Y}_4}]         &(0, -1, 1, 0) \cr
				&b_0\delta_4^2\bar{\zeta}^2w^3\tilde{z}^5+b_2\delta_4\bar{\zeta}w^2\tilde{x}\tilde{z}^3+b_4w\tilde{x}^2\tilde{z}=0  &&\cr\hline
\Sigma^{D_6}_{4} & \delta_4 =0 & [\delta_4]\cdot ([b_5] - [\delta_3]) \cdot [b_3]   \cdot    [\tilde{Y}_4]& 1\cr
						& \bar{y} =0 & - [b_3]\cdot [b_5] \cdot \bar{\zeta} \cdot [\tilde{Y}_4] &   (1, 0, 0, -1)\cr
						& b_4 w \tilde{x}^2 \tilde{z} + \bar\alpha \delta_3 \tilde{x}^3 =0 && \cr\hline								
\Sigma^{D_6}_{\tilde{z}} & \tilde{z} =0 & [b_3] \cdot [b_5] \cdot [\tilde{z}] \cdot [\tilde{Y}_4]	& 1 \cr
					& 	 \bar{\alpha}^2\bar{\zeta}\delta_3\tilde{x}^3 + w\bar{y}\hat{y} = 0  &&   (1, 0,0 , 1) \cr
					&  (\text{where }\hat{y}\delta_4+\bar{y}\delta_3=0)  && \cr
					& b_3=0 &				
\end{array}
\end{equation}
A few remarks about this: the multiplicities of the curves are all inherited from the $D_5$ matter surfaces, except for the case of $\Sigma_3^{D_6}$. This curve is defined by 
$\delta_3= \hat{y} =b_3=0$ and  
\be\label{QuadratricEq}
b_0\delta_4^2\bar{\zeta}^2w^3\tilde{z}^5+b_2\delta_4\bar{\zeta}w^2\tilde{x}\tilde{z}^3+b_4w\tilde{x}^2\tilde{z}=0  \,.
\ee
These imply automatically $b_5=0$. 
The restriction of this equation to $\delta_3=0$ makes $b_{0}, b_2, b_4$ vary only over $S_2$, and further restriction to $b_3=b_5=0$ restricts them to a point. So, as explained in \cite{Esole:2011sm} we can treat (\ref{QuadratricEq}) as a quadratic equation  in $\tilde{x}$ and $\bar{\zeta}$ with constant coefficients, which has two roots. These two solutions give rise to the multiplicity $2$ for $\Sigma_{3}^{D_6}$ -- the Cartan charges reflect this, as the curve $ [\delta_3]\cdot ([b_5] - [\delta_4]) \cdot [b_3]\cdot [\tilde{Y}_4]  - [b_3] \cdot [b_5] \cdot \bar{\alpha} \cdot [{\tilde{Y}_4}] $ has charge $(0, -2, 2, 0)$ but is reducible. 

Finally, we use the tables of section \ref{app:subsec:weights} to identify each set of Cartan charges with a particular $SU(5)$ weight vector in the standard notation
\begin{equation}\begin{array}{c|c|c|c}
\text{Curve} & \text{Cartan charges} & \text{Weight} & \text{Multiplicity} \\ \hline
\Sigma_{\bar{\zeta}}^{D6} & (-2,1,0,0) & -\alpha_1 & 2 \\
\Sigma_{\bar{\alpha}}^{D6} & (0,1,-2,1) & -\alpha_3 & 2 \\
\Sigma_{34}^{D6} & (1,-1,1,-1) & -(\mu_{10}-\alpha_1-\alpha_2-\alpha_3) & 2 \\
\Sigma_3^{D6} & (0,-1,1,0) & \mu_5-\alpha_1-\alpha_2 & 2 \\
\Sigma_4^{D6} & (1,0,0,-1) & \mu_{10}-\alpha_2-\alpha_3-\alpha_4 & 1
\end{array}\end{equation}

\subsection{$SU(5)$ weights and roots}
\label{app:subsec:weights}

For convenience we list the weights and roots of the $\mathbf{24}$, $\mathbf{10}$, and $\mathbf{5}$ representations of $SU(5)$ in this subsection.

\begin{equation}\begin{array}{c|c}\hline
\text{Cartan charges of }\mathbf{24}& \text{Root} \\\hline
(2,-1,0,0) & \alpha_1 \\
(-1,2,-1,0) & \alpha_2 \\
(0,-1,2,-1) & \alpha_3 \\\
(0,0,-1,2) & \alpha_4 \\
(1,1,-1,0) & \alpha_1+\alpha_2 \\
(-1,1,1,-1) & \alpha_2+\alpha_3 \\
(0,-1,1,1) & \alpha_3+\alpha_4 \\
(1,0,1,-1) & \alpha_1+\alpha_2+\alpha_3 \\
(-1,1,0,1) & \alpha_2+\alpha_3+\alpha_4 \\
(1,0,0,1) & \alpha_1+\alpha_2+\alpha_3+\alpha_4 \\
(-2,1,0,0) & -\alpha_1 \\
(1,-2,1,0) & -\alpha_2 \\
(0,1,-2,1) & -\alpha_3 \\
(0,0,1,-2) & -\alpha_4 \\
(-1,-1,1,0) & -\alpha_1-\alpha_2 \\
(1,-1,-1,1) & -\alpha_2-\alpha_3 \\
(0,1,-1,-1) & -\alpha_3-\alpha_4 \\
(-1,0,-1,1) & -\alpha_1-\alpha_2-\alpha_3 \\
(1,-1,0,-1) & -\alpha_2-\alpha_3-\alpha_4 \\
(-1,0,0,-1) & -\alpha_1-\alpha_2-\alpha_3-\alpha_4\\ 
 \\ \hline

\text{Cartan charges of }\mathbf{5} & \text{Weight} \\ \hline
(1,0,0,0) & \mu_5 \\
(-1,1,0,0) & \mu_5-\alpha_1 \\
(0,-1,1,0) & \mu_5-\alpha_1-\alpha_2 \\
(0,0,-1,1) & \mu_5-\alpha_1-\alpha_2-\alpha_3 \\
(0,0,0,-1) & \mu_5-\alpha_1-\alpha_2-\alpha_3-\alpha_4\\ 
 \\ \hline

\text{Cartan chages of }\mathbf{10} & \text{Weight} \\ \hline
(0,1,0,0) & \mu_{10} \\
(1,-1,1,0) & \mu_{10}-\alpha_2 \\
(-1,0,1,0) & \mu_{10}-\alpha_1-\alpha_2 \\
(1,0,-1,1) & \mu_{10}-\alpha_2-\alpha_3 \\
(-1,1,-1,1) & \mu_{10}-\alpha_1-\alpha_2-\alpha_3 \\
(1,0,0,-1) & \mu_{10}-\alpha_2-\alpha_3-\alpha_4 \\
(0,-1,0,1) & \mu_{10}-\alpha_1-2\alpha_2-\alpha_3 \\
(-1,1,0,-1) & \mu_{10}-\alpha_1-\alpha_2-\alpha_3-\alpha_4 \\
(0,-1,1,-1) & \mu_{10}-\alpha_1-2\alpha_2-\alpha_3-\alpha_4 \\
(0,0,-1,0) & \mu_{10}-\alpha_1-2\alpha_2-2\alpha_3-\alpha_4
\end{array}\end{equation}


\section{Some properties of $\tilde{Y}_4$ and $\tilde{X}_5$}
\label{app:properties}

In this Appendix, we review some useful properties of $\tilde{Y}_4$ and $\tilde{X}_5$ along with simple ways to perform intersection computations.  We focus on the specific resolution studied in detail in sections \ref{sec:yukawas} and \ref{sec:Gflux}.

\subsection{Intersection relations in $\tilde{X}_5$ and $\tilde{Y}_4$}

We first describe a number of nontrivial relations that one encounters in the study of divisor intersections in $\tilde{X}_5$ and $\tilde{Y}_4$.

\subsubsection{Intersection relations in $\tilde{X}_5$}

Let us start with nontrivial relations involving divisor intersections in $\tilde{X}_5$.  We have described all of these at various points in section \ref{sec:generalities} but for completeness we list them all in one place here
\begin{equation}\begin{split} 0 &= \sigma(\sigma+2c_1)(\sigma+3c_1) \\
0&= \sigma\cdot E_i \\
0&= (\sigma+2c_1-E_1)(\sigma+3c_1-E_1)(S_2-E_1) \\
0&= (\sigma+2c_1-E_1-E_2)(\sigma+3c_1-E_1-E_2)(E_1-E_2) \\
0&= (\sigma+3c_1-E_1-E_2-E_3)(E_2-E_3) \\
0&= (\sigma+3c_1-E_1-E_2-E_4)(E_1-E_2-E_4) \,.
\end{split}\label{blowuprels}\end{equation}
The first of these represents the failure of the sections $w$, $x$, and $y$ that were homogeneous coordinates on the $\mathbb{P}^2$ fiber of $X_5$ to vanish.  The second acknowledges that all blow-ups are performed in the $w=1$ coordinate patch so that the divisor $w=0$ in the class $\sigma$ does not intersect any exceptional divisors.  The remaining 4 are the ``blow-up" relations that follow from the fact that we cannot have simultaneous vanishing of all homogeneous coordinates on any $\mathbb{P}^2$ or $\mathbb{P}^1$ that is grown when we do a blow-up.

In addition to these we have a few more relations.  To see them, let us explicitly list the sets of homogeneous coordinates associated to each blow-up that cannot simultaneously vanish
\begin{equation}\begin{array}{ll}\text{Blow-up 1} & [x_1,y_1,\tilde{z}] = [\bar{\alpha}\tilde{x}\delta_3,\bar{\alpha}\bar{y}\delta_3^2,\tilde{z}] \\
\text{Blow-up 2} & [\tilde{x},\tilde{y},\tilde{\zeta}] = [\tilde{x},\bar{y}\delta_3,\bar{\zeta}\delta_4] \\
\text{Blow-up 3} & [\bar{y},\bar{\alpha}] \\
\text{Blow-up 4} & [\hat{y},\bar{\zeta}]
\end{array}\label{blowupcoords}\end{equation}
From this we see that there cannot be simultaneous solutions to
\begin{equation}\tilde{z}=\delta_3=0\text{  or  }\tilde{z}=\bar{\alpha}=0\,.\end{equation}
In other words
\begin{equation}E_2\cdot (S_2-E_1) = E_3\cdot (S_2-E_1) = 0\,.\label{ztilderels}\end{equation}

\subsection{Intersection relations in $\tilde{Y}_4$}

When we restrict divisors on $\tilde{X}_5$ to $\tilde{Y}_4$ we find a number of new relations.  To see these, it is helpful to recall the defining equation of $\tilde{Y}_4$
\begin{equation}w\bar{y}\hat{y} + \bar{\alpha}\bar{\zeta}\left(b_0w^3\tilde{z}^5\delta_4^2\bar{\zeta}^2 + b_2w^2\tilde{x}\tilde{z}^3\delta_4\bar{\zeta} + b_4w\tilde{z}\tilde{x}^2 + \bar{\alpha}\delta_3\tilde{x}^3\right)\label{appY4eq}\end{equation}
and also the nontrivial relation connecting some of the sections that we get from the final blow-up
\begin{equation}b_5\tilde{x}=\hat{y}\delta_4 + \bar{y}\delta_3 - b_3w\tilde{z}^2\bar{\zeta}\delta_4 \,.
\label{appY4rel}\end{equation}
Combining this information with \eqref{blowupcoords} it is easy to see explicitly that there are a number of pairs of sections that cannot simultaneously vanish on $\tilde{Y}_4$.  We list them in the following table
\begin{equation}\begin{array}{c|c}\text{Pair of sections that do not vanish simult. on }\tilde{Y}_4 & \text{Relation} \\ \hline
\tilde{z},\,\,\,\hat{y} & (\sigma+3c_1-E_1-E_2-E_4)\cdot_{\tilde{Y}_4}(S_2-E_1)=0 \\
\tilde{x},\,\,\,\delta_4 & E_4\cdot_{\tilde{Y}_4}(\sigma+2c_1-E_1-E_2)=0 \\
\tilde{x},\,\,\,\bar{\zeta} & (\sigma+2c_1-E_1-E_2)\cdot_{\tilde{Y}_4}(E_1-E_2-E_4)=0 \\
\bar{\alpha},\,\,\,\bar{\zeta} & (E_2-E_3)\cdot_{\tilde{Y}_4}(E_1-E_2-E_4)=0 \\ \hline
\bar{y},\,\,\,\bar{\alpha} & (\sigma+3c_1-E_1-E_2-E_3)\cdot_{\tilde{Y}_4}(E_2-E_3)=0 \\
\hat{y},\,\,\,\bar{\zeta} & (\sigma+3c_1-E_1-E_2-E_4)\cdot_{\tilde{Y}_4}(E_1-E_2-E_4)=0 \\
\tilde{z},\,\,\,\delta_3 & E_3\cdot_{\tilde{Y}_4}(S_2-E_1)=0 \\
\tilde{z},\,\,\,\bar{\alpha} & (E_2-E_3)\cdot_{\tilde{Y}_4}(S_2-E_1)=0 \\
\end{array}\label{pairrels}\end{equation}
The last four are just the restriction of relations that hold on $\tilde{X}_5$ to $\tilde{Y}_4$.  The first two of these four are the blow-up relations \eqref{blowuprels} associated to the third and fourth blow-ups while the second two are the restriction of \eqref{ztilderels} to $\tilde{Y}_4$.

There is one more relation that we will find useful in the text.  The first entry of \eqref{pairrels} is not very helpful for our discussion of $G$-fluxes in section \ref{sec:Gflux}, where we really want to eliminate trivial combinations of surfaces to avoid clutter and confusion.  The reason is that it involves $\sigma\cdot S_2$ which we manifestly do not use in that discussion.  The fact that $\tilde{z}$ and $\hat{y}$ do not simultaneously vanish, however, does have the following implication.  Consider the two surfaces obtained by restricting
\begin{equation}\tilde{z} = \bar{y}\delta_3+b_5\tilde{x}=0\end{equation}
and
\begin{equation}\tilde{z} = \bar{y}\delta_3=0\label{surf1}\end{equation}
to $\tilde{Y}_4$.  Both of these describe surfaces in the class
\begin{equation}(S_2-E_1)\cdot_{\tilde{Y}_4} (\sigma +3c_1-E_1-E_2)\,.
\label{surf2}\end{equation}
Let us use \eqref{appY4eq} and \eqref{appY4rel} to describe \eqref{surf1} and \eqref{surf2} more explicitly.  In so doing we will see some of the logic that produced \eqref{pairrels}.

Looking first at \eqref{surf1}, note that $\tilde{z}=0$ and $\bar{y}\delta_3+b_5\tilde{x}=0$ implies by \eqref{appY4rel} that
\begin{equation}\hat{y}\delta_4=0\,.\end{equation}
We know from \eqref{pairrels} that $\hat{y}$ cannot vanish when $\tilde{z}$ does so this surface is really equivalent to the restriction of
\begin{equation}\tilde{z}=\delta_4=0\end{equation}
to $\tilde{Y}_4$.  This establishes that
\begin{equation}(S_2-E_1)\cdot_{\tilde{Y}_4}(\sigma+3c_1-E_1-E_2) = (S_2-E_1)\cdot_{\tilde{Y}_4}E_4\,.\label{newcond1}\end{equation}
Now let's look at \eqref{surf2}.  This naively has two components but we know from \eqref{pairrels} that $\tilde{z}$ and $\delta_3$ cannot simultaneously vanish.  As a result, we only get the component $\tilde{z}=\bar{y}=0$.  From \eqref{appY4eq} we see that whenever $\tilde{z}=\bar{y}=0$ we also have $\bar{\alpha}^2\bar{\zeta}\delta_3\tilde{x}^3=0$.  The only one of these that can vanish when $\tilde{z}$ does is $\bar{\zeta}$.  It is now easy to check that the restriction of $\tilde{z}=\bar{y}=0$ to $\tilde{Y}_4$ is in fact equivalent to the restriction of
\begin{equation}\tilde{z}=\bar{\zeta}=0\,.\end{equation}
This establishes that
\begin{equation}(S_2-E_1)\cdot_{\tilde{Y}_4}(\sigma+3c_1-E_1-E_2) = (S_2-E_1)\cdot_{\tilde{Y}_4}(E_1-E_2-E_4) \,.
\label{newcond2}\end{equation}
We can now take a difference of \eqref{newcond1} and \eqref{newcond2} to get a relation that does not involve $\sigma\cdot S_2$
\begin{equation}(S_2-E_1)\cdot_{\tilde{Y}_4}(E_1-E_2-2E_4)=0\,.\label{newnewrel}\end{equation}
Combined with all but the first relation in \eqref{pairrels} we have a set of 8 relations on the intersections
\begin{equation}E_i\cdot_{\tilde{Y}_4},\quad E_i\cdot_{\tilde{Y}_4}c_1,\quad E_i\cdot_{\tilde{Y}_4}S_2 \,.
\label{appints}\end{equation}
These are exactly the building blocks that we used to construct $G$-flux in section \ref{subsubsec:buildingblocks}.  There are 18 of these and 8 relations so in fact only a 10-dimensional space in the family of surfaces parametrized by \eqref{appints} is nontrivial.  A useful parametrization of that space is in terms of the Cartan surfaces
\begin{equation}E_i\cdot_{\tilde{Y}_4}c_1\quad\text{and}\quad E_i\cdot_{\tilde{Y}_4}S_2 \,.\label{surfbas1}\end{equation}
and two of the $E_i\cdot_{\tilde{Y}_4}E_j$'s.  A convenient choice for two $E_i\cdot_{\tilde{Y}_4}E_j$'s to keep is
\begin{equation}E_2\cdot_{\tilde{Y}_4}\text{ and }E_3\cdot_{\tilde{Y}_4}E_4\,.
\label{surfbas2}\end{equation}
Using the relations \eqref{pairrels} and \eqref{newnewrel} we can derive the following formulae that allow us to replace any other $E_i\cdot_{\tilde{Y}_4}E_j$'s with a linear combination of the surfaces in \eqref{surfbas1} and \eqref{surfbas2}
\begin{equation}\begin{split}\label{Y4Rels}
E_1\cdot_{\tilde{Y}_4}E_1 &= 4c_1\cdot_{\tilde{Y}_4}E_4 - 2E_2\cdot_{\tilde{Y}_4}E_4 + (E_1-2E_4)\cdot_{\tilde{Y}_4}S_2 \\
E_1\cdot_{\tilde{Y}_4}E_2 &= E_2\doty S_2 \\
E_1\cdot_{\tilde{Y}_4}E_3 &= E_3\doty S_2 \\
E_1\cdot_{\tilde{Y}_4}E_4 &= (2c_1-E_2)\doty E_4\\
E_2\cdot_{\tilde{Y}_4}E_2 &= -2E_2\doty E_4 -2c_1\doty (E_1-E_2-2E_4) + (E_1-2E_4)\doty S_2\\
E_2\cdot_{\tilde{Y}_4}E_3 &= -2c_1\doty (E_1-E_2-2E_4)+S_2\doty (E_1-E_2+E_3-2E_4) -(E_2+E_3)\doty E_4\\
E_3\cdot_{\tilde{Y}_4}E_3 &= -2E_2\doty E_4 + (E_1+E_2-E_3-2E_4)\doty S_2 - c_1\doty (2E_1+E_2-3E_3-4E_4) \\
E_4\cdot_{\tilde{Y}_4}E_4 &= -c_1\doty (E_1-E_2-3E_4)-2E_2\doty E_4\,.
\end{split}\end{equation}

\subsection{Method of computing}

In the rest of this Appendix, we describe a simple way to perform some of the computations that we need in $\tilde{X}_5$.  There is nothing particularly novel or insightful here; rather it is a description of how to easily do some calculations that we found useful.

For many purposes we are interested in studying complete intersections of divisors in $\tilde{X}_5$.  In our original space $X_5$ this is rather straightforward.  The only nonzero complete intersections are of the form $\sigma^2 D_1D_2D_3$ for $D_i$ three divisors in $B_3$ and the result of this is
\begin{equation}\sigma^2 D_1D_2D_3 = D_1\cdot_{B_3}D_2\cdot_{B_3}D_3\,.\end{equation}
In $\tilde{X}_5$ we have a number of exceptional divisors to deal with but we can keep all that straight by remembering the following simple facts.  Suppose we blow up a manifold $X$ of dimension $n$ along a variety $A$ of codimension $d$.  In doing this we obtain an exceptional divisor $E$ by growing a $\mathbb{P}^{d-1}$ along the variety $A$.  This $\mathbb{P}^{d-1}$ is parametrized by $d$ projective coordinates that are in fact sections of bundles on the blown-up space of the form ${\cal{O}}(D-E)$.  The fact that they cannot simultaneously vanish will give us a relation of the form
\begin{equation}E^d = \sum_{m=0}^{d-1} D^{n-i}E^m\label{blowuprelationgen}\end{equation}
that allows us to replace any power of $E$ that is greater than $d-1$ with a series of terms that have degree $d-1$ or less.

Now, the exceptional divisor fails to intersect the total transform of any variety of the original space $X$ that has codimension greater than $n-d$.  This means that
\begin{equation}E\cdot \prod_{a=1}^{n-d+1}{\cal{D}}_a=0\label{Evanishing}\end{equation}
for any divisors ${\cal{D}}_a$ that are pulled back from $X$.  If we are interested in computing complete intersections of $n$ divisors, this means that
\begin{equation}E^p\prod_{a=1}^{n-p}{\cal{D}}_a=0\,,\qquad p<d\,.
\end{equation}
So, if we have a complete intersection of $n$ divisors, any powers of $E$ that are $d$ or larger can be replaced using \eqref{blowuprelationgen} with terms of degree less than $d$.  Once we rewrite everything as a polynomial in $E$ of degree less than $d$, any term with a nonzero power of $E$ vanishes by \eqref{Evanishing}.  This means that we can conveniently compute $n$-fold intersections in the once blown-up space by using the replacement \eqref{blowuprelationgen} as many times as possible and then simply replacing $E$ with zero.  This procedure is easily extended to spaces constructed using multiple blow-ups as long as we are careful to treat all intersections involving the newest exceptional divisor before moving backwards in the sequence.  This last subtlety is necessary because in everything above we assumed the ${\cal{D}}_a$'s were total transforms.

\section{Connections with Esole-Yau}
\label{app:EsoleYau}

In this Appendix we collect a few results connecting our notation in terms of global holomorphic sections on $\tilde{X}_5$ with the coordinates used by Esole and Yau \cite{Esole:2011sm}.  None of this is particularly important for the results of this paper but is included for convenience of the reader.

\subsection{The curves $C_{1\pm}$, $C_{2\pm}$, and $C_0$ after the first two blow-ups}

The first two blow-ups successfully resolve the $A_4$ singularity of \eqref{SU5tate} that occurs above the codimension 1 locus $z=0$ in $B_3$.  Singularities remain above loci of codimension 2 in $B_3$ but already at that point it is possible to see explicitly the resolved 2-cycles of the $A_4$ Dynkin diagram.  Esole and Yau explicitly describe these in patches \cite{Esole:2011sm}.  In our approach, we describe them in terms of the global holomorphic sections $w$, $\tilde{x}$, $\tilde{y}$, $\tilde{z}$, $\tilde{\zeta}$, and $\alpha$ on the twice resolved 5-fold $X_5^{(2)}$.  To do this, we simply restrict the equation for $Y_2^{(2)}$ from \eqref{Y42} to the singular fiber $z=\tilde{z}\tilde{\zeta}\alpha=0$.  This will give us a 3-fold that represents the resolution of the singular fiber over $S_2$.  The curves are obtained by supplementing this with a pair of equations on $B_3$ that single out a curve $\Sigma$ on $B_3$ that meets $S_2$ in a single point $p_0$.  We can therefore describe the resolved 2-cycles homologically as 2-cycles inside $\tilde{X}_5$.  We find five components in all
\begin{equation}\begin{array}{c|c|c}\text{2-cycle} & \text{Equations} & \text{Class in }X_5^{(2)} \\ \hline
C_0 & w\tilde{y}^2 - \left[\tilde{x}^3\tilde{\zeta}\alpha^2 + b_5\tilde{x}\tilde{y}w\right]=\tilde{z}=0 & (3\sigma + 6c_1-2E_1-2E_2)\cdot (S_2-E_1)\cdot [\pi_X^*\Sigma] \\
C_{1+} & \tilde{y}=\tilde{\zeta}=0 & (\sigma+3c_1-E_1-E_2)\cdot (E_1-E_2)\cdot [\pi_X^*\Sigma] \\
C_{1-} & \tilde{y}-b_5\tilde{x} = \tilde{\zeta}=0 & (\sigma+3c_1-E_1-E_2)\cdot (E_1-E_2)\cdot [\pi_X^*\Sigma \\
C_{2+} & \tilde{y}=\alpha=0 & (\sigma+3c_1-E_1-E_2)\cdot (E_1-E_2)\cdot [\pi_X^*\Sigma] \\
C_{2-} & \tilde{y}-\left[b_3\tilde{z}^2w\tilde{\zeta}+b_5\tilde{x}\right]=\alpha=0 & (\sigma+3c_1-E_1-E_2)\cdot (E_1-E_2)\cdot [\pi_X^*\Sigma]
\end{array}\end{equation}
We have labeled these curves according to the notation of \cite{Esole:2011sm} for ease of comparison.  Notice that the 5 curves are not homologically distinct inside $X_5^{(2)}$ or $Y_4^{(2)}$.  

\subsection{Two Final Blow-ups}

The last step of the resolution procedure described in section \ref{subsubsec:finalblowups} is a pair of blow-ups along the loci \eqref{FinalBlowupLoci}.  There are 6 choices for this pair, each of which gives rise to a different resolution \cite{Esole:2011sm}.  To faciliate comparison with \cite{Esole:2011sm}, we list here how the global sections we use to define the factored form \eqref{factstruct} are related to the local coordinates of \cite{Esole:2011sm} in the patch ${\cal{U}}_{31}$ on which they focus for the final two blow-ups.
\begin{equation}\begin{array}{c|c}\text{Holomorphic Section} & \text{Local Coordinate in \cite{Esole:2011sm}} \\ \hline
\tilde{y} & y \\
\tilde{y}-b_3w\tilde{z}^2\tilde{\zeta}-b_5\tilde{x} & s \\
\alpha & x \\
\tilde{\zeta} & w \\
b_0 w^3\tilde{z}^5\tilde{\zeta}^2 + b_2\tilde{z}^3\tilde{\zeta}w^2\tilde{x} + b_4\tilde{z}w\tilde{x}^2 + \alpha\tilde{x}^3 & t
\end{array}\end{equation}
In \cite{Esole:2011sm}, the notation ${\cal{E}}_{AB}$ is used to denote the resolution obtained by blowing up $X_5^{(2)}$ along $A=y=0$ and $B=s=0$.  We can translate this to our language as
\begin{equation}\begin{array}{c|cc}\text{Resolution} & u_a & u_b \\ \hline
{\cal{E}}_{xw} & \alpha & \tilde{\zeta} \\
{\cal{E}}_{wx} & \tilde{\zeta} & \alpha \\
{\cal{E}}_{xt} & \alpha & b_0 w^3\tilde{z}^5\tilde{\zeta}^2 + b_2\tilde{z}^3\tilde{\zeta}w^2\tilde{x} + b_4\tilde{z}w\tilde{x}^2 + \alpha\tilde{x}^3 \\
{\cal{E}}_{tx} & b_0 w^3\tilde{z}^5\tilde{\zeta}^2 + b_2\tilde{z}^3\tilde{\zeta}w^2\tilde{x} + b_4\tilde{z}w\tilde{x}^2 + \alpha\tilde{x}^3  & \alpha \\
{\cal{E}}_{wt} & \tilde{\zeta} & b_0 w^3\tilde{z}^5\tilde{\zeta}^2 + b_2\tilde{z}^3\tilde{\zeta}w^2\tilde{x} + b_4\tilde{z}w\tilde{x}^2 + \alpha\tilde{x}^3  \\
{\cal{E}}_{tw} & b_0 w^3\tilde{z}^5\tilde{\zeta}^2 + b_2\tilde{z}^3\tilde{\zeta}w^2\tilde{x} + b_4\tilde{z}w\tilde{x}^2 + \alpha\tilde{x}^3  & \tilde{\zeta}
\end{array}\end{equation}
The resolution that we study explicitly in the text corresponds to ${\cal{E}}_{xw}$.

\section{Connection to Local Models}
\label{app:local}

Many of the results in this paper can be written in terms of intersections that take place inside the surface $S_2$.  This is not surprising and reflects the ability of local models, which rely only on the geometry near $S_2$, to capture many important global features of $\tilde{Y}_4$.  To facilitate the connection of our results to those of local models, we review here some elementary facts about the spectral cover formalism \cite{Donagi:2009ra}, encountering standard notation along the way.  This discussion is not meant to be exhaustive in any way.  The interested reader is referred to the sizeable literature in \cite{Weigand:2010wm} and references contained therein.

In this section only we will use local model notation.  Formulae should not be directly compared with those from the main text without care.

\subsection{Spectral Data}

Local models are based on a worldvolume description of the physics associated with singularities in F-theory compactifications on elliptically fibered Calabi-Yau's above codimension 1 loci in the base $B_3$.  In the oft studied case of F-theory compactifications on Calabi-Yau 4-folds with a surface of $A_4$ singularities, the worldvolume description is an 8-dimensional gauge theory associated to the stack of 7-branes described by the codimension 1 singularity.  The holomorphic data of this theory can nicely be organized into the structure of an underlying $E_8$ gauge theory in which $E_8$ is broken to $SU(5)_{\rm GUT}$ by a nontrivial Higgs bundle.  The Higgs bundle describes two crucial ingredients: a nontrivial expectation value for an adjoint scalar field $\phi$ that varies along $S_2$ and a nontrivial gauge field background on the surface $S_2$.  To break $E_8\rightarrow SU(5)_{\rm GUT}$ both of these should take values in the $SU(5)_{\perp}$ commutant of $SU(5)_{\rm GUT}$ inside $E_8$.

In any coordinate patch on $S_2$, $\phi$ can be diagonalized so that its behavior is locally captured by that of its eigenvalues.  As they vary over $S_2$, the five eigenvalues of $\phi$ will trace out a five-sheeted cover of $S_2$ that must be holomorphic to solve the equations of motion.  The field $\phi$ is a section of the canonical bundle $K_{S_2}$ of $S_2$ so this cover naturally lives in the total space of that bundle
\begin{equation}{\cal{C}}_{\text{Higgs,local}} = b_{0,loc} s^5 + b_{2,loc}s^3 + b_{3,loc}s^2 + b_{4,loc} s + b_{5,loc}\sim b_0\prod_{i=1}^5 (s_i - \lambda_i)\end{equation}
with $s$ a section of $K_{S_2}$.  The $b_{m,loc}$ specify the Casimirs of $\phi$ and, when chosen generically so that ${\cal{C}}_{\text{Higgs,local}}$ is smooth, uniquely specify the $\phi$ configuration \cite{Cecotti:2010bp}.

For computational purposes it is often convenient to compactify the canonical bundle $K_{S_2}$ into its projectivization $Z=\mathbb{P}({\cal{O}}\oplus K_{S_2})$ \cite{Donagi:2009ra}.  Divisors on $Z$ correspond to pullbacks of divisors in $S_2$ as well as a new divisor $\sigma$ that descends from the hyperplane of the $\mathbb{P}^1$ fiber.  As a hyperplane of $\mathbb{P}^1$ is a single point, $\sigma$ is just a section of the fibration.  The projective coordinates of the $\mathbb{P}^1$ fiber are typically denoted by $U$ and $V$ and are sections of the bundles
\begin{equation}\begin{array}{c|c}\text{Section} & \text{Bundle} \\ \hline 
U & {\cal{O}}(\sigma) \\
V & {\cal{O}}(\sigma+c_1)
\end{array}\end{equation}
where $c_1$ in this section is shorthand for $c_1(S_2)$
\begin{equation}c_1\sim c_1(S_2)\text{ in this section only}\end{equation}
The (compactification of the) spectral cover is now a divisor in $Z$ defined by
\begin{equation}{\cal{C}}_{\text{Higgs,loc}}:\,\,\,b_{0,loc}U^5 + b_{2,loc}U^3V^2 + b_{3,loc}U^2V^3 + b_{4,loc}UV^4 + b_{5,loc}V^5=0\,.
\end{equation}
It is conventional in the local model literature to use $\pi$ for the projection
\begin{equation}\pi:Z\rightarrow S_2\end{equation}
and $p_C$ for the projection of ${\cal{C}}_{\text{Higgs,loc}}$ to $S_2$ induced by $\pi$
\begin{equation}p_C:{\cal{C}}_{\text{Higgs,loc}}\rightarrow S_2\,.\end{equation}
The divisor class of ${\cal{C}}_{\text{Higgs,loc}}$ depends on the classes of the $b_{m,loc}$ in $S_2$.  It is conventional to denote the divisor in $S_2$ defined by $b_{0,loc}=0$ by $\eta$ so that
\begin{equation}{\cal{C}}_{\text{Higgs,loc}}=5\sigma + \pi^*\eta\,.
\end{equation}
It is also common to define $t$ implicitly by
\begin{equation}\eta = 6c_1-t\,.\end{equation}
The configuration of the adjoint scalar $\phi$ is recovered from ${\cal{C}}_{\text{Higgs}}$ as a push-forward
\begin{equation}\phi = p_{C*}{\cal{C}}_{\text{Higgs}}\end{equation}

The spectral data $b_{m,loc}$ of the gauge theory is identified with the data of the local Calabi-Yau geometry in the singular 4-fold $Y_4$ near $S_2$.  More specifically, the $b_{m,loc}$ are just restrictions of the sections $b_m$ appearing in \eqref{SU5tate} to $S_2$
\begin{equation}b_{m,loc} = b_m|_{S_2}\,.
\end{equation}
From this, it is easy to see that $-t$ becomes identified with the normal bundle of $S_2$ inside $B_3$
\begin{equation}-t\leftrightarrow {\cal{O}}(S_2)|_{S_2}\,.
\end{equation}

\subsection{Matter Curves}

The matter content of the 8-dimensional gauge theory originates from the $\mathbf{248}$-dimensional $E_8$ adjoint.  In the presence of the $\phi$ expectation value that breaks $E_8$ to $SU(5)_{\rm GUT}$ this representation splits, with bifundamentals acquiring masses from their couplings to $\phi$.  The decomposition of the $\mathbf{248}$ under $E_8\rightarrow SU(5)_{\rm GUT}\times SU(5)_{\perp}$ is standard
\begin{equation}\mathbf{248}\rightarrow (\mathbf{24},\mathbf{1})\oplus (\mathbf{1},\mathbf{24})\oplus \left[(\mathbf{10},\mathbf{5})\oplus\text{cc}\right]\oplus \left[(\mathbf{\overline{5}},\mathbf{10})\oplus\text{cc}\right]\,.
\end{equation}
Roughly speaking, the $\mathbf{10}$'s get mass from eigenvalues $t_i$ of $\phi$ while the $\mathbf{\overline{5}}$'s get mass from sums of eigenvalues $t_i+t_j$ of $\phi$ with $i\ne j$.  More properly, matter fields in the $\mathbf{10}$ and $\mathbf{\overline{5}}$ of $SU(5)_{\rm GUT}$ should be thought of as sections of suitable bundles on matter curves inside ${\cal{C}}_{\text{Higgs,loc}}$.  The $\mathbf{10}$ matter curve is the intersection of ${\cal{C}}_{\text{Higgs,loc}}$ with the section $\sigma$
\begin{equation}\Sigma_{\mathbf{10}} = {\cal{C}}_{\text{Higgs,loc}}\cdot \sigma\,,\end{equation}
which is just
\begin{equation}U=b_5=0\end{equation}
in equations.  This projects to the curve $b_5=0$ in $S_2$ above which the singularity type enhances to ``$D_5$" in \eqref{SU5tate}.  The $\mathbf{\overline{5}}$ matter curve is the intersection of ${\cal{C}}_{\text{Higgs,loc}}$ with its image under the $\mathbb{Z}_2$ involution $\tau$ on the $\mathbb{P}^1$ fiber that takes $V\rightarrow -V$ less components that are part of the fixed locus of $\tau$ on $Z$.  Homologically this is
\begin{equation}\Sigma_{\mathbf{\overline{5}}}: {\cal{C}}\cdot {\cal{C}} - {\cal{C}}\cdot (3\sigma_{\infty}+\sigma)\end{equation}
where
\begin{equation}\sigma_{\infty}=\sigma+\pi^*c_1\,.\end{equation}
In equations, this curve is described by
\begin{equation} b_0U^4 + b_2U^2V^2 + b_4V^4 = 0\,,\qquad b_3U^2 + b_5 V^2 = 0\,.\end{equation}
This projects to the curve $b_0b_5^2-b_2b_3b_5+b_3^2b_4=0$ inside $S_2$ above which the singularity type enhances to $A_5$ in \eqref{SU5tate}.

As described in \cite{Marsano:2011nn}, the spectral divisor formalism connects to the Higgs bundle spectral cover in a simple way.  Near the surface of $A_4$ singularities, the limiting behavior of the spectral divisor is captured by the Higgs bundle spectral cover and the matter curves inside ${\cal{C}}_{\text{Higgs,loc}}$ correspond to the loci where the spectral divisor intersects the matter surfaces where wrapped M2's corresponding to $\mathbf{10}$'s and $\mathbf{\overline{5}}$'s propagate.  In the main text, this identification is made completely explicit by looking at the proper transform of the spectral divisor in the resolved Calabi-Yau 4-fold $\tilde{Y}_4$.

\subsection{Gauge Bundle}

The spectral cover ${\cal{C}}_{\text{Higgs,loc}}$ nicely describes the field configuration of $\phi$ and provides us with a description of curves where $\mathbf{10}$'s and $\mathbf{\overline{5}}$'s propagate.  It remains, however, to say what the gauge bundle is doing.  In general we can have a nontrivial $SU(5)_{\perp}$ gauge bundle configuration and this is specified by choosing a line bundle $L$ on ${\cal{C}}_{\text{Higgs,loc}}$.  The gauge bundle on $S_2$ is reconstructed from this data by again using the push-forward
\begin{equation}E = p_{C*}L\,.\end{equation}
We require this to be an $SU(5)_{\perp}$ bundle so any choice of $L$ will not do.  We need a particular choice that satisfies
\begin{equation}c_1(p_{C*}L)=0\end{equation}
or equivalently by Grothendieck-Riemann-Roch
\begin{equation}p_{C*}c_1(L)-\frac{1}{2}p_{C*}r=0\,,\label{traceless}\end{equation}
where $r$ is the ramification divisor of the covering $p_C$.  Explicitly, $r$ is the restriction to ${\cal{C}}_{\text{Higgs,loc}}$ of a divisor in the class
\begin{equation}r = \left[{\cal{C}}_{\text{Higgs,loc}}-\sigma-\sigma_{\infty}\right]|_{ {\cal{C}}_{\text{Higgs,loc}}}\,.
\end{equation}
Because of the condition \eqref{traceless}, it is conventional to construct $c_1(L)$ by first writing
\begin{equation}c_1(L) = \frac{1}{2}r + \gamma\end{equation}
and to specify a divisor $\gamma$ that satisfies
\begin{equation}p_{C*}\gamma=0\,.\label{tracelessgamma}\end{equation}
There are not many choices for a ``traceless" class $\gamma$ inside ${\cal{C}}_{\text{Higgs,loc}}$.  When ${\cal{C}}_{\text{Higgs,loc}}$ is smooth and well-behaved as we assume in this paper, all of its divisors are inherited from those of the ambient space $Z$.  The choices are
\begin{equation}{\cal{C}}_{\text{Higgs,loc}}\cdot \sigma\qquad\text{and}\qquad {\cal{C}}_{\text{Higgs,loc}}\cdot \pi^*\Sigma\end{equation}
for curve classes $\Sigma$ in $S_2$.  There is in general only one combination satisfying \eqref{tracelessgamma}.  This leads to a 1-parameter family of $\gamma$'s \cite{Donagi:2009ra}
\begin{equation}\gamma = a\left(5\sigma - \pi^*(\eta-5c_1)\right)\cdot {\cal{C}}_{\text{Higgs,loc}}\end{equation}
for some suitable $a$.  The object $\gamma$ is particularly natural because its restriction to the matter curves $\Sigma_{\mathbf{10}}$ and $\Sigma_{\mathbf{\overline{5}}}$ determines the chiral spectrum.  Explicit computations yield
\begin{equation}\begin{split}n_{\mathbf{10}}-n_{\mathbf{\overline{10}}} &= \gamma\cdot \Sigma_{\mathbf{10}}\\
&= -a\eta\cdot_{S_2}(\eta-c_1) \\ 
n_{\mathbf{\overline{5}}}-n_{\mathbf{5}} &=\Sigma_{\mathbf{\overline{5}}}\\
&=-a\eta\cdot_{S_2} (\eta-c_1)\,,\end{split}\end{equation}
where we have used the fact that
\begin{equation}\sigma\cdot \sigma_{\infty}=0\end{equation}
with
\begin{equation}\sigma_{\infty}\equiv \sigma+c_1\,.\end{equation}
We must be careful about one crucial thing when dealing with $\gamma$.  It is $L$ that must be an integer quantized bundle so $\gamma$ is not necessarily an integer divisor class.  Rather, it satisfies the quantization condition
\begin{equation}\gamma+\frac{1}{2}r\in H^2({\cal{C}}_{\text{Higgs,loc}},\mathbb{Z})\,.\label{localquantization}\end{equation}
This looks reminiscent of the $G$-flux quantization condition \eqref{Gfluxquant} and the similarity is even more striking in light of the fact the net chirality of matter fields is determined by an index involving $\gamma$.  This suggests that $\gamma$ is the object of the local model that corresponds to $G$-flux in the global setting.  The correspondence of $\gamma$ with $G$ has been expected for some time from the perspective of Heterotic/F-theory duality \cite{Curio:1998bva,Andreas:1999ng} and directly in local model building \cite{Donagi:2009ra}.  In the spectral divisor framework \cite{Marsano:2010ix,Marsano:2011nn}, this identification comes about by realizing $G$ as a holomorphic surface inside the spectral divisor which limits to a curve $\gamma$ in ${\cal{C}}_{\text{Higgs,loc}}$ near the surface of $A_4$ singularities.  Indeed, every curve in ${\cal{C}}_{\text{Higgs,loc}}$ can be viewed as a surface constructed from the spectral divisor and hence in $\tilde{Y}_4$.  The Higgs bundle spectral cover should capture intersections of these surfaces as in \cite{Marsano:2011nn} and we provide an explicit dictionary in the main text that allows us to verify this fact.

Returning to the local model, we can use \eqref{localquantization} to properly quantize $\gamma$.  Recalling that
\begin{equation}\begin{split}r &= {\cal{C}}_{\text{Higgs,loc}}\cdot ({\cal{C}}_{\text{Higgs,loc}}-\sigma-\sigma_{\infty})\\
&= {\cal{C}}_{\text{Higgs,loc}}\cdot \left(3\sigma + \eta-c_1\right) \,,\end{split}\end{equation}
we see that $\gamma+\frac{1}{2}r$ is indeed an integral class provided $a$ is an odd half-integer.  We can make this explicit by writing $a=\frac{1}{2}(2n+1)$ for $n\in\mathbb{Z}$ and writing $\gamma$ as
\begin{equation}\gamma = \frac{1}{2}(2n+1)\left(5\sigma-\pi^*(\eta-5c_1)\right)\cdot {\cal{C}}_{\text{Higgs,loc}}\,.\end{equation}

As a final remark we note that $G$ induces a 3-brane charge
\begin{equation}n_{D3,G} = \frac{1}{2}G^2\,.\end{equation}
As we said above, the spectral divisor formalism describes $G$ as a holomorphic surface that is constructed in a particular way from its ``limit" $\gamma$ near the surface of $A_4$ singularities.  Further, we expect that intersections of those surfaces should be computable in the local language as described in more detail in \cite{Marsano:2011nn}.  It is therefore not surprising that the 3-brane charge induced by the $G$-flux corresponding to $\gamma$ can be computed within the local model.  The local expression is
\begin{equation}n_{D3} = -\frac{1}{2}\gamma\cdot_{ {\cal{C}}_{\text{Higgs,loc}}}\gamma = \frac{5}{8}\left(2n+1\right)^2\eta (\eta-5c_1)\,.\end{equation}
It should be noted that the appearance of essentially this expression in computations with spectral covers for $SU(5)_{\perp}$ bundles in Heterotic compactifications on elliptically fibered Calabi-Yau 3-folds (which admit an F-theory dual) provided one of the first pieces of evidence that $\gamma$ was essentially the $G$-flux \cite{Andreas:1999ng}.  The connection between the Heterotic spectral cover and Higgs bundle spectral cover in models with a Heterotic dual is reviewed in the spectral divisor language in \cite{Marsano:2010ix}.


\newpage

\bibliographystyle{JHEP}

\begin{thebibliography}{10}

\bibitem{Esole:2011sm}
M.~Esole and S.-T. Yau, {\it {Small resolutions of SU(5)-models in F-theory}},
  \href{http://xxx.lanl.gov/abs/1107.0733}{{\tt 1107.0733}}.

\bibitem{Blumenhagen:2009yv}
R.~Blumenhagen, T.~W. Grimm, B.~Jurke, and T.~Weigand, {\it {Global F-theory
  GUTs}},  {\em Nucl.Phys.} {\bf B829} (2010) 325--369,
  [\href{http://xxx.lanl.gov/abs/0908.1784}{{\tt 0908.1784}}].

\bibitem{Marsano:2011nn}
J.~Marsano, N.~Saulina, and S.~Schafer-Nameki, {\it {On G-flux, M5 instantons,
  and U(1)s in F-theory}},  \href{http://xxx.lanl.gov/abs/1107.1718}{{\tt
  1107.1718}}.

\bibitem{Donagi:2009ra}
R.~Donagi and M.~Wijnholt, {\it {Higgs Bundles and UV Completion in F-Theory}},
   \href{http://xxx.lanl.gov/abs/0904.1218}{{\tt 0904.1218}}.

\bibitem{Andreas:2009uf}
B.~Andreas and G.~Curio, {\it {From Local to Global in F-Theory Model
  Building}},  {\em J.Geom.Phys.} {\bf 60} (2010) 1089--1102,
  [\href{http://xxx.lanl.gov/abs/0902.4143}{{\tt 0902.4143}}].

\bibitem{Marsano:2009ym}
J.~Marsano, N.~Saulina, and S.~Schafer-Nameki, {\it {F-theory Compactifications
  for Supersymmetric GUTs}},  {\em JHEP} {\bf 08} (2009) 030,
  [\href{http://xxx.lanl.gov/abs/0904.3932}{{\tt 0904.3932}}].

\bibitem{Collinucci:2009uh}
A.~Collinucci, {\it {New F-theory lifts. II. Permutation orientifolds and
  enhanced singularities}},  {\em JHEP} {\bf 1004} (2010) 076,
  [\href{http://xxx.lanl.gov/abs/0906.0003}{{\tt 0906.0003}}].

\bibitem{Blumenhagen:2009up}
R.~Blumenhagen, T.~W. Grimm, B.~Jurke, and T.~Weigand, {\it {F-theory uplifts
  and GUTs}},  {\em JHEP} {\bf 0909} (2009) 053,
  [\href{http://xxx.lanl.gov/abs/0906.0013}{{\tt 0906.0013}}].

\bibitem{Marsano:2009gv}
J.~Marsano, N.~Saulina, and S.~Schafer-Nameki, {\it {Monodromies, Fluxes, and
  Compact Three-Generation F-theory GUTs}},  {\em JHEP} {\bf 08} (2009) 046,
  [\href{http://xxx.lanl.gov/abs/0906.4672}{{\tt 0906.4672}}].

\bibitem{Marsano:2009wr}
J.~Marsano, N.~Saulina, and S.~Schafer-Nameki, {\it {Compact F-theory GUTs with
  $U(1)_{PQ}$}},  {\em JHEP} {\bf 04} (2010) 095,
  [\href{http://xxx.lanl.gov/abs/0912.0272}{{\tt 0912.0272}}].

\bibitem{Grimm:2009yu}
T.~W. Grimm, S.~Krause, and T.~Weigand, {\it {F-Theory GUT Vacua on Compact
  Calabi-Yau Fourfolds}},  {\em JHEP} {\bf 1007} (2010) 037,
  [\href{http://xxx.lanl.gov/abs/0912.3524}{{\tt 0912.3524}}].

\bibitem{Cvetic:2010rq}
M.~Cvetic, I.~Garcia-Etxebarria, and J.~Halverson, {\it {Global F-theory
  Models: Instantons and Gauge Dynamics}},  {\em JHEP} {\bf 1101} (2011) 073,
  [\href{http://xxx.lanl.gov/abs/1003.5337}{{\tt 1003.5337}}].

\bibitem{Chen:2010ts}
C.-M. Chen, J.~Knapp, M.~Kreuzer, and C.~Mayrhofer, {\it {Global SO(10)
  F-theory GUTs}},  {\em JHEP} {\bf 1010} (2010) 057,
  [\href{http://xxx.lanl.gov/abs/1005.5735}{{\tt 1005.5735}}].

\bibitem{Chen:2010tp}
C.-M. Chen and Y.-C. Chung, {\it {Flipped SU(5) GUTs from $E_8$ Singularities
  in F-theory}},  {\em JHEP} {\bf 1103} (2011) 049,
  [\href{http://xxx.lanl.gov/abs/1005.5728}{{\tt 1005.5728}}].

\bibitem{Chung:2010bn}
Y.-C. Chung, {\it {On Global Flipped SU(5) GUTs in F-theory}},
  \href{http://xxx.lanl.gov/abs/1008.2506}{{\tt 1008.2506}}.

\bibitem{Chen:2010tg}
C.-M. Chen and Y.-C. Chung, {\it {On F-theory $E_6$ GUTs}},  {\em JHEP} {\bf
  1103} (2011) 129, [\href{http://xxx.lanl.gov/abs/1010.5536}{{\tt
  1010.5536}}].

\bibitem{Knapp:2011wk}
J.~Knapp, M.~Kreuzer, C.~Mayrhofer, and N.-O. Walliser, {\it {Toric
  Construction of Global F-Theory GUTs}},
  \href{http://xxx.lanl.gov/abs/1101.4908}{{\tt 1101.4908}}.

\bibitem{Knapp:2011ip}
J.~Knapp and M.~Kreuzer, {\it {Toric Methods in F-theory Model Building}},
  \href{http://xxx.lanl.gov/abs/1103.3358}{{\tt 1103.3358}}.

\bibitem{Vafa:1996xn}
C.~Vafa, {\it {Evidence for F-Theory}},  {\em Nucl. Phys.} {\bf B469} (1996)
  403--418, [\href{http://xxx.lanl.gov/abs/hep-th/9602022}{{\tt
  hep-th/9602022}}].

\bibitem{Morrison:1996na}
D.~R. Morrison and C.~Vafa, {\it {Compactifications of F-Theory on Calabi--Yau
  Threefolds -- I}},  {\em Nucl. Phys.} {\bf B473} (1996) 74--92,
  [\href{http://xxx.lanl.gov/abs/hep-th/9602114}{{\tt hep-th/9602114}}].

\bibitem{Morrison:1996pp}
D.~R. Morrison and C.~Vafa, {\it {Compactifications of F-Theory on Calabi--Yau
  Threefolds -- II}},  {\em Nucl. Phys.} {\bf B476} (1996) 437--469,
  [\href{http://xxx.lanl.gov/abs/hep-th/9603161}{{\tt hep-th/9603161}}].

\bibitem{Friedman:1997yq}
R.~Friedman, J.~Morgan, and E.~Witten, {\it {Vector bundles and F theory}},
  {\em Commun.Math.Phys.} {\bf 187} (1997) 679--743,
  [\href{http://xxx.lanl.gov/abs/hep-th/9701162}{{\tt hep-th/9701162}}].

\bibitem{Bershadsky:1996nh}
M.~Bershadsky, K.~A. Intriligator, S.~Kachru, D.~R. Morrison, V.~Sadov, {\em
  et.~al.}, {\it {Geometric singularities and enhanced gauge symmetries}},
  {\em Nucl.Phys.} {\bf B481} (1996) 215--252,
  [\href{http://xxx.lanl.gov/abs/hep-th/9605200}{{\tt hep-th/9605200}}].

\bibitem{Bershadsky:1997zs}
M.~Bershadsky, A.~Johansen, T.~Pantev, and V.~Sadov, {\it {On four-dimensional
  compactifications of F theory}},  {\em Nucl.Phys.} {\bf B505} (1997)
  165--201, [\href{http://xxx.lanl.gov/abs/hep-th/9701165}{{\tt
  hep-th/9701165}}].

\bibitem{Curio:1998bva}
G.~Curio and R.~Y. Donagi, {\it {Moduli in N=1 heterotic / F theory duality}},
  {\em Nucl.Phys.} {\bf B518} (1998) 603--631,
  [\href{http://xxx.lanl.gov/abs/hep-th/9801057}{{\tt hep-th/9801057}}].

\bibitem{Donagi:2008ca}
R.~Donagi and M.~Wijnholt, {\it {Model Building with F-Theory}},
  \href{http://xxx.lanl.gov/abs/0802.2969}{{\tt 0802.2969}}.

\bibitem{Hayashi:2008ba}
H.~Hayashi, R.~Tatar, Y.~Toda, T.~Watari, and M.~Yamazaki, {\it {New Aspects of
  Heterotic--F Theory Duality}},  {\em Nucl. Phys.} {\bf B806} (2009) 224--299,
  [\href{http://xxx.lanl.gov/abs/0805.1057}{{\tt 0805.1057}}].

\bibitem{Donagi:2008kj}
R.~Donagi and M.~Wijnholt, {\it {Breaking GUT Groups in F-Theory}},
  \href{http://xxx.lanl.gov/abs/0808.2223}{{\tt 0808.2223}}.

\bibitem{Hayashi:2009ge}
H.~Hayashi, T.~Kawano, R.~Tatar, and T.~Watari, {\it {Codimension-3
  Singularities and Yukawa Couplings in F-theory}},  {\em Nucl.Phys.} {\bf
  B823} (2009) 47--115, [\href{http://xxx.lanl.gov/abs/0901.4941}{{\tt
  0901.4941}}].

\bibitem{Tatar:2009jk}
R.~Tatar, Y.~Tsuchiya, and T.~Watari, {\it {Right-handed Neutrinos in F-theory
  Compactifications}},  {\em Nucl.Phys.} {\bf B823} (2009) 1--46,
  [\href{http://xxx.lanl.gov/abs/0905.2289}{{\tt 0905.2289}}].

\bibitem{Hayashi:2009bt}
H.~Hayashi, T.~Kawano, Y.~Tsuchiya, and T.~Watari, {\it {Flavor Structure in
  F-theory Compactifications}},  \href{http://xxx.lanl.gov/abs/0910.2762}{{\tt
  0910.2762}}.

\bibitem{Braun:2011zm}
A.~P. Braun, A.~Collinucci, and R.~Valandro, {\it {G-flux in F-theory and
  algebraic cycles}},  \href{http://xxx.lanl.gov/abs/1107.5337}{{\tt
  1107.5337}}.

\bibitem{Marsano:2010ix}
J.~Marsano, N.~Saulina, and S.~Schafer-Nameki, {\it {A Note on G-Fluxes for
  F-theory Model Building}},  {\em JHEP} {\bf 11} (2010) 088,
  [\href{http://xxx.lanl.gov/abs/1006.0483}{{\tt 1006.0483}}].

\bibitem{Andreas:1999ng}
B.~Andreas and G.~Curio, {\it {On discrete twist and four flux in N=1 heterotic
  / F theory compactifications}},  {\em Adv.Theor.Math.Phys.} {\bf 3} (1999)
  1325--1413, [\href{http://xxx.lanl.gov/abs/hep-th/9908193}{{\tt
  hep-th/9908193}}].

\bibitem{Dolan:2011iu}
M.~J. Dolan, J.~Marsano, N.~Saulina, and S.~Schafer-Nameki, {\it {F-theory GUTs
  with U(1) Symmetries: Generalities and Survey}},
  \href{http://xxx.lanl.gov/abs/1102.0290}{{\tt 1102.0290}}.

\bibitem{Hayashi:2010zp}
H.~Hayashi, T.~Kawano, Y.~Tsuchiya, and T.~Watari, {\it {More on Dimension-4
  Proton Decay Problem in F-theory -- Spectral Surface, Discriminant Locus and
  Monodromy}},  {\em Nucl.Phys.} {\bf B840} (2010) 304--348,
  [\href{http://xxx.lanl.gov/abs/1004.3870}{{\tt 1004.3870}}].

\bibitem{Grimm:2010ez}
T.~W. Grimm and T.~Weigand, {\it {On Abelian Gauge Symmetries and Proton Decay
  in Global F-theory GUTs}},  {\em Phys.Rev.} {\bf D82} (2010) 086009,
  [\href{http://xxx.lanl.gov/abs/1006.0226}{{\tt 1006.0226}}].

\bibitem{Intriligator:1997pq}
  K.~A.~Intriligator, D.~R.~Morrison and N.~Seiberg,
  ``Five-dimensional supersymmetric gauge theories and degenerations of
  Calabi-Yau spaces,''
  Nucl.\ Phys.\  B {\bf 497}, 56 (1997)
  [arXiv:hep-th/9702198].

\bibitem{Morrison:2011mb}
  D.~R.~Morrison, W.~Taylor,
  ``Matter and singularities,''
    [arXiv:1106.3563 [hep-th]].

\bibitem{Witten:1996md}
E.~Witten, {\it {On flux quantization in M-theory and the effective action}},
  {\em J. Geom. Phys.} {\bf 22} (1997) 1--13,
  [\href{http://xxx.lanl.gov/abs/hep-th/9609122}{{\tt hep-th/9609122}}].

\bibitem{Sethi:1996es}
S.~Sethi, C.~Vafa, and E.~Witten, {\it {Constraints on low-dimensional string
  compactifications}},  {\em Nucl. Phys.} {\bf B480} (1996) 213--224,
  [\href{http://xxx.lanl.gov/abs/hep-th/9606122}{{\tt hep-th/9606122}}].

\bibitem{Collinucci:2010gz}
A.~Collinucci and R.~Savelli, {\it {On Flux Quantization in F-Theory}},
  \href{http://xxx.lanl.gov/abs/1011.6388}{{\tt 1011.6388}}.

\bibitem{Katz:2011qp}
S.~Katz, D.~R. Morrison, S.~Schafer-Nameki, and J.~Sully, {\it {Tate's
  algorithm and F-theory}},  \href{http://xxx.lanl.gov/abs/1106.3854}{{\tt
  1106.3854}}.

\bibitem{Marsano:2008jq}
J.~Marsano, N.~Saulina, and S.~Schafer-Nameki, {\it {Gauge Mediation in
  F-Theory GUT Models}},  {\em Phys. Rev.} {\bf D80} (2009) 046006,
  [\href{http://xxx.lanl.gov/abs/0808.1571}{{\tt 0808.1571}}].

\bibitem{Heckman:2008qt}
J.~J. Heckman and C.~Vafa, {\it {F-theory, GUTs, and the Weak Scale}},  {\em
  JHEP} {\bf 0909} (2009) 079, [\href{http://xxx.lanl.gov/abs/0809.1098}{{\tt
  0809.1098}}].

\bibitem{aluffi-blowups}
P.~{Aluffi}, {\it {Chern classes of blow-ups}},  {\em Mathematical Proceedings
  of the Cambridge Philosophical Society} {\bf 148} (2010) 227--242,
  [\href{http://xxx.lanl.gov/abs/0809.2425}{{\tt 0809.2425}}].

\bibitem{Fulton}
W.~Fulton, {\em Intersection theory}.
\newblock Springer-Verlag, 1984.

\bibitem{Katz:1996xe}
S.~H. Katz and C.~Vafa, {\it {Matter from geometry}},  {\em Nucl.Phys.} {\bf
  B497} (1997) 146--154, [\href{http://xxx.lanl.gov/abs/hep-th/9606086}{{\tt
  hep-th/9606086}}].

\bibitem{Weigand:2010wm}
T.~Weigand, {\it {Lectures on F-theory compactifications and model building}},
  {\em Class.Quant.Grav.} {\bf 27} (2010) 214004,
  [\href{http://xxx.lanl.gov/abs/1009.3497}{{\tt 1009.3497}}].

\bibitem{Cecotti:2010bp}
S.~Cecotti, C.~Cordova, J.~J. Heckman, and C.~Vafa, {\it {T-Branes and
  Monodromy}},  \href{http://xxx.lanl.gov/abs/1010.5780}{{\tt 1010.5780}}.

\end{thebibliography}
\renewcommand{\refname}{Bibliography}

\providecommand{\href}[2]{#2}\begingroup\raggedright\endgroup

\end{document}